\documentclass{emulateapj}
\usepackage{natbib}
\usepackage{graphicx}
\usepackage{graphics,epsfig}
\usepackage{tabularx}
\usepackage{adjustbox,lipsum}
\usepackage{subfigure}
\usepackage{amsmath}
\usepackage{float}
\usepackage{courier}
\usepackage{xcolor}
\usepackage[normalem]{ulem}
\usepackage{colortbl}
\usepackage[colorlinks=true,
                     linkcolor=red,
                     citecolor=blue]{hyperref}
\definecolor{green(html/cssgreen)}{rgb}{0.0, 0.65, 0.31}
\definecolor{green(html/cssgreen)}{rgb}{0.0, 0.5, 0.0}

\usepackage{rotating}

\singlespace

\newcommand{\ba}{\begin{eqnarray}}
\newcommand{\ea}{\end{eqnarray}}

\newcommand{\gsim}   {\mbox{$_>\atop^{\sim}$}}
\newcommand{\lt}     {\mbox{$<$}}
\newcommand{\gt}     {\mbox{$>$}}

\newcommand{\freqa}   {$98\,$}
\newcommand{\freqb}   {$150\,$}

\newcommand  \beq    {\begin{equation}}

\newcommand  \eeq    {\end{equation}}
\newcommand  \gtsim  {\lower.5ex\hbox{$\; \buildrel > \over \sim \;$}} 
\newcommand  \ltsim  {\lower.5ex\hbox{$\; \buildrel < \over \sim \;$}}
\newcommand{\lap} {$\stackrel{<}{_\sim}$}

\begin{document}

\title{The Atacama Cosmology Telescope: A Measurement of the Cosmic Microwave Background Power Spectra at \freqa and \freqb GHz}


\author{
Steve~K.~Choi\altaffilmark{1,2,3},
Matthew~Hasselfield\altaffilmark{4,5,6},
Shuay-Pwu~Patty~Ho\altaffilmark{3},
Brian~Koopman\altaffilmark{7},
Marius~Lungu\altaffilmark{3,8},
Maximilian~H.~Abitbol\altaffilmark{9},
Graeme~E.~Addison\altaffilmark{10},
Peter~A.~R.~Ade\altaffilmark{11},
Simone~Aiola\altaffilmark{4,3},
David~Alonso\altaffilmark{9}
% ...
}
\altaffiltext{1}{Department of Physics, Cornell University, Ithaca, NY, USA 14853}
\altaffiltext{2}{Department of Astronomy, Cornell University, Ithaca, NY 14853, USA}
\altaffiltext{3}{Joseph Henry Laboratories of Physics, Jadwin Hall,
Princeton University, Princeton, NJ, USA 08544}
\altaffiltext{4}{Center for Computational Astrophysics, Flatiron Institute, 162 5th Avenue, New York, NY 10010 USA}
\altaffiltext{5}{Department of Astronomy and Astrophysics, The Pennsylvania State University, University Park, PA 16802, USA}
\altaffiltext{6}{Institute for Gravitation and the Cosmos, The Pennsylvania State University, University Park, PA 16802, USA}
\altaffiltext{7}{Department of Physics, Yale University, 217 Prospect St, New Haven, CT 06511}
\altaffiltext{8}{Department of Physics and Astronomy, University of
Pennsylvania, 209 South 33rd Street, Philadelphia, PA, USA 19104}
\altaffiltext{9}{Department of Physics, University of Oxford, Keble Road, Oxford, UK OX1 3RH}
\altaffiltext{10}{Dept. of Physics and Astronomy, The Johns Hopkins University, 3400 N. Charles St., Baltimore, MD, USA 21218-2686}
\altaffiltext{11}{School of Physics and Astronomy, Cardiff University, The Parade, 
Cardiff, Wales, UK CF24 3AA}
% ...


\date{\today}

\begin{abstract}
We present the temperature and polarization angular power spectra of the CMB measured by the Atacama Cosmology Telescope (ACT) from 5400 deg$^2$ of the 2013--2016 survey, which covers $>$15000 deg$^2$ at 98 and 150 GHz. For this analysis we adopt a blinding strategy to help avoid confirmation bias and, related to this, show numerous checks for systematic error done before unblinding. Using the likelihood for the cosmological analysis we constrain secondary sources of anisotropy and foreground emission, and derive a ``CMB-only" spectrum that extends to $\ell=4000$. At large angular scales, foreground emission at \freqb\,GHz is $\sim$1\% of TT and EE  within our selected regions and consistent with that found by {\sl Planck}. Using the same likelihood, we obtain the cosmological parameters for $\Lambda$CDM for the ACT data alone with a prior on the optical depth of $\tau=0.065\pm0.015$. $\Lambda$CDM is a good fit. The best-fit model has a reduced $\chi^2$ of 1.07 ($\rm{PTE}=0.07$) with $H_0=67.9\pm1.5$ km/s/Mpc. 
We show that the lensing BB signal is consistent with $\Lambda$CDM and limit the celestial EB polarization angle to $\psi_P =-0.07^{\circ}\pm0.09^{\circ}$. We directly cross correlate ACT with {\sl Planck} and observe generally good agreement but with some discrepancies in TE. All data on which this analysis is based will be publicly released.

\bigskip

\end{abstract}

\maketitle

\section{Introduction}
\label{sec:act_intro}
\setcounter{footnote}{0}
The Atacama Cosmology Telescope (ACT), described in \citet{fowler/etal:2007} and \citet{thornton/2016}, observes the mm-wave sky from northern Chile with arcminute resolution.  Its primary goal is to make maps of the CMB temperature anisotropy and polarization at angular scales and sensitivities that complement those of the {\sl WMAP} and {\sl Planck} satellites. This paper and a companion paper, \citet{aiola/etal:2020} (hereafter A20), present results from ACT's 2013--2016 nighttime sky maps.

The six-parameter $\Lambda$CDM standard model of cosmology is now well established, yet there remain ``tensions" both within the CMB sector \citep[e.g.,][]{bennett/etal:2014, addison/etal:2016, henning/etal:2018} and between the CMB and other data sets, most notably with measurements of $H_0$ at 
z\lap 1 [e.g., \citet{riess/etal:2019, wong/etal:2019, shajib/etal:2019}, although not significantly with \citet{freedman/etal:2019}. See also e.g., \citet{knox/millea:2019}]. Here and in A20 we present a significant step toward addressing the tensions with a new precise measurement 
with much of the weight of the parameter determination coming from the CMB's 
polarization and its correlation with temperature as opposed to its temperature anisotropy. Any residual experimental systematic errors in the ACT data set, apart from an overall calibration factor, are independent of those in {\sl WMAP} and 
{\sl Planck}. Thus the data set, on its own and in combination with {\sl WMAP} (or {\sl Planck} at $\ell < 800$), provides an important independent assessment of the standard model. 


This paper covers the power spectra from the 2013--2016 nighttime sky maps, covariance matrices for the spectra, data consistency and null checks, the level of foreground emission in the maps, the likelihood for determining the cosmological parameters, the ACT-only $\Lambda$CDM cosmological parameters, and finally the coadded foreground-cleaned CMB power spectra. A20 
describes the data selection, maps, and presents more extensive constraints on the cosmological parameters derived from the spectra and likelihood presented here in combination with {\sl WMAP} and {\sl Planck}.

This paper and A20 are part of ACT's fourth data release, DR4.
Previous releases\footnote{All data are released through NASA's LAMBDA site. https://lambda.gsfc.nasa.gov/product/act/} are DR1, which covered a southern region (centered on $\rm{RA}=60^\circ$ $\rm{dec}=-52.7^\circ$) in 2008 at 148 GHz \citep[e.g.,][]{dunner/etal:2013, dunkley/etal:2011}; DR2, which covered the south and the ``SDSS Stripe 82" equatorial region in 2008--2010, and added 217 GHz and 277 GHz data \citep[e.g.,][]{das/etal:2011,sievers/etal:2013,gralla/etal:2019}; and DR3,
which covered a number of regions on the equator in 2013--14 at 150 GHz \citep[e.g.,][]{louis/2017}, hereafter L17, and \citet{naess/2014}). DR4 includes DR3 as a subset. Both DR1 and DR2 used data from the unpolarized millimeter bolometric array camera (MBAC) \citep{swetz/etal:2011} while DR3 and DR4 are based on ACTPol, a polarization sensitive bolometric receiver \citep{thornton/2016}.

The methods for analyzing CMB data are now quite mature. Nevertheless, the analysis presented here entails a considerable jump in complexity over what we have reported in the past. The data comprise a heterogeneous set of observations from eleven regions of the sky with different sizes and depths. Some of the regions are observed over multiple years under different configurations of the receiver and at different elevations. Over 11 TB of raw data are projected into $\gsim 2\times10^8$ map pixels using a maximum-likelihood mapmaking approach. Roughly 80\% of the maps for DR4 are reduced to two sets of ten power spectra that enter the likelihood along with an accounting for systematic errors, foreground emission, and the correlations between spectra.

%
%
%
%


We begin in Section~\ref{sec:new} with an overview of changes in the analysis since DR3. These are expanded on throughout the rest of the paper.
In Section~\ref{sec:inst}, we briefly describe the instrument.
Sections~\ref{sec:obs} and \ref{sec:data_select} describe the observations and data selection for the cosmological analysis. 
The covariance matrix and computation of the coadded CMB power 
spectra are outlined in Section~\ref{sec:pipeline}. In Section~\ref{sec:cal_etc} we present the calibration, instrument polarization angle, and mapmaking transfer function. Following that we consider checks for different types of systematic error  
in Sections~\ref{sec:consistency}, \ref{sec:null_tests}, and \ref{sec:misc_syst}. In Section~\ref{sec:fg_diff} we assess the level of diffuse Galactic foreground emission in the maps after which, 
in Section~\ref{sec:likelihood}, we present the likelihood function calculation and results on the cosmological, secondary foreground, and ``nuisance" parameters. We expand on these and other results in  
Section~\ref{sec:results} and conclude in Section~\ref{sec:conclude}.

\section{Changes in the analysis since DR3}
\label{sec:new}

We have made significant improvements in analysis methodology and algorithms since the last ACT data release. 
Although this analysis builds on that in L17, almost all the software has been rewritten. In spite of these significant changes, the new spectra are consistent with those in L17\footnote{This statement is based on the $\chi^2$ of the simple difference between spectra using L17 and its uncertainties as the ``data" and this spectrum as the ``model." We have not yet done a map-level comparison although Figure~\ref{fig:CLlcdm} compares cosmological parameters between L17 and this paper.} but with more data the error bars are now typically 2.3 times smaller. Updates to the data selection and mapmaking are discussed in A20. We list complementary improvements below.
%
%

\subsection{Blinding}

For DR4 we adopted a blinding strategy to help shield us from confirmation bias on the cosmological parameters, especially $H_0$. Before ``opening the box" we ($a$) required that the maps and spectra pass a series of null and consistency tests described later in this paper; ($b$) did not compare to cosmological models of the data; ($c$) compared the EB\footnote{We use the notation ``XY" to refer to power spectra of the form
${\cal D}^{XY}_\ell = \ell(\ell+1)C^{XY}_\ell/2\pi$ where $X$ and $Y$ denote $T$, $E$, or $B$, the temperature, E-mode, and B-mode spectra respectively; and $C_\ell$ is the angular power spectrum for spherical harmonic index $\ell$. For E and B modes we use the conventions in \citet{kamionkowski/etal:1997,zaldarriaga/seljak:1997}. Where it is necessary to indicate an average over a band of $\ell$s we use $C_b$.} spectrum to null only after applying all known instrumental effects that could rotate the polarization angle; ($d$) assessed the dust and synchrotron contamination through cross-correlation with the {\sl Planck} 353 GHz and {\sl WMAP} K-band maps to establish expectations for foreground contamination; ($e$) selected the range $350.5\leq\ell\leq7525.5$;\footnote{We report band centers. The band boundaries depend on $\ell$ and range from $\ell=326$ to $\ell=7925$. See Table~\ref{tab:specs}. } and (f) computed only parameter differences from different partitions of the data with the parameter likelihood, for example we computed the parameters for \freqa\,GHz minus the parameters for \freqb\,GHz spectra, etc. After following this sequence, and running a full set of simulations for the power spectrum and likelihood, we extracted the cosmological parameters and compared the spectra to the best-fit model. One of the benefits of the blinding strategy was that it imposed discipline on assessing potential systematic errors before looking at the results. Nevertheless, the post-unblinding analysis resulted in an additional cut of the TT power spectrum below $\ell<600$, 
as we discuss in Section\,\ref{sec:misc_syst}, and a reassessment of the temperature to polarization leakage as discuss in A20 and below.


\subsection{Improved planet mapping, beam modeling, and window functions}
\label{sec:planet_mapping}
The pipeline for mapping the planets is new, resulting in cleaner maps for assessing the beam profiles. Season average radial profiles,
shown in Section~\ref{sec:inst}, extend to roughly $-40$ dB of the peak or 40 dBi (35 dBi) at \freqa GHz (\freqb GHz).  The primary improvement comes from how the atmospheric contribution to the planet map is assessed and subtracted. Multiple atmospheric eigenmodes are fitted to data in a region that does not contain the planet, in contrast to fitting a single common mode as previously done, and then subtracted from the region containing the planet. In addition, we now inverse variance weight the detectors. 

Our improved beam mapping resulted in a more detailed understanding of the temperature to polarization leakage than given in L17. For $\ell<4000$ as much as 0.2\% of the temperature signal can leak into the polarization, and cause non-celestial correlation. The effect is described in A20 and below.

The pipeline for producing the beam window functions has been substantially rewritten and enhanced in multiple ways but is still based on \citet{hasselfield/etal:2013}. Our modeling now includes a scattering term from the primary reflector surface deformations. 

\subsection{New power spectrum code and covariance matrix}
Our power spectrum code (see Section~\ref{sec:pipeline}) is based on the curved sky as opposed to the flat sky, the spatial window functions are now customized and not necessarily rectangular, cross-linking is assessed, the source mask is apodized, and the mapmaking transfer function is accounted for,
whereas in L17 it was negligibly different from unity.\footnote{There are four different transfer functions due to: the mapmaking pipeline software (here and in Section~\ref{sec:map_tfunc}), the Fourier-space filtering (Section~\ref{sec:fspace_tfunc}), the beam window function (A20), and the pixel window function. } The maps are calibrated to {\sl Planck} using $600<\ell <1800$ in contrast to L17 which used $350<\ell <2000$ for one array (PA2, see Section~\ref{sec:inst}) followed by calibrating the second array (PA1) to it. 

The covariance matrix includes noise simulations for the 
diagonal, pseudo-diagonal,\footnote{These are the diagonal terms in sub blocks of the matrix.} and ``diagonal-plus-one'' terms, as well as analytic terms for the lensing, super-sample lensing variance, and Poisson point sources (see Section~\ref{sec:act_covmat}).

\section{The Instrument}
\label{sec:inst}
ACT is a 6\,m off-axis aplanatic Gregorian telescope that scans in azimuth as the sky drifts through the field of view. There have been three generations of receivers, MBAC \citep{swetz/etal:2011} which observed at 150, 220, and 277 GHz,
ACT's first polarization-sensitive receiver, ACTPol [\citet{thornton/2016}, see also Appendix~\ref{appen:fgcalcs} for updated band centers]  which observed at \freqa\,GHz and 
\freqb\,GHz, and the Advanced ACTPol (AdvACT) receiver \citep{henderson/etal:2016,ho/2017,choi/etal:2018,li/etal:2018} which is currently configured with detector arrays at 30, 40, 97, 149, and 225 GHz. This paper presents results from ACTPol. 

The instrument characteristics are summarized in Table~\ref{tab:inst}. There are three separate polarized arrays (PAs) of NIST-fabricated feedhorn-coupled MoCu TES detectors \citep{grace/etal:2014,datta/etal:2014,ho/etal:2016}, PA1, PA2, PA3, each in an ``optics tube" with its own set of filters and lenses. All operate near 100~mK. PA3, added in the 2015 season (s15), is dichroic, which means it simultaneously measures \freqa and \freqb GHz polarizations at the output of one feed horn. In the analysis we account for changes in calibration, pointing, time constants, and beamwidth over time. As such, the table reports typical characteristics.

\begin{table*}[tp!]
\caption{Instrument Characteristics}
\vspace{-0.1in}
\begin{center}
\begin{tabular}{|l|c|c|c|c|}
\hline
Observing season & s13 & s14 & s15 & s16 \\
\hline
PA1 (149.6 GHz)$^a$ &  &  &  &     $\cdots$   \\
Array sensitivity$^b$ ($\mu$Ks$^{1/2}$)& 15 & 23  &23 & \\
Median time const. ($f_{\rm 3dB}$) $^c$ (ms, Hz) & 2.1 (76)  & 3.9 (41) & 5.4 (29)&     $\cdots$   \\
Main beam solid angle$^d$ (nsr)& 202 & 199  &197&\\
$\theta_{\rm{FWHM}}$ $^e$ (arcmin)& 1.35 & 1.35  & 1.35&\\
Aspect ratio$^f$ & 1.04 & 1.03  & 1.04&\\
\hline
PA2 (149.9 GHz)$^a$  &        $\cdots$           &  &   &      \\
Array sensitivity ($\mu$Ks$^{1/2}$) && 13 & 16& 16\\
Median time const. ($f_{\rm 3dB}$) (ms, Hz) &  & 1.9 (84) & 2.3 (69) & 1.7 (94)  \\
Main beam solid angle (nsr)& &183 & 188  &185\\
$\theta_{\rm{FWHM}}$  (arcmin)& & 1.32 & 1.33  & 1.33\\
Aspect ratio && 1.01 & 1.03  & 1.02\\
\hline
PA3 (97.9 GHz)$^{a}$ &   $\cdots$    &  $\cdots$ &  &   \\
Array sensitivity ($\mu$Ks$^{1/2}$) &&& 14 & 14\\
Median time const. ($f_{\rm 3dB}$) (ms, Hz) &  &  & 1.1 (140) & 0.98 (160)   \\
Main beam solid angle (nsr)& & & 504  &476\\
$\theta_{\rm{FWHM}}$ (arcmin) &  & & 2.06&2.06 \\
Aspect ratio & && 1.18  & 1.12\\
PA3 (147.6 GHz)$^a$ &     $\cdots$&     $\cdots$    &   &   \\
Array sensitivity ($\mu$Ks$^{1/2}$) &&& 20 & 20\\
Median time const. ($f_{\rm 3dB}$) (ms, Hz) & && 1.2 (130) & 1.1 (140)   \\
Main beam solid angle (nsr)& && 270  &238\\
$\theta_{\rm{FWHM}}$ (arcmin)&  &   & 1.49& 1.46 \\
Aspect ratio & &  & 1.08&1.08\\
\hline
\end{tabular}
\end{center}
\label{tab:inst}
{\small Notes: $a$) The effective frequencies are for a CMB source. The uncertainty is 2.4\,GHz as discussed in Appendix~\ref{appen:fgcalcs}. The total number of detectors, regardless of whether they are operating or dark, is 1024 for each of PA1, PA2, and PA3. Each feedhorn in PA1 and PA2 couples to two detectors while each in PA3 couples to four detectors. $b$) All sensitivities are NET on the sky relative to the CMB for a precipitable water vapor (PWV) of 
$W_v=1$~mm. They are derived from the time series during a planet calibration, and rounded to the nearest $\mu$K. In a given year the sensitivities may be combined in inverse quadrature. For example, the net sensitivity on the sky in 2015 was $8.6\,\mu$Ks$^{1/2}$. For comparison, the combined measured white noise levels for the {\sl Planck} satellite HFI instrument in the 100 and 143 GHz bands is 40 and 17.3 $\mu$Ks$^{1/2}$ relative to the CMB, or 15.9 $\mu$Ks$^{1/2}$ with frequencies combined  \citep{planck_hfi:2016}. 
These arrays were replaced by the even more sensitive AdvACT's PA4, PA5, and PA6 for observations in 2017/18/19. 
$c$) The time constants, $\tau$, depend on loading and base temperature. We report them for $0.5<W_v<1$\,mm. For nominal observations 1 Hz corresponds roughly to $\ell=400$, thus 
$f_{\rm 3dB}=35$~Hz maps to $\ell=14000$.
$d$) ``Instantaneous'' solid angles rounded to the nearest nsr. These are increased by jitter in the pointing.
$e$) Full width at half maximum.
$f$) The aspect ratio is defined as the ratio of the maximum to minimum $\theta_{\rm{FWHM}}$ as measured in perpendicular directions.
}
\end{table*}

One of the most challenging aspects of characterizing the instrument is quantifying the optical response or ``beam." At the precision of the current generation of experiments, unaccounted for solid angle near the main beam can have a noticeable effect on the shape of the beam window function. Our primary source for measuring the beam is Uranus. Its effective antenna temperature is 120--180 mK depending on the orbital parameters \citep{weiland/etal:2011,hasselfield/etal:2013, planck_planets:2017}.
Saturn is also useful but, due to its brightness of 3--6 K (antenna temperature), it has the potential to saturate detectors depending on their saturation powers.\footnote{Detector non-linearity possibly contributed to the $\sim 7$\% discrepancy between the {\sl Planck} measurement of Saturn's temperature and ACT's DR2 measurement \citep{planck_planets:2017, hasselfield/etal:2013}. }

\begin{figure}[tp!]
\centering
\includegraphics[width=0.5\textwidth]{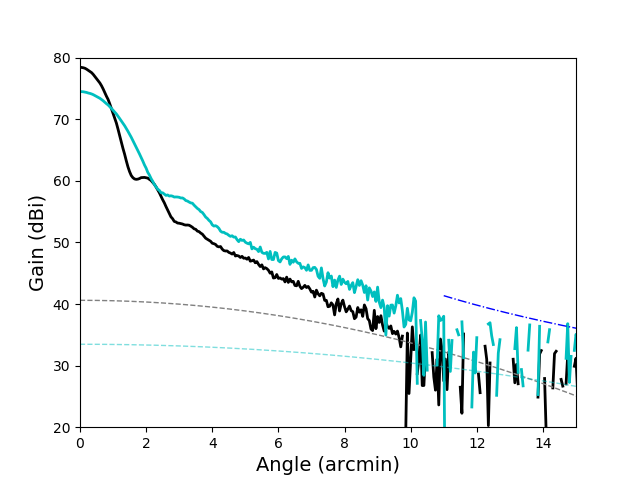}
\caption{\small The average \freqa GHz (cyan) and \freqb GHz (black) beam profiles in ``gain above isotropic'' ($4\pi/\Omega_B$). The forward gains are 74.5 and 78.4 dBi respectively. The two dashed curves on the bottom show the scattering beam due to the surface roughness. For reference, the blue dash-dot line, offset for clarity, shows the slope of a $1/\theta^3$ profile. Negative values due to noise fluctuations are not plotted.}
\label{fig:beam_profiles}
\end{figure}

Figure~\ref{fig:beam_profiles} shows the beam profiles for \freqa\,and \freqb\,GHz from the combination of multiple measurements of Uranus. The maps are made from data within $\theta_p=45^\prime$ of Uranus. The data from $12^\prime <\theta_p < 45^\prime$ are used to solve for the contribution from atmospheric fluctuations inside $12^\prime$ and subtract it. One unavoidable consequence of subtracting the atmospheric contribution is that the profiles have an unknown offset. Two other contributions to the profile near $12^\prime$ are the diffraction from the image of the cold stop on the primary, which falls as $1/\theta^3$, and the scattering from the irregular primary surface. All three terms enter the beam model and thus the window function as described in A20. 

We routinely measure the primary reflector shape with targets at the corners of the 71 panels that compose it. Our model shows that the surface roughness derived from these targets is 1.5 times the average surface deviation. Our measurements show that the full surface is well described by an rms fluctuation level of 20\,$\mu$m with a correlation length of 28~cm. The gain from such a surface is given by Equation 8 in \citet{ruze:1966}\footnote{The equation has a typo. Inside the summation, the variance should be raised to the $n^{th}$ power and not simply squared. The correlation length, $c$, is defined through the correlation function for deviations from the ideal surface in the perpendicular direction, $C(r)=\sum_{i,j}z(\vec{r}_i)z(\vec{r}_j)/N=(rms)^2\exp(-r^2/c^2)$ where $\vec{r}$ is the position on the surface, $r=|\vec{r}_i-\vec{r}_j|$, $z$ is the deviation, and the sum is over $N$ measurement pairs on the surface for some $\Delta r$. } and shown in Figure~\ref{fig:beam_profiles} for \freqa\,and \freqb\,GHz. 
The scattered beam has roughly 1.5\% the solid angle of the main beam at \freqb\,GHz and thus extrapolating the main beam profile with $1/\theta^3$ underestimates the main beam solid angle, $\Omega_B$.

As shown in L17, there are polarized sidelobes of the main beam produced by elements in the optics tubes at \freqb\,GHz. These ghosts are located roughly $15^\prime$ from the optical axis at roughly the noise level shown in Figure~\ref{fig:beam_profiles} (they are clearly seen in maps made with Saturn) and accounted for in the mapmaking as described in A20.

The beams for PA3 are 10--20\% elliptical as shown in 
Table~\ref{tab:inst}. The beam scale of roughly $2^\prime$ corresponds to $\ell=10800/\theta\approx5000$, with $\theta$ in arcminutes, which is well above the cosmological signal. In addition, the observing strategy partially rotationally averages the beam further reducing the effect of its intrinsic ellipticity. Our modeling shows that the effect results in an additive bias for TE and TB that is approximately constant over our multipole range with an amplitude that is no more than 0.2 sigma away from zero.
Any residual ellipticity will have a negligible effect on the cosmology presented in DR4 and in this release we make no corrections for it. Based on the galactic center temperature and the level of our beam sidelobes, we also find the stray light from the galaxy to be negligible in the frequency bands and angular scales of interest. Upcoming publications will consider the beam analysis at more depth.

\section{Observations}
\label{sec:obs}
The DR4 observations span four years and cover roughly half of the sky with three different detector arrays and observing strategies.\footnote{This section has significant overlap with a similar one in A20 and is provided here for continuity.}
While data were taken throughout the day, in this paper we present only the nighttime data which we define to be between 2300 and 1100 UTC.\footnote{Daytime data constitutes roughly $\sim$45\% of the total (2013--2016) data volume and will be analyzed separately. Chilean time, CLT, is UTC-4 but daylight savings time leads to departures from this. In May, for example, 2300 in the UK is 1800 local at the telescope.}
The heterogeneity of the set requires significant bookkeeping but carries with it built-in cross checks for depth, scan length, scan elevation, pointing repeatability, and detector characteristics.

A typical night of observations begins by pointing to the selected azimuth and elevation. We then measure the current-voltage characteristics (an IV curve) of all detectors to determine their transition profiles and select the bias for groups of 32 to 96 detectors. Followup IV curves are taken roughly every two hours.

\begin{figure*}[tp!]
\epsscale{1.2} 
\includegraphics[width=\textwidth]{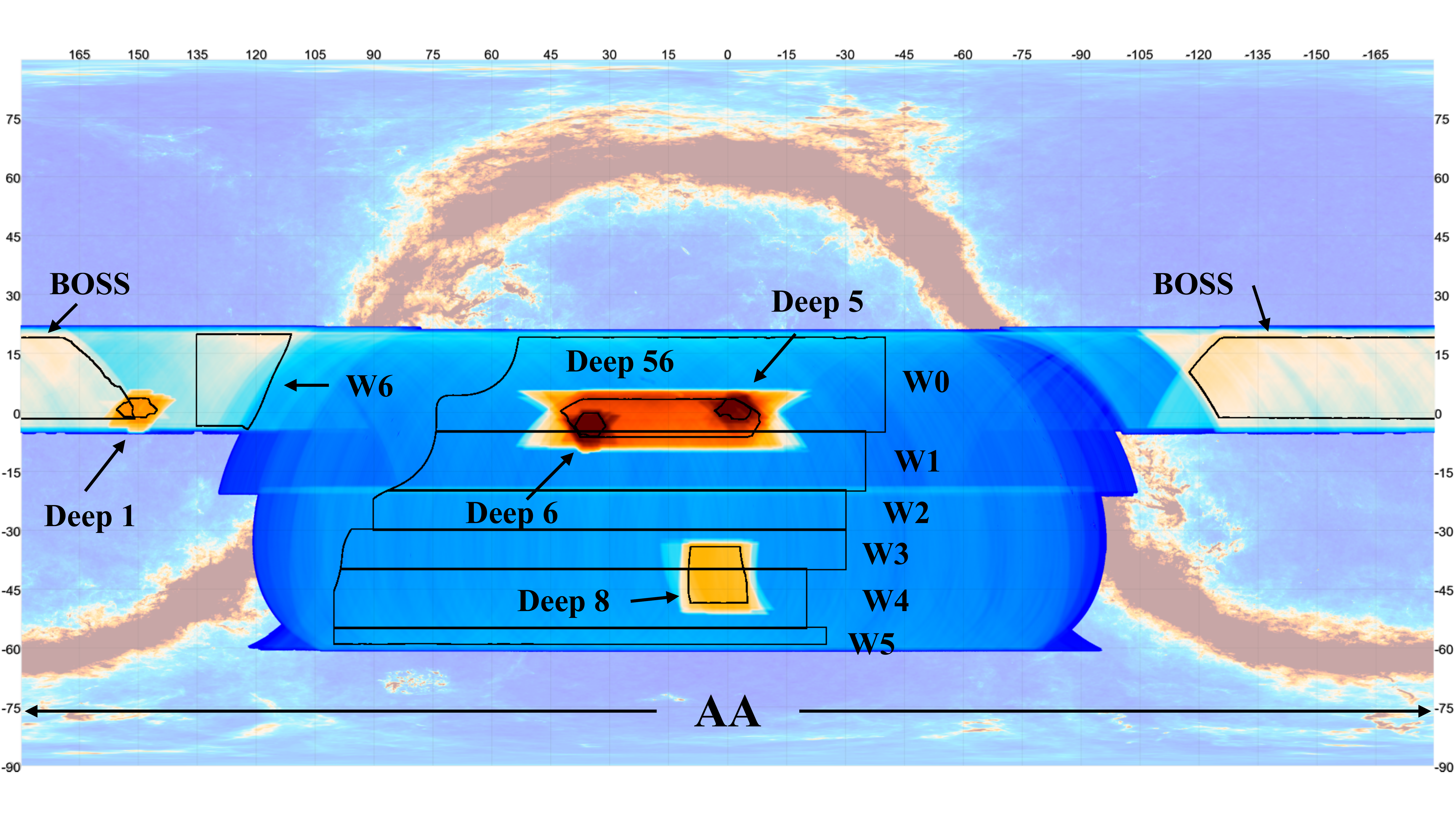}
\vspace{-0.7cm}
\caption{The cumulative ACT DR4 coverage shown in equatorial coordinates for observations between 2013--2016. The background is the {\sl Planck} 353 GHz intensity map. The x-axis (y-axis) shows the RA (dec.) coordinates in degrees. The color scale indicates the depth of the map. The noise levels are reported in Table~\ref{tab:obs}. Regions w2, w6, and D8 are not part of the cosmological analysis. }
\label{fig:actpol_coverage}
\end{figure*}

Over the span of observations for DR4, the trend has been to cover more and more sky as the instrument sensitivity improves and as we learn how to observe and map large areas. Observations took place between Sept. 11 and Dec. 14 in 2013 (s13, 94d (94 days)) covering ``Deep 1" (D1), ``Deep 5" (D5), ``Deep 6" (D6); between Aug. 20 and  Dec. 31 in 2014 (s14, 133d) covering  ``Deep 56" (D56) which encompasses D5 and D6; between Apr. 21 2015 and Feb. 1 2016 (s15, 286d\footnote{In both 2014 and 2015, we also observed D5 and D6 when circumstances permitted.}) covering in addition to D56, ``Deep 8" (D8), and an area that overlaps part of the Baryon Oscillation Spectroscopic Survey (BOSS[BN], \citet{boss_dr13:2017}); and lastly between May 24 and Dec 27 in 2016 (s16, 217d) covering nearly half the sky in what has become the nominal scan pattern for ACT since fielding the new AdvACT (AA) arrays. The AA region is divided into seven subregions or spatial windows, w0 through w6, to optimize the power spectrum analysis. The observations are summarized in Table~\ref{tab:obs} and the overall footprint is shown in Figure~\ref{fig:actpol_coverage}. 

\begin{table*}[tp!]
\caption{Summary of Nighttime Observations}
\begin{center}
\begin{tabular}{|l|c|c|c|c|}
\hline
  & s13 & s14 & s15 & s16 \\
\hline

Region$^a$ & D1/D5/D6 & D56 & D56/D8/BN & AA \\
Area $^b$ (deg$^2$)& 66/64/61 & 565 & 565/197/1837& 11920\\
Area PS  (deg$^2$)& 23/20/20 & 340 & 340/120/1400& 3600\\
Noise threshold $^c$  & 0.23/0.3/0.3 & 0.2 & 0.2/0.04/0.08& $\cdots$\\
Cross-linking threshold $^d$ & 0.96/0.72/0.72 & 0.8 & 0.8/0.99/0.9& $\cdots$\\
\hline
PA1 (150 GHz)$^e$ &  &  &  &      \\
Noise level$^f$ ($\mu$K-arcmin) &15/12/9 & 27 &27/35/67&  $\cdots$ \\
\hline
PA2 (150 GHz)&        $\cdots$           &  &   &      \\
Noise level ($\mu$K-arcmin) &&19 &18/18/35&47--80\\
\hline
PA3 (98 GHz) &  $\cdots$ & $\cdots$    &   &   \\
Noise level ($\mu$K-arcmin) &  &&17/20/33&60--100\\
\hline
PA3 (148 GHz)&     $\cdots$            &     $\cdots$    &   &   \\
Noise level ($\mu$K-arcmin) &&& 27/29/49&86--168\\
\hline
\end{tabular}
\end{center}
\label{tab:obs}
{\small Notes: $a$) The regions are shown in Figure~\ref{fig:actpol_coverage}. Table~\ref{tab:scans} gives the scanning parameters. $b$) The top line gives the area assuming uniform weighting out to the edge of the spatial apodization. This corresponds to the visual impression. The area denoted ``PS" is that used for power spectrum estimation after following the procedure in Section~\ref{sec:spatial_window}. $c$) The noise threshold for selecting the spatial window as described in Section~\ref{sec:spatial_window}. For example, for D1 the 23\% highest noise pixels are dropped. There are no thresholds for AA, because the regions were hand picked by visually examining the noise and cross-linking maps. $d$) The cross-linking upper bound for selecting the spatial window as described in Section~\ref{sec:spatial_window}. Uniform cross-linking corresponds to an index of zero and no cross-linking corresponds to unity. $e$) In s16, PA4, a dichroic array measuring at 150 and 
220 GHz \citep{henderson/2016b, ho/2017}, replaced PA1 but data from it are not part of DR4. $f$) These noise levels are based on the ``white noise" or $\ell>3000$ region of the power spectra (see Figure~\ref{fig:noise_ps}). A20 reports noise levels based on the noise maps, which weight the data differently, and include regions that may be excluded by the noise and cross-linking thresholds imposed here.
Table values differ slightly from those in L17 due to improved selection criteria.
For AA we give the range of noise levels in the six regions (spatial windows) we analyzed along with the total area of the six regions.}
\end{table*}

\section{Data selected for cosmological analysis}
\label{sec:data_select}
Before turning to the analysis, we present the different observing regions, describe the suite of power spectra that are computed, and give the dimensions of the basic covariance matrices. 
The maps used in power spectra in DR4 are from BN, D1, D5, D6, D56, D8, and w0, w1, w2, w3, w4, w5, w6 of the AA region. For the cosmological analysis we omit D8 due to its poor cross-linking (Section~\ref{sec:spatial_window}), omit w2 because it failed a null test, and omit w6 due to insufficient noise modeling. However, these regions are still useful for galaxy cluster and point source studies. Regions D5 and D6 are part of D56; we treat them separately in part of the analysis although we coadd them with the parent region for the final product. Thus, there are eight distinct regions in the cosmological analysis. Although w0 and w1 overlap with D56, they are larger and shallower so the correlations can be ignored.

Our power spectra are computed in 59 bands with centers spanning from $\ell=21$ to $\ell=7525.5$ as described in the next section. Our cosmological analysis is based on the $n_{\ell,c}=52$ bands from $\ell=350.5$ to $\ell=7525.5$ in the TE and EE spectra, and from $\ell=600.5$ to 
$\ell=7525.5$ for TT as discussed in Section~\ref{sec:misc_syst}. Here, the subscript ``{\it c}" is for ``cosmology." The lower bound was selected as part of the blinding procedure motivated in part by a $k$-space cutoff in the maps corresponding to $\ell\approx90$, which is due to the Fourier-space transfer function as described in Section~\ref{sec:fspace_tfunc}, and in part by our experience in L17 where the lower bound was $\ell=500$ for TT and $\ell=350$ for TE/EE. There is evidence that in polarization the ACT maps are well converged to $\ell\sim100$ \citep{li/etal:2020} so we show in our compilation plot, but do not use, preliminary data in EE for $\ell<350$. The maximum $\ell$ is determined by the signal-to-noise ratio. We process TB, EB, and BB along with the rest of the spectra. The first two provide built-in null checks of the spectra. With BB we show consistency with the lensing signal.

Table~\ref{tab:num_ps} lists all the spectra for DR4. As an example of the different combinations of spectra, consider the D56 region. It was observed in s14 with PA1 and PA2 and then again in s15 with PA1, PA2, and PA3. Accounting for different seasons and different arrays there is 1(1) TT(TE) spectrum at \freqa\,GHz, 5(10) at \freqa$\times$\freqb\,GHz, and 15(25)  at \freqb GHz. We keep \freqa$\times$\freqb\,GHz and \freqb$\times$\freqa\,GHz separate for TE and TB but combine them for EB. For the full data set, there are a total of 570 spectra because $N_{TT}=N_{EE}=N_{BB}=N_{EB}$ and $N_{TE}=N_{TB}$ where $N$ is the number of spectra. Of the total, the subset used for the cosmological analysis includes 228 separate power spectra as broken out in the table.

\begin{table*}[thb!]
\caption{Summary of Nighttime Spectra for DR4 at \freqa, \freqa$\times$\freqb  and \freqb\,GHz}
\begin{center}
\begin{tabular}{|l|c|c|c|}
\hline
{\small Region} & {\small TT/TE Spectra} & 
{\small $N_{TT}$} & {\small $N_{TE}$}\\
\hline
\freqa\,GHz &  &   &  \\
{\small D56}     &    {\small s15-3}     & 1 & 1        \\
{\small D8}    &       {\small  {\color{gray} s15-3}    }      &    1    & 1  \\
{\small BN} &   {\small  s15-3    }       &         1   &  1\\
{\small AA}      & {\small  s16-3-w0, s16-3-w1, {\color{gray}s16-3-w2}, s16-3-w3, s16-3-w4, s16-3-w5} & 6  & 6 \\
Total cosmo     && 7     & 7 \\
Total      && 9     & 9 \\
\hline
\hline
\freqa $\times$ \freqb\,GHz &  &   &  \\
D56   &    {\small s14-1-150$\times$s15-3-98, s14-2-150$\times$s15-3-98, s15-3-98$\times$s15-1-150,}     &&          \\
&    {\small   s15-3-98$\times$s15-2-150, s15-3-98$\times$s15-3-150}      &   5      &  10  \\
D8  &       {\small {\color{gray} Same as for BN}}     &    3   & 6  \\
BN &   {\small  s15-3-98$\times$1-150, s15-3-98$\times$2-150, s15-3-98$\times$3-150}  &    3  & 6 \\ 
AA$^a$ &  {\small s16-3-98$\times$2-150-w0, s16-3-98$\times$3-150-w0, s16-3-98$\times$2-150-w1,  s16-3-98$\times$3-150-w1, } &      &  \\
& {\small {\color{gray} s16-3-98$\times$2-150-w2, s16-3-98$\times$3-150-w2},
s16-3-98$\times$3-150-w3, s16-3-98$\times$2-150-w3, } &      &  \\
& {\small s16-3-98$\times$2-150-w4, s16-3-98$\times$3-150-w4, s16-3-98$\times$2-150-w5, s16-3-98$\times$3-150-w5} & 12  & 24 \\
Total cosmo     && 18    & 36 \\
Total      && 23    & 46 \\
\hline
\hline
\freqb\,GHz &  &   &  \\
{\small D1} &  {\small s13-1 } &  1 & 1 \\
{\small D5}     &  {\small s13-1 } & 1  & 1\\
{\small D6}   & {\small s13-1 }       & 1& 1\\
{\small D56}     &    {\small s14-1, s14-1$\times$s14-2, s14-1$\times$s15-1, s14-1$\times$s15-2, s14-1$\times$s15-3, s14-2,}     &&          \\
         &    {\small   s14-2$\times$s14-1,  s14-2$\times$s15-2, s14-2$\times$s15-3, s15-1, s15-1$\times$s15-2,  s15-1$\times$s15-3, }      &&           \\
         &    {\small   s15-2, s15-2$\times$s15-3, s15-3}      &   15      &  25  \\
{\small D8}  &       {\small {\color{gray} Same as for BN}}     &    6   & 9  \\
{\small BN}  &  {\small  s15-1,  s15-1$\times$s15-2, s15-1$\times$s15-3, s15-2, s15-2$\times$s15-3,  s15-3 }       &    6   &  9\\
AA$^a$ &  {\small s16-2-w0, s16-2-w1, {\color{gray}s16-2-w2}, s16-2-w3, s16-2-w4, s16-2-w5, } &      &  \\
& {\small s16-3-w0, s16-3-w0$\times$s16-2-w0, s16-3-w1, s16-3-w1$\times$s16-2-w1, } &      &  \\
 &  {\small {\color{gray} s16-3-w2, s16-3-w2$\times$s16-2-w2}, s16-3-w3, s16-3-w3$\times$s16-2-w3, s16-3-w4,} &      &  \\
 & {\small       s16-3-w4$\times$s16-2-w4, s16-3-w5, s16-3-w5$\times$s16-2-w5} &  18    & 24 \\
 Total cosmo     && 39    & 57 \\
 Total      && 49    & 70 \\
\hline
\hline
Total cosmo     && 64     & 100 \\
Total      && 80     & 125 \\
\hline
\end{tabular}
\begin{flushleft}
{\small To save space we use a slimmed notation of (season)-(array number)-(frequency). For spectra within the same region and year we use, for example, s13-1 to denote s13-1$\times$s13-1.  For the AA region there are six independent spatial windows that are denoted as (season)-(array number)-(window). The total number of spectra used for the cosmological analysis is $N_{TT}+N_{EE}+N_{TE}=228$. $a$) Of the 18/24 \freqb\,GHz TT/TE spectra in AA, 15/20 are part of the cosmology data set; of the 12/24 for \freqa$\times$\freqb\,GHz spectra, there are 10/20. For all entries, $N_{TT}=N_{EE}=N_{BB}=N_{EB}$ and $N_{TE}=N_{TB}$. The entries in gray are part of DR4 but not part of the cosmological analysis. }
\end{flushleft}
\end{center}
\label{tab:num_ps}
\end{table*}%

The spectra for cosmology from each of the eight separate regions are coadded over array and season into ten groups consisting of \freqa, \freqa$\times$\freqb, and \freqb\,GHz for TT and EE and \freqa , \freqa$\times$\freqb, \freqb$\times$\freqa, and \freqb\,GHz for TE. This coaddition, or projection, is done using the full covariance matrix for each region. 

The covariance matrix for one spectrum is $n_{\ell,c} \times n_{\ell,c}$, and so for a single frequency for TT, TE, and EE it is $156 \times 156$. However, TE has double the number of spectra when it is made from two different frequencies because $E_{98} T_{150}$ is different from $E_{150} T_{98}$. Thus, for the three frequency combinations the matrix is $3\times156 + n_{\ell,c}=520=10n_{\ell,c}$ on a side. In summary, there are eight independent $10n_{\ell,c}\times10n_{\ell,c}$ matrices (D56+D5+D6, D1, BN, w0, w1, w3, w4, w5). To make the shape of the D1 covariance matrix match the others, its diagonal elements in the \freqa\,GHz sector are filled with large numbers, because D1 is observed only at \freqb\,GHz. The correlations between non-overlapping regions can be ignored. As noted above, we do not account for the small correlation due to the overlap of w0 and w1 with D56+D5+D6.

For the likelihood analysis, D56+D5+D6 is combined with D1 for the ``deep" regions and BN, w0, w1, w3, w4, and w5 are combined for the ``wide" regions. Each subset consists of $10n_{\ell,c}$ coadded spectra and their associated $10n_{\ell,c}\times10n_{\ell,c}$ covariance matrix.  
These are the inputs for the likelihood. 
The separation into two groups is driven by the different detection thresholds for point sources as described in Section~\ref{sec:results}. For plotting and presenting the CMB spectrum, we coadd spectra from \freqa and \freqb\,GHz as described in the next section.

\section{The power spectrum pipeline}
\label{sec:pipeline}
Here we outline the steps used to compute the power spectra from the maps, their covariance matrices, the band power window functions, and the transfer function from Fourier-space filtering. 

\subsection{Enumeration of the spectra}

For each set (season/region/array) of maps, the data are split temporally to have $n_d=4$ maps each for $I$, $Q$, and $U$ Stokes parameters. This is done so that we only compute cross-spectra and thus avoid noise bias.\footnote{If two maps with the same noise are cross-correlated, the resulting power spectrum contains the noise power. Cross spectra avoid this bias.} For the AA region, due to its shallow depth, $n_d=2$. In the same season, regions observed by different arrays have the same temporal intervals for the data splits. We compute the cross power spectrum of each pair of the data-split maps, but perform the averaging differently depending on the array and season.
Specifically, a single-array power spectrum at one frequency in one season is computed from the unweighted average of the $n_d(n_d-1)/2$ cross data-split power spectra. For different arrays in one season, we only exclude the cross spectrum between the data split maps of the same temporal period and average the $n_d^2 - n_d$ cross data-split spectra. For spectra from different seasons,
we average all $n_d^2$ cross data-split spectra. The spectra resulting from these different averages for different combinations are named in Table~\ref{tab:num_ps}.

\subsection{Angular power spectrum estimation}

The power spectrum code\footnote{The pipeline for computing the spectra was originally written for \cite{choi/2015}. Its accuracy was confirmed by comparing it to an independent code from Kendrick Smith.}
uses the now-standard curved sky pseudo-$C_\ell$ approach to account for the incomplete and nonuniform coverage of the sky and beam smoothing \citep{hivon/2002, kogut/2003, brown/2005}. It was tested against the power spectrum estimator code used in L17  in the flat sky limit, against a suite of simulations, and against the publicly available Simons Observatory curved sky power spectrum pipeline  \texttt{PSpipe}.\footnote{Available through Github at \href{https://github.com/simonsobs/PSpipe}{PSpipe}.} The different codes are in excellent agreement, and the remaining difference between the curved sky codes is $<0.01\sigma$.

The power spectrum estimation is intimately tied to the mapping projection. The maps from previous ACT data releases were made in the cylindrical equal-area (CEA) projection, which changes resolution in latitude as a function of distance from the equator. Since the AdvACT survey covers a large range in declination, $-61^{\circ}<\delta<21^{\circ}$, the CEA projection would require oversampling near the equator. We have thus adopted the plate carr$\mathrm{\acute{e}}$e (CAR) pixelization. Although it is a rectangular projection and equi-spaced in latitude, it is not an equal-area projection. In each latitude ring, pixels are equi-spaced in longitude such that there are the same number of pixels per ring. This means that the physical distance between pixel centers for rings near the equator is greater than that for rings away from the equator, thus Fourier transforming the map and simply binning the Fourier modes at the same $\ell$, as in the usual flat-sky approximation, would result in a bias. However, computing the spherical harmonic transforms (SHTs) with the Clenshaw-Curtis quadrature in the \texttt{libsharp} library, our baseline procedure, gives an unbiased estimate of the SHT at any declination \citep{reinecke/2013}.

\subsubsection{Masking the maps}
Different foreground components in both intensity and polarization enter at different angular scales. At large angular scales, we apply the {\sl Planck} ``100 GHz cosmology mask'' to mask regions containing large Galactic foregrounds \citep{planck_overview:2018} and then fit for residuals as described in Section~\ref{sec:results}. At smaller angular scales, bright point sources dominate. We coadd \freqb\,GHz maps in the deep regions (D56, D1, and D8) and find point sources with a $5\sigma$ flux greater than 15 mJy in intensity (A20). These are then masked both in the intensity and polarization maps at $5^\prime$ ($8^\prime$) radius and apodized beyond the mask edge with a sine function that extends over $10^\prime$ ($15^\prime$) at each source position for the \freqb (\freqa) GHz maps. For the shallower and wider regions, AA and BN, we do the same but with a flux cut of 100 mJy (A20). 
As explained in A20, there are roughly 400 extended sources over the full region that are identified with external catalogs that are also masked. However, we do not mask Sunyaev-Zel'dovich (SZ) clusters \citep{sunyaev/zeldovich:1972} and instead include them in our foreground model (Appendix~\ref{appen:fgcalcs}).

\subsubsection{Cross-linking and the spatial window function}
\label{sec:spatial_window}

We select regions with good noise properties as follows.
ACT's constant elevation scan trajectories project into the maps as almost straight lines. When the same region is observed at different elevations or while setting as opposed to rising, scan lines are rotated with respect to the original direction and the target region is said to be cross-linked.

In general, the better the cross-linking the better our map solutions reflect the true sky. One way to understand this is that the noise in the scanning direction of a single TOD, a roughly 10 min stretch of time-ordered data, is large and localized in 2D Fourier space. Observations at multiple cross-linking angles improve the rotational symmetry of the noise in the Fourier plane, with the improvement related to the amount of cross-linking. We account for the degree of cross-linking in the simulations as described in Section~\ref{sec:noise_sim} and in the spatial window as described next.

To parametrize the degree of cross-linking in a region, we make ``cross-linking maps" by summing up the number of observations at each pixel by representing the projected scan angle as a polarization angle. For example, scans that project to horizontal (e.g., RA) or vertical directions (e.g., dec) on the sky result in a $+Q_\times$ or $-Q_\times$ cross-linking map. Stokes $I_\times$ in this case corresponds to the usual hit-count map. We then compute the level of cross-linking from $P_\times = \sqrt{Q_\times^2+U_\times^2}/I_\times$, where $P_\times=1$ means no cross-linking (just one scan direction). We set thresholds to select regions with a minimum amount of cross-linking for each region. For instance, a threshold $<0.7$ for D56 retains most of the regions observed with two orthogonal scans. For D8, which is located in a particular declination where sky rotation does not allow orthogonal scans, we investigated a threshold of 0.99 but eventually dropped the region from the cosmological analysis due to its poor cross-linking. We set the same threshold for the cross-linking maps from all seasons and arrays for each region, set all pixels below (above) the threshold to be 1 (0), then multiply the maps together. The cross-linking thresholds are given in Table~\ref{tab:obs}.

\begin{figure}
\centering
\epsscale{0.8} 
\includegraphics[width=0.4\textwidth]{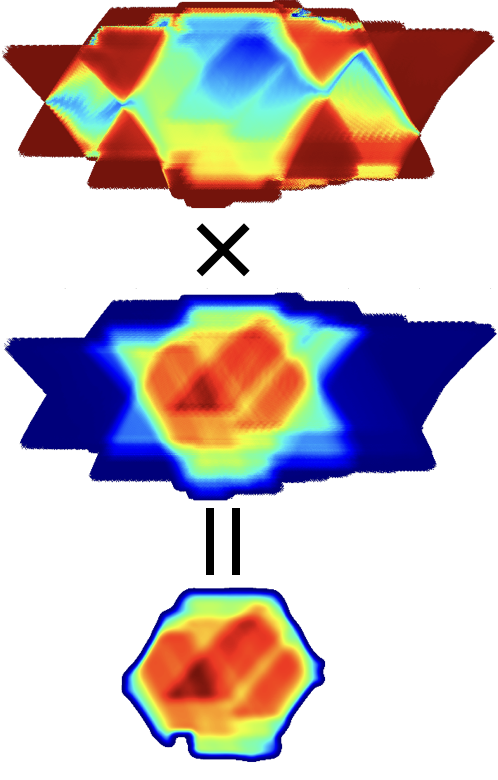}
\caption{Graphical depiction of obtaining the spatial window function for the D6 region. The top panel shows the cross-linking map; the middle panel shows the normalized inverse noise variance of the coverage; and the bottom panel shows the spatial window function. The color scales are from 0 to 1 shown in blue to red. After applying the cross-linking threshold to the top map and the noise threshold to the middle map, they are multiplied together to obtain the bottom map. The source mask and {\sl Planck} Galactic mask, not shown, are also applied. See Figure~\ref{fig:actpol_coverage} for the size and location of the region. The outline of the bottom map represents the ``Area" in 
Table~\ref{tab:obs}, whereas ``Area PS" corresponds to the effective area after inverse noise variance weighting.}
\label{fig:spatial_window}
\end{figure}

The second step in determining the boundary is to threshold the noise maps in percentile to exclude the noisiest regions. The thresholds are also given in Table~\ref{tab:obs}. Finally we take the common boundary mask for each region, apodize over $\sim1^{\circ}$ around the edge with a sine function, then multiply the corresponding inverse variance map to get the spatial window function. The procedure is shown graphically in Figure~\ref{fig:spatial_window}. 
This process ensures the maps of a given region from different seasons and arrays are each weighted with the corresponding inverse variance weights while sharing the same overall boundary.\footnote{The maps of a given region from different seasons and arrays start with slightly different boundaries due to the small pointing offsets between arrays on the telescope.}

\smallskip

\subsubsection{Ground pickup and the Fourier-space mask} 

ACT scans horizontally at different azimuths at different times of the day and year. The contamination from the ground is projected as constant declination stripes in the sky maps. In \cite{das/etal:2011} and L17, Fourier modes with $|\ell_x| < 90$ and $|\ell_y| < 50$ were masked to remove this ground contamination in the data. An exact mode coupling matrix was computed accounting for this Fourier mask in the flat-sky power spectrum estimator code used in L17. Because the ground contamination is projected horizontally on the equatorial coordinates (in RA direction), we continue to mask these contaminated modes in Fourier space, the space in which the modes stay localized. Then we estimate the power spectra of the filtered maps with the curved-sky code, then correct for the loss of power due to filtering with a one-dimensional transfer function determined with simulations as described in Section~\ref{sec:fspace_tfunc}.

\subsection{Simulations}
\label{sec:sims}

We use simulations to compute some elements of the covariance matrix, 
find the probabilities for the consistency checks and null tests, assess the transfer function from the Fourier space filtering, determine the uncertainties on foreground parameters and B-modes, and test the likelihood. The simulated maps include three components: the CMB, foreground emission, and noise as we detail below. As part of DR4, we provide code that generates the simulations used for this work and related papers on, for example, component separation \citep{madhavacheril/etal:2019} and lensing \citep{darwish/etal:2020,han/2020}.

\subsubsection{CMB and foreground emission} 
\label{subsec:sig_sim}

We generate 500 CMB Gaussian realizations of the full sky based on {\sl Planck}'s best fit model 
\citep{planck_cosmo:2019}.\footnote{We use $\Omega_b h^2=0.02219$, $\Omega_c h^2=0.1203$, $h=0.6702$, optical depth $\tau=0.066$, amplitude of scalar perturbations $A_s=2.151\times 10^{-9}$, and scalar spectral index of $n_s=0.9625$. We take $k_0 = 0.05$ Mpc$^{-1}$ as the pivot scale and the total mass of neutrinos of 0.06 eV.} 

These simulations are then lensed by a Gaussian realization of the lensing 
field \citep{naess/louis:2013} with the following algorithm.
Each pixel in the lensed map is given by the value of an unlensed Gaussian map at a position deflected by the local value of the gradient of the lensing potential. This deflected position will in general not correspond to a pixel center in the unlensed map, so interpolation is needed. We do this by generating the unlensed map on a CAR grid at twice the target resolution, and then interpolating to the deflected positions using bicubic spline interpolation. We also take into account the small change in $Q/U$ caused by parallel transport of the polarization vectors along these short displacements. The lensing operation is performed at $1^\prime$ resolution and agrees with theory to better than 1\% up to $\ell = 5000$. 

The aberration due to our motion with respect to the CMB is accounted for in the data power spectra before entering the likelihood (the simulations are not aberrated). This treatment is not exact because aberration distorts the maps in a way that does not translate simply to a power spectrum, but it is sufficient for the current level of sensitivity. Figure 14 in L17 shows that the effect is $\sim$1\% in amplitude in EE.  We also correct a factor of approximately two in a subdominant component of the aberration correction in L17.\footnote{Equation 8 of
\citet{planck_aberration:2014} shows the frequency dependent part of the boosting. In L17 $b_\nu=1$ as opposed to
the correct $b_\nu=2.04$ for \freqb\,GHz.}

Foreground emission from extragalactic sources is simulated with Gaussian random fields on the full sky as well. This means that the amplitudes are drawn from a Gaussian distribution around the expectation. In general, foreground emission is non-Gaussian but the Gaussian approximation is sufficient for our needs. The simulation package includes components from radio galaxies, thermal SZ clusters, and dusty, star-forming galaxies with power spectra given by models from \citet{dunkley/etal:2013}.
The simulations are done for \freqa and \freqb \,GHz accounting for the covariance between frequencies. Diffuse components of the foregrounds, such as from Galactic dust, synchrotron, and anomalous microwave emission, are not included in the simulations.

For both the CMB and foreground emission, we extract the given sky region (e.g., D56, BN, etc.) from the beam-convolved full-sky simulation and then convolve by the appropriate pixel window function.

\subsubsection{Noise} 
\label{sec:noise_sim}
We define ``noise'' to be any source of power in the maps that is not nulled when subtracting splits of the data.  In DR4, the noise properties vary considerably between regions. Understanding and being able to simulate this is essential for interpreting the significance of the results. To this end, we build an empirical noise model from splits of the data in a way analogous to how the spatial window was determined (Section~\ref{sec:spatial_window}). 

The ACT maps are diverse in depth, area, cross-linking, and detector properties (Table~\ref{tab:inst}).  There are multiple characteristics of the noise that we would like to capture in simulations: ($a$) it has a strong $1/f$ character due to the atmosphere; ($b$) its white noise level (at least) is inhomogeneous in real space because some areas are observed more often than others;  ($c$) it is anisotropic in 2D Fourier space due to detector correlations, sky curvature, and especially, imperfect cross-linking (see Section~\ref{sec:spatial_window}); ($d$) its 2D Fourier space properties are inhomogeneous in real space (see Appendix~\ref{appen:noise_prescription}) due to the scan strategy; and ($e$) it exhibits correlations between Stokes $I$, $Q$, and $U$, and between
\freqa and \freqb\,GHz. In the case of the dichroic array PA3, there are correlations between the \freqa and \freqb\,GHz channels as large as 40\% at low and intermediate multipoles due to the common atmosphere. This correlation is captured in the simulations as described below. Correlation coefficients of $\sim$10\% at $\ell=500$ are seen between the PA1 and PA2 arrays in the BN and D8 regions, but we do not include these in the simulations.

The noise simulations are done in 2D Fourier space. For the power spectrum analysis, we simulate 28 maps individually (for regions D1, D5, D6, D56, BN, w0, w1, w3, w4 and w5 for each array/frequency/season, as tabulated in Table~\ref{tab:inst}).  To generate them efficiently, we make two approximations. The first is that the real-space inhomogeneity of the 2D noise spectrum  (point $d$ above) can be ignored within a simulated map.  (The sub-regions in AA were chosen with this criterion in mind.) The second is that each noise map can be modeled as a realization of a Gaussian random field (with an anisotropic 2D Fourier power spectrum) modulated in amplitude by a function of sky position (to account for point ($b$) above). These approximations do not affect the mean estimate of the CMB and foreground band powers, but they do affect the covariance estimates. The approximations are valid in the deep regions within the spatial windows defined in Section~\ref{sec:spatial_window} but are not fully descriptive of the wide regions, especially in the AA region. Nevertheless, based on the consistency checks described in Section~\ref{sec:consistency} we find them sufficient for the present analysis.

With the above two approximations, we simulate the noise for each combination of array, frequency and season in Table~\ref{tab:inst}. Our prescription, described in more detail in Appendix~\ref{appen:noise_prescription}, allows us to capture the large range of noise properties including the large correlations described in ($e$) above. The end product of the simulation pipeline is a set of 500 simulated maps, each of which has a common CMB plus foreground realization for all regions but the noise characteristics appropriate for each individual region.

\subsection{Covariance matrix}
\label{sec:act_covmat}

There are three levels of covariance matrices in the analysis. At the first level, the individual cross spectra in a given region form the elements. For D56, this matrix has 21 TT terms, 36 TE terms, and 21 EE terms for the combined \freqa, \freqa$\times$\freqb, and \freqb\,GHz entries in Table~\ref{tab:num_ps}. Thus the full matrix is $78\times n_{\ell,c}$ by $78\times n_{\ell,c}$. At the next level, this is reduced to one $10n_{\ell,c}\times 10n_{\ell,c}$ matrix for each of the eight regions, by coadding over seasons and arrays. \footnote{D5 and D6 are at first separate from D56 but then combined with it to make eight regions. See Section~\ref{sec:data_select}.} It is at this stage that the window function (A20) and calibration uncertainties are added to the covariance matrix.
Lastly, these matrices are combined into two $10n_{\ell,c}\times 10n_{\ell,c}$ matrices, one each for the 15 mJy and 100 mJy source cut as described above. We next describe the constituents of the full covariance matrix and how we use simulations to arrive at the form that enters the likelihood. While the description focuses on the TT/TE/EE matrix, we use a similar construction for TB/EB/BB. 

We use the basic form of the covariance as outlined in, for example, \citet{louis/etal:2013} but update it to account for advances in quantifying the lensing. There are six different types of components:

1) The diagonal elements\footnote{For the covariance matrix, $\Sigma$, we use the notation $(\Delta C_b)^2$ for the diagonal elements and Cov$_{bb^\prime}$ for the off-diagonal elements.} are primarily instrument noise and cosmic variance. For TT, for example, the auto-spectrum has the form:
\begin{equation}
(\Delta C^{XX}_b)^2 = \frac{2}{\nu_b}\big((C^{XX}_b)^2
+2C^{XX}_b N_b^{XX} + (N_b^{XX})^2\big)
\label{eq:1}
\end{equation} 
where $b$ stands for a bin in $\ell$, $\nu_{b}$ is the number of modes or $(2\ell+1)\Delta\ell_b f^{sky} t_{b}$, and $N_b$ is the noise in the band. The last term, $t_b$, is the transfer function due to the mapping process and Fourier-space filter. The first term on the right in Equation~\ref{eq:1} is the cosmic variance. The superscript 
$X$ denotes $T$, $E$, or $B$.

For cross spectra, which we use exclusively in our analysis, the above becomes
\begin{eqnarray}
(&\Delta& C_b^{XX})^2 = \\  & &\frac{1}{\nu_b}\Bigg[2 (C_b^{XX})^2+ 4\frac{C_b^{XX}}{n_d}N_b^{XX}+   \frac{2}{n_d(n_d-1)}\big(N_b^{XX}\big)^2 \Bigg],\nonumber
\end{eqnarray}
where $n_d$ is the number of splits of the data and $N_b^{XX}$ is the noise in each split. 
For expressions of the form $(\Delta C_b^{XY})^2$ and a derivation of the above see \citet{louis/etal:2019}. Signal plus noise simulations are used to get accurate estimations of $\Delta C^{XX}_b$ and $\Delta C^{XY}_b$.

2) The ``pseudo-diagonal'' terms come from correlations between $C_{b}^{XX}$ and $C_{b}^{YY}$, for example, and are dominated by sample variance for the power spectra in the same region coming from different array combinations. These are found with simulations and checked analytically. 

3) The ``diagonal-plus-one" terms are the correlations between $C_{b}^{XX}$ and $C_{b\pm1}^{XX}$. These depend on filtering, masking, and the spatial window. We determine these from the simulation as well. We do not account for the off-diagonal terms on the pseudo-diagonals or the $C_{b}^{XX}$ to $C_{b\pm2}^{XX}$ correlations. These latter terms are $<$3\% for D56 and $<$0.5\% for BN.

4) There are off-diagonal correlations from lensing that arise from a single lensing $L$-mode fluctuation inside a region
\citep{benoit-levy/etal:2012, peloton/etal:2017} simultaneously affecting many $\ell$ values. They are given by:

\begin{equation}
\mathrm{Cov}^{XY,WZ}_{bb'} = \frac{4 \pi}{\Omega} \sum_{L,\ell,\ell'} \mathcal{U}^{XY}_{b\ell}\left[ \frac{\partial C^{XY}_\ell}{\partial C^{\phi \phi}_L}  \frac{2(C_L^{\phi \phi})^2}{(2L+1)}  \frac{\partial C^{WZ}_{\ell'}}{\partial C^{\phi \phi}_L} \right]\mathcal{U}^{WZ}_{b'\ell'}
\end{equation}
where $L$ is a lensing mode, $\phi$ is the deflection field, $\Omega$ is the effective area of the region, and $\mathcal{U}_{b\ell}$ is the band power window function described in Section~\ref{subsec:bpwf}. These terms are computed analytically as in \citet{motloch/hu:2017}.

5) As pointed out by \citet{manzotti/etal:2014} and \citet{pavel/hu:2019}, there is a lensing super-sample variance, which arises from the variation of the mean convergence over the survey footprint. It is represented as
\begin{equation}
\mathrm{Cov}^{XY,WZ}_{bb'} = \sum_{\ell,\ell'} \mathcal{U}^{XY}_{b\ell}\left[ \frac{\partial \ell^2 C^{XY}_\ell}{\partial \ln \ell}  \frac{\sigma_\kappa^2}{\ell^2 \ell'^2}  \frac{\partial \ell'^2C^{WZ}_{\ell'}}{\partial \ln \ell'} \right]\mathcal{U}^{WZ}_{b'\ell'}
\end{equation}
where $\sigma^2_\kappa$ is the variance of the convergence field, $\kappa = -\nabla^2 \phi /2$, in the survey footprint. These are computed analytically.

6) For sources in the Poisson regime (i.e., neglecting clustering), the power spectra and trispectra are given in terms of the number counts by
\begin{eqnarray}
    C_\ell&=&g_2^2(\nu)\int_{S_{min}}^{S_{max}} S^2 \frac{dN}{dS\,d\Omega}dS\\
 {\cal T} &=&g_2^4(\nu)\int_{S_{min}}^{S_{max}} S^4 \frac{dN}{dS\,d\Omega}dS
\end{eqnarray}
where $g_2(\nu)$ is a factor to convert from Jy/Sr to $\mu$K relative to the CMB (see Appendix~\ref{appen:fgcalcs}). The first of these terms is included in the simulations but the second is added analytically. The band power covariances for the combination are given by:
\begin{equation}
\mathrm{Cov}_{bb^\prime} = \frac{2C_b^2\delta_{bb^\prime}}{(2\ell+1)\Delta\ell_b(\Omega/4\pi)} +
\frac{\cal T}{\Omega},
\end{equation}
where $\Omega$ is the effective area of the region (e.g., \citet{komatsu/seljak:2002}).

In the first reduction step between the three levels of covariance matrices, all elements of the full-region covariance matrix (e.g., 4056$\times$4056 for D56) except for the diagonal and pseudo-diagonal terms are zeroed out and the groups of spectra, say 15 TT spectra at \freqb\,GHz for D56, are combined into one. 
Formally, the calculation is done with 
\begin{equation}
\begin{split}
C_{ca}^{XY} &= (P^T \Sigma^{-1} P)^{-1}P^T\Sigma^{-1}C^{XY} \\
\end{split}
\label{eq:coadd}
\end{equation}
where $C^{XY}$ is the vector that includes, say, 4056 elements for D56, $\Sigma$ is the covariance matrix with all but the diagonals and pseudo-diagonals zeroed out, $P$ is the projection matrix populated with 1s and 0s, and $C_{ca}^{XY}$ is the coadded ($ca$) vector of power spectra.

This same procedure is repeated for 500 simulations each with a different signal and noise. From the distribution of the simulations, we compute the diagonal, pseudo-diagonal, and diagonal-plus-one terms of the $10n_{\ell,c}\times10n_{\ell,c}$ matrix $\Sigma_{ca}$, the coadded covariance matrix for $C_{ca}^{XY}$. Our approach is to use simulations where a robust estimate can be obtained and to use analytic expressions elsewhere. A typical diagonal-plus-one term has a correlation value of $-0.05$ for the wide regions and $-0.1$ for the deep regions. Then for each region we add calibration and beam covariance matrices to $\Sigma_{ca}$, computed with Gaussian errors in calibration to {\sl Planck} (Section~\ref{subsec:cal}) and the beam errors from Uranus measurements respectively (A20, L17, \cite{das/etal:2011}).

In the second reduction step, 
after checking the data consistency as described in Section~\ref{sec:consistency}, we use the ten separate covariance matrices and Equation~\ref{eq:coadd} to inverse variance weight and coadd all power spectra from the different regions into a single deep and single wide power spectrum with their associated covariance matrices.

\subsection{Band power window function}
\label{subsec:bpwf}
The band power window functions are used to bin the theory power spectra to compare to the data. They depend on the mode coupling matrix and $\ell$-space binning scheme and are slightly different for each region. Band power window functions are coadded using the power spectrum covariance matrix, which takes into account the weight variations among different regions. This coadded band power window function is used in the likelihood. 

\subsection{Fourier-space filter transfer function}
\label{sec:fspace_tfunc}

We estimate this transfer function by comparing the power spectra of the simulated maps before and after applying the Fourier-space filter. In principle, a full transfer matrix describing the possible bin-to-bin power transfer is needed. We test the necessity of this level of complexity by examining the consistency between the transfer functions estimated with two differently shaped spectra, $\Lambda$CDM TT and EE power spectra, and find that a simple 1D implementation is sufficient for our needs. We also count the number of modes removed by our filter in the 2D Fourier plane to check the transfer function analytically.

\begin{figure}
\begin{center}
\includegraphics[width=0.5\textwidth]{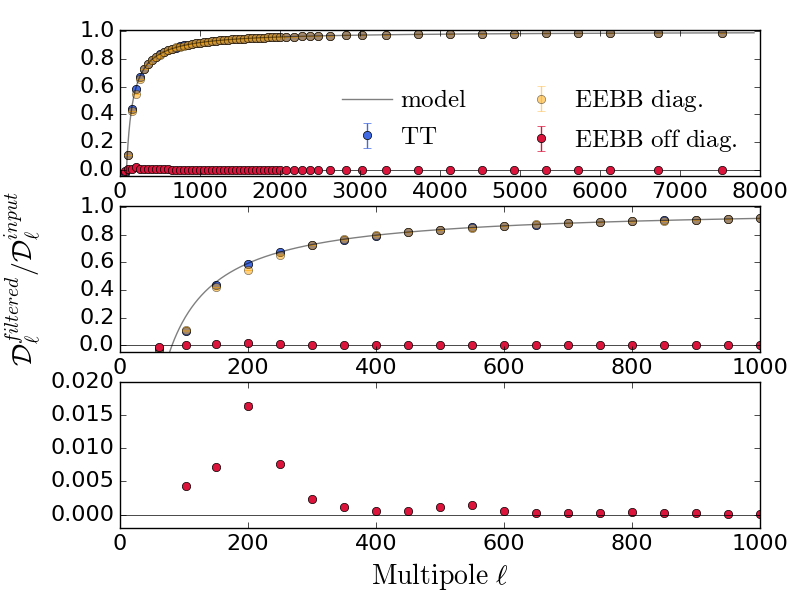}
\caption{A comparison of the modeled Fourier-Space filter transfer function to simulations. Using signal-only CMB maps of the D56 region, we compute the power spectra before and after applying the Fourier space filter (cutting $|\ell_x| < 90$ and $|\ell_y| < 50$). The binned transfer function for TT, and the EE/BB transfer matrix elements at each $\ell$ bin are computed using Equation~\ref{eq:tfunc}, shown in all three panels. Note that the bottom two panels show an expanded view of the $\ell<1000$ region in the top panel. The analytic estimate of the transfer function is shown in gray solid line. The model describes the simulations to within 0.5\% for $\ell>300$. The bottom panel shows the EE/BB mixing term, which is $<0.15$\% for $\ell>350$. This mixing correction was necessary only for the sensitivity levels achieved in D56.}
\label{fig:tfunc}
\end{center}
\end{figure}

In addition to directly acting on the TT, TE, and EE spectra, Fourier-space filtering of the Stokes $Q$ and $U$ maps can lead to mixing of the E and B modes.\footnote{We note this differs from the usual E-B mixing due to incomplete sky coverage in the pseudo-$C_\ell$ approach, which is analytically corrected with the mode coupling matrix.} We characterize this with a transfer matrix for each $\ell$ given by,
\[ \left (
  \begin{tabular}{c}
  $\widetilde{T}_{\vec{\ell}}$ \\
  $\widetilde{E}_{\vec{\ell}}$ \\
  $\widetilde{B}_{\vec{\ell}}$ \\
  \end{tabular}
\right )
 = 
\left (
  \begin{tabular}{ccc}
$t_{\vec{\ell}}$ & 0 & 0  \\
0 & $d_{\vec{\ell}}$ & $o_{\vec{\ell}}$  \\
0 & $o_{\vec{\ell}}$ & $d_{\vec{\ell}}$  \\
  \end{tabular}
\right )
\left (
  \begin{tabular}{c}
  $T_{\vec{\ell}}$ \\
  $E_{\vec{\ell}}$ \\
  $B_{\vec{\ell}}$ \\
  \end{tabular}
\right )
, \]
where $\vec{\ell}$ denotes a 2D $\ell$, $X_{\vec{\ell}}$ is the 2D Fourier transform of the $T$, $E$, or $B$ map, $\widetilde{X}_{\vec{\ell}}$ is the 2D Fourier transform of the respective filtered map, $t_{\vec{\ell}}$ is the 2D filter for temperature, $d_{\vec{\ell}}$ is the 2D filter for polarization, and $o_{\vec{\ell}}$ is the 2D mixing kernel for polarization. The resulting binned power spectra on the filtered maps are given by,
\begin{equation}
\begin{split}
\widetilde{C}_b^{TT} &= t_b^2 C_b^{TT} \\
\widetilde{C}_b^{EE} &= d_b^2 C_b^{EE} + o_b^2 C_b^{BB} + 2 d_b o_b C_b^{EB}  \\
\widetilde{C}_b^{BB} &= d_b^2 C_b^{BB} + o_b^2 C_b^{EE} + 2 d_b o_b C_b^{EB}  \\
\widetilde{C}_b^{TE} &= t_b d_b C_b^{TE} + t_b o_b C_b^{TB}  \\
\widetilde{C}_b^{TB} &= t_b d_b C_b^{TB} + t_b o_b C_b^{TE}  \\
\widetilde{C}_b^{EB} &= (d_b^2 + o_b^2) C_b^{EB} + d_b o_b (C_b^{EE} + C_b^{BB}), \\
\end{split}
\label{eq:tfunc}
\end{equation}
where $\widetilde{C}_b$ ($C_b$) denotes the (un)filtered map power spectra, and $t_b^2$ and $d_b^2$ represent the 1D binned transfer function, and $o_b^2$ is the E-B mixing term. We find that the diagonal element of the EE/BB matrix $d_b^2$ is consistent with $t_b^2$ from the TT transfer function within 1\% shown in Figure~\ref{fig:tfunc}. The mixing term $o_b^2$ is estimated to be $<1\%$ at $\ell\geq200$ (and $<0.1\%$ for $\ell\geq600$). This small mixing correction was necessary for only the sensitivity levels achieved in D56.

In summary, the effects of this transfer function are well understood.
The transfer function for the TT and EE power spectra are consistent with each other to $<$1\%, both are consistent with the analytic estimate of the transfer function to $<$1\%, and for $\ell>300$
the transfer functions estimated from simulations are consistent with the analytic estimate to $<0.5\%$. 

\section{Temperature calibration, polarization Angle, and mapping transfer function}
\label{sec:cal_etc}

There are four general areas of systematic error that would not be uncovered in the internal consistency and null tests described in Sections~\ref{sec:consistency} and~\ref{sec:null_tests} below. They are the overall calibration, the instrumental polarization angle, the mapping transfer functions,
and the beam window functions. The first three of these are described in the following and the last is addressed in A20.

\subsection{Calibration}
\label{subsec:cal}
Before combining spectra, we calibrate them for each region/array/season to the {\sl Planck} temperature maps, weighted by the ACT spatial window (Section~\ref{sec:spatial_window}), over the range $600< \ell<1800$ at \freqa\,GHz and \freqb\,GHz. There is no separate calibration for TE or EE. 

We calibrate through cross correlation as described in \citet{hajian/etal:2011} and \citet{louis/etal:2014}. In keeping with the blinding protocol, we do not plot any ACT$\times${\sl Planck} power spectra while calibrating. As a check for possible systematic errors we compare the calibration in two ranges, $600< \ell<1200$ (low) and $1200< \ell<1800$ (high). The weighted mean of the low and high calibrations, for the power spectrum, agrees within 0.007 with the overall calibration. The ratio of the high/low calibrations for all spectra used for cosmology is 0.992 at \freqb\,GHz in the sense that relative to {\sl Planck} there is slightly less power at $\sim 1\sigma$ significance in the low $\ell$ range. The same ratio is 1.002 at \freqa\,GHz. The overall calibration error for the full data set is 1\% in the power spectrum. We note that this value is only for the CMB and does not apply to foregrounds or compact sources.

\subsection{Polarization angle}
\label{subsec:pol_angle}
A critical calibration parameter is the polarization angle $\psi_P$, which describes the rotation in Stokes parameter space of the polarization signal in the maps relative to the sky. It is determined with a combination of metrology, modeling, and planet observations.\footnote{We report polarization angles that follow the IAU convention \citep{hamaker/bregman:1996} by computing $\psi_P=(1/2){\rm arg}(Q-iU)$.}  The alignment of optical elements with respect to the detector wafers and the cryostat position relative to the primary and secondary reflectors can introduce a source of rotation of the entire detector array when projected onto the sky. In addition, the orientation of the orthogonal pick-up antennas in each detector and in relation to all other detectors within an optics tube must be considered while constraining $\psi_P$. In the wafer fabrication process, this orientation is held to $<0.001^\circ$. These combined effects may be characterized by a single angle. However, because the optical elements can rotate the polarization, this constraint alone cannot determine the polarization angle. Our model of the full optical system, reflectors plus lenses, shows that the polarization angle rotates continuously across an array by up to $\sim 1.7^{\circ}$ near the edge of the focal plane as one moves away from the primary optical axis of the telescope \citep{koopman/etal:2016}. We incorporate this effect in our model for the polarization angle. 

Observation of planets and bright sources determine the pointing angles of the detectors to an accuracy of $\theta_{point}\sim 4 ^{\prime\prime}$ for the $\sim20$ sources (including planets) with S/N$>600$. The constraint on the array orientation 
limits the contribution to the polarization angle to $2\theta_{point}/\phi_{FOV}=0.1^\circ$, where $\phi_{FOV}\approx 1^\circ$ is the field of view of the array and $\theta_{point}$ is the maximum rotation at its periphery. The optical model plus the measurements of the pointing set the polarization angle for each detector.

\begin{figure}[t]
    \centering
    \includegraphics[width=0.5\textwidth]{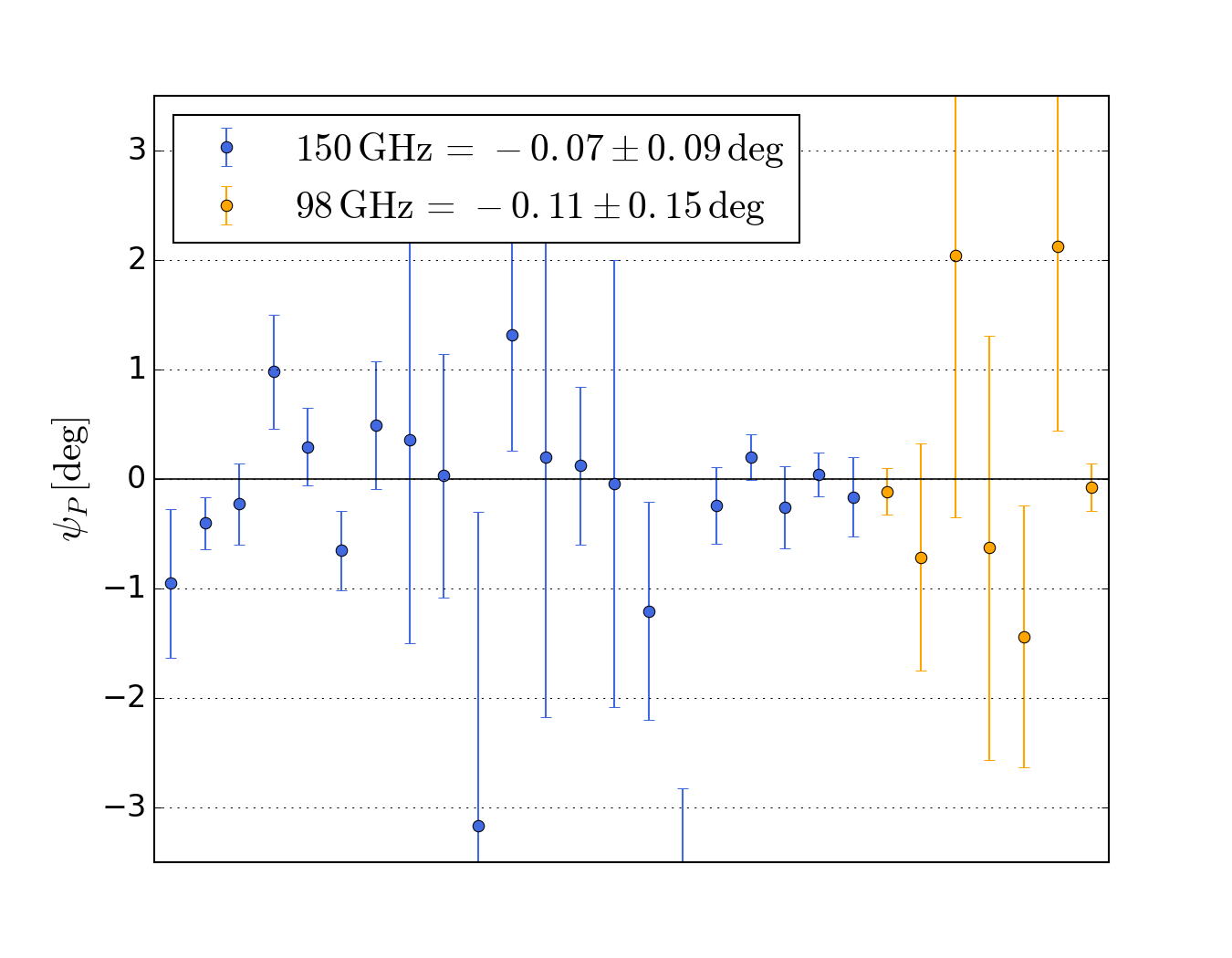}
    \caption{The EB null angle $\psi_P$ is shown for all fields/seasons/arrays of data. The weighted mean of 150 GHz (98 GHz) angles is $-0.07^{\circ}\pm0.09^{\circ}$ ($-0.11^{\circ}\pm0.15^{\circ}$), and $\chi^2/\mathrm{dof} = 1.20$ (0.68).}
\label{fig:angle_all}
\end{figure}

After we accept the solution for $\psi_P$, we compute $\psi_P$ based on the EB cross-spectra \citep[e.g.,][]{keating/etal:2013}.
In the absence of parity violating physics such as that produced by axion and magnetic fields in the primordial perturbations or during the propagation of CMB photons from the CMB last scattering, the EB spectrum should be null. After accounting for aberration, we compute $\psi_P$ for each array and season. Although there is a distribution, with the largest outlier 2.2$\sigma$ away from zero, there are no clear trends. The reduced $\chi^2$ for 28 different measurements is 1.07. Restricting the data set to eight representative values that sample all four seasons, $\chi^2/\nu=0.8$. Given that the cryostat was removed, worked on, replaced, and repositioned each season, this suggests that the determination of $\psi_P$ is robust.
A weighted mean of all \freqb\,GHz (\freqa\,GHz) measurements, shown in Figure~\ref{fig:angle_all}, gives 
$\psi_P =-0.07^{\circ}\pm0.09^{\circ}$ ($-0.11^{\circ}\pm0.15^{\circ}$), with $\chi^2/\mathrm{dof} = 1.20$ (0.68). 
Although one can determine the instrumental $\psi_P$ by nulling the EB signal \citep{keating/etal:2013}, we make no such correction.\footnote{A rule of thumb for measuring the tensor-to-scalar ratio $r$ is $r_{bias}=\sigma_{\psi_P}^2/125$ where $\sigma_{\psi_P}$ is in degrees \citep{abitbol/etal:2016,nati/etal:2017,minami/etal:2019}. 
Our results suggest it may be possible to achieve $r_{bias}<0.001$ through a combination of modeling and pointing, combined with cross correlation of large and small aperture instruments \citep{li/etal:2020}.} 

Because we do not use EB to set the polarization angle, $\psi_P =-0.07^{\circ}\pm0.09^{\circ}$
may be interpreted as a limit on Chern-Simons models as a source of cosmic birefringence \citep{carroll/field/jackiw:1990}.
For example, if the cosmic birefringence is generated by the uniform misalignment of the ultra-light axions \citep[e.g.,][]{marsh:2016}
then the $\psi_P$ constraint on the polarization angle leads 
to a constraint of $\phi_i g_{a\gamma} = (-2.5 \pm 3.2 )\times 10^{-3}$, where $\phi_i$ is a field value of the axion field before the axion starts to oscillate, and $g_{a\gamma}$ is the coupling. Our constraint on the isotropic cosmic birefringence improves on results obtained in previous works \citep[e.g.,][]{mei/etal:2015,zhai/etal:2020}. Relatedly, following 
\citet{Sigl/Trivedi:2018}, our $2\sigma$ limit of $|\psi_P|<0.25^\circ$ may be interpreted as a limit
on the axion-like particle coupling constant of $g_{a\gamma}< 0.4\times10^{-15}$ (GeV)$^{-1}$ for an axion mass of $m_a=3\times10^{-26}$eV.
 
 An independent determination by \citet{namikawa/etal:2020} gives $\psi_P=0.12^{\circ}\pm0.06^{\circ}$.  The difference is due to using only s14 and s15 for D56, noise debiasing of the polarization spectra, simplified treatment of covariances without beam effects, and analyzing $200 <\ell< 2048$. When the technique used in this paper is limited to this subset of data, we find $\psi_P= 0.095^\circ\pm0.087^\circ$ (for \freqa +\freqb\,GHz), consistent with the \citet{namikawa/etal:2020} result (using their opposite sign convention). We note that the ACT analysis in \citet{namikawa/etal:2020} focused on a constraint on the anisotropic birefringence power spectrum, in contrast to the constraints on isotropic birefringence discussed here. Based on SPT data, \citet{bianchini/etal:2020} also measure the anisotropic cosmic birefringence power spectrum, extending the results in \citet{namikawa/etal:2020}, and derive a similar upper limit on a scale-invariant anisotropic birefringence spectrum.

\subsection{Mapping transfer function}
\label{sec:map_tfunc}
One of the attractive features of maximum likelihood mapmaking is that it produces unbiased maps. In other words, the power spectrum of an unbiased map should not need to be corrected for the mapping process. However, we add one operation to our mapmaking that does slightly bias the power spectrum. As described in A20, maximum-likelihood ground template maps are made in azimuth-elevation coordinates, and deprojected from the TODs. This deprojection removes the strongest ground signals and results in a flat transfer function of $>\sim 0.997$ for most regions at $350<\ell<8000$ (and smaller at $\ell<350$, which we do not consider in this analysis). Smaller regions (D1, D5, and D6), which contribute small weight in the total statistics of the coadded spectra, have a more complex shape with a dip to $\sim 0.98$ near $\ell= 4000$.\footnote{A more efficient ground deprojection transfer function, which is deferred to future analyses, can be done in 2D Fourier space.}
Each spectrum is divided by the appropriate transfer function. The uncertainties on the transfer functions were investigated with simulations and found to be negligible ($<0.02\sigma$).

\section{Internal Consistency checks}
\label{sec:consistency}
With multiple seasons of observations made with multiple arrays, there are many possible pair-wise combinations of data that may be tested for consistency. There are three broad classes of checks as discussed, for example, in \citet{louis/etal:2019}. One entails consistency between two maps, a second between the power spectra of the two maps and a third between the power spectrum of one map and its cross spectrum with the other. We consider only the first two of these. In general, when there is a measurable signal in the maps, the consistency between maps is the most stringent test to pass.
However, depending on the systematic effect, for example a multiplicative bias, the power spectrum null can be more stringent. Thus we present both.

The power spectrum of the difference, or ``null," between maps $A$ and $B$ is,
\begin{equation}
\label{eq:null_spec}
\hat{C}_b^{A-B} = \hat{C}_b^{AA} + \hat{C}_b^{BB} - 2\hat{C}_b^{AB}.
\end{equation}
where $\hat{C}_b=C_b+N_b$ is the measured quantity for example for an auto spectrum, $C_b$ is the underlying power spectrum, and $N_b$ is the noise.  For this to be a true map-level null power spectrum, the same spatial window function needs to be used in estimating each of the power spectra on the right hand side. We have constructed our spatial windows, as described in Section~\ref{sec:spatial_window}, to be roughly the same for the different seasons/arrays for one region even though it required cutting otherwise good data. As a result, the same spatial window functions and maps that are used for the consistency tests are used for the cosmological analysis.

The general case for the covariance matrix of $\hat{C}_b^{A-B}$ is worked out in \citet{das/etal:2011} and
\citet{louis/etal:2019}. When assessing the covariance, we work directly with power spectra after the $n_d$ splits, described in the introduction of Section~\ref{sec:pipeline}, have been combined. In the limit that $C^{AA}_b$ and $C^{BB}_b$ cancel in Equation~\ref{eq:null_spec}, in other words that the underlying spectra and spatial windows match, and that the noise in each of the $n_d$ splits is the same, the variance for the power spectrum of the null map for each bin $b$ is given by:
\begin{eqnarray}
\label{eq:map_null_err}
(\Delta \hat{C}_b^{A-B})^2 &=& \frac{1}{\nu_b} \Bigg[\frac{2}{n_d(n_d-1)}\big((N_b^{AA})^2 
+ (N_b^{BB})^2\big) \notag \\
&+& 4\frac{N_b^{AA}N_b^{BB}}{n_d^2} \Bigg],
\end{eqnarray}
where $N^{AA}$ is the noise in one of the $n_d$ splits that goes into determining the spectrum of map $A$ and likewise $N^{BB}$ for map $B$.

In contrast, the variance for the difference between power spectra is given by 
\begin{eqnarray}
\label{eq:ps_null_err}
(\Delta[\hat{C}_b^{AA}-\hat{C}_b^{BB}])^2 &=& \frac{1}{\nu_b} \Bigg[\frac{2}{n_d(n_d-1)}\big((N_b^{AA})^2 + 
(N_b^{BB})^2\big) \notag \\
&+& \frac{4}{n_d}C_b(N_b^{AA}+N_b^{BB}) \Bigg].
\end{eqnarray}
Here again we have assumed that the power spectra of the underlying signal and spatial windows match so that $C_b=C_b^{AA}=C_b^{BB}$. 

In practice, as opposed to computing Equation~\ref{eq:map_null_err} directly, we compute the null power spectrum covariance with the covariance matrix $\Sigma$ from Equation~\ref{eq:coadd} and a projection matrix with elements of the form $P=(1,1,-2)$.
Similarly, for the power spectrum difference we use 
$P=(1,-1)$.\footnote{These projections are appropriate for TT and EE. The form for TE for the map spectra difference is $(1,1,-1,-1)$.} This formalism provides a general and compact way to compute the multiple different combinations of elements of the covariance matrix that enter the consistency tests. Also, using the full covariance matrix, it can be generalized to consistency checks between power spectra of non-overlapping regions as the null power spectrum error bars then contain the representative signal variance.
Lastly, we correct for the different levels of point source contamination when comparing the deep and wide regions.

\begin{figure}[tp!]
\centering
\includegraphics[width=\columnwidth]{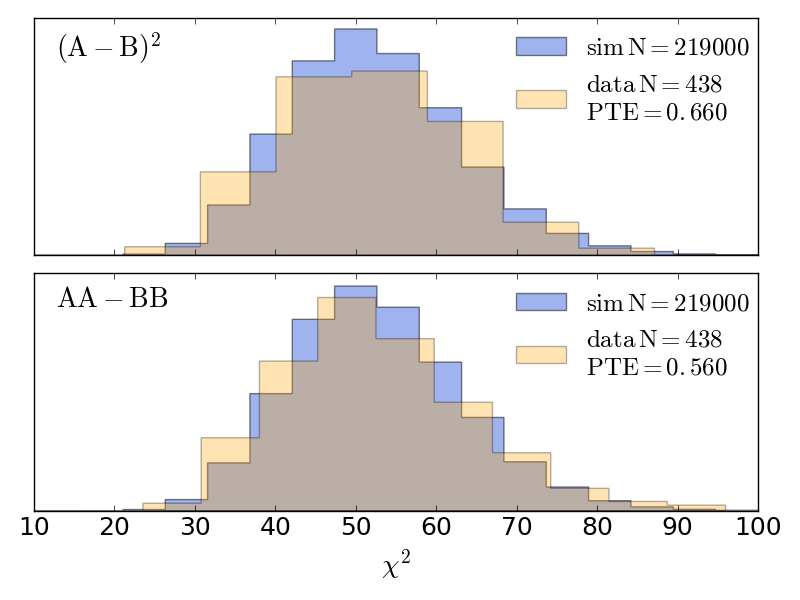}
\vspace{-0.2in}
\caption{Summary of pair-wise consistency checks for all combinations of spectra from D5, D6, D56, D8, BN, and AA. D1 has one spectrum so differences cannot be formed. The top panel shows the map-level differences and the bottom panel shows the power spectrum differences. The expectation is $\chi^2=n_{\ell,c}$, the number of $\ell$ bins. The probability to exceed (PTE), the probability of obtaining a higher $\chi^2$ than the measurement, is computed with the simulations. The histogram integrals are normalized to unity. Some of the elements of this are shown in Figure~\ref{fig:consist-d5-d6-d56}.}
\label{fig:consist-all}
\end{figure}

Once the null spectrum is found, either from individual spectra or combinations of spectra, we compute $\chi^2$ of the $n_{\ell,c}$ bins. Figure~\ref{fig:consist-all} shows the distribution of this
$\chi^2$ for the 438 difference spectra (the sum of the degrees of freedom in column 3 of Table~\ref{tab:consistency_summary} divided by $n_{\ell,c}$) compared to the distribution of nulls with 500 simulations for each. Additional comparisons of the data and inter-patch tests are shown in Appendix~\ref{appen:consistency_checks}. In summary, the spectra nulls within each region and the spectra nulls between regions are consistent with the nulls of simulations for both types of tests.

In a third form of consistency check we compute $\chi^2$ for each $\ell$ bin for the combination of \freqa, \freqa$\times$\freqb, and \freqb\,GHz spectra (where available) from all of D56, BN, D8, and AA. Each of these regions has more than one spectrum so we can subtract the mean spectrum and examine the distribution around it. Because we remove the mean, residual foreground emission will be removed to some degree. This test is done using only the diagonal elements of the covariance matrix, although the cosmic variance and noise terms are appropriately separated, thus it is less rigorous than the others. For AA, we combine spectra from the five different windows into one spectrum per frequency combination. We also include D8 in this test. Then for one $\ell$ bin in TT, there are 20 D56, 9 BN, 9 D8, and 15 AA spectra. With 9 degrees of freedom per $\ell$ bin the expectation is $\chi^2=44$.
A similar treatment for TE gives $\chi^2=71$.
In the last step, we sum over the $\ell$ bins in each spectra and find a reduced $\chi^2$ of 1.03, 0.94, 1.03, 0.86, 0.96, 1.13 for TT, TE, EE, TB, EB, and BB respectively. This shows that the distributions about the means are well behaved when many spectra are averaged together. The power of this test is that one checks for consistency with uncertainties of comparable size to those used in the cosmological analysis.

\section{Null tests}
\label{sec:null_tests}

We use null tests to target particular systematic effects. Specifically, we check that when the data are split roughly in half based on fast versus slow time constant, high versus low scan elevation, or high versus low precipitable water vapor (PWV), the two splits are consistent.  

Performing the null tests requires making new maps. We follow the same procedure as for the primary science maps (A20). In each of the three tests, the data are split at the TOD level to maximize the systematic in question while giving roughly equal statistical weight to each subset. From the ``null maps"  we compute the power spectrum of the difference. The error bars for these spectra are estimated analytically because generating simulations for all the null tests is computationally prohibitive. 

\subsection{Time constants}
\label{sec:tau_test}
The time response of each detector is limited by its electrothermal properties and in the low-inductance limit can be modeled as a one-pole filter with time constant $\tau=1/2\pi f_{\rm 3dB}$ \citep{irwin/hilton:2005}. 
The finite response time results in a small shift in the measured position of a point on the sky, depending on the scan direction. If not properly corrected, they can lead to a low-pass filtering of the data. The time constant null maps are designed to assess this effect.

We split the data so that ``low'' corresponds to 
$\tau$ below the median value in Table~\ref{tab:inst} and ``high" is above. The results of the test are given in Table~\ref{tab:null_results_tau} and Figure~\ref{fig:null_summary}. There is good consistency between the low and high detector time-constant data.

\begin{table}[tp!]
\caption{Summary of the Time Constant Null Tests}
\vspace{-0.15in}
\begin{center}
\begin{adjustbox}{max width=\columnwidth}
\begin{tabular}{c c c}
\hline
\hline
Array & Frequency & $\chi^2/\nu$ (PTE)\\
\hline
PA1 & \freqb\,GHz & 1285/1248 (0.23) \\
PA2 & \freqb\,GHz & 595/624 (0.79) \\
PA3 & \freqa\,GHz & 291/312 (0.80) \\
 & \freqb\,GHz & 294/312 (0.76) \\
\hline
\end{tabular}
\end{adjustbox} 
\end{center}
\vspace{-0.1in}
{\small For each array we report $\chi^2$/dof (PTE, probability to exceed).\\}
\label{tab:null_results_tau}
\end{table}

\subsection{Elevation of observations}
\label{sec:el_test}
Maps made from scans at low and high elevations ($\alpha$, see Table~\ref{tab:scans}) will have different levels of ground and atmosphere contamination. The elevation split is designed to search for this contamination. 

The elevation at which we split is computed separately for each region and varies between $\alpha=37^\circ$ and $\alpha=55^\circ$. In the BN and D8 regions it is not possible to split the TODs into a high-elevation and a low-elevation group while maintaining enough coverage and cross-linking to make a proper map of each. In these cases we make a map with 80\% of the high-elevation data and 20\% of the low-elevation data and compare this to a map with the percentages reversed. These ratios were chosen to ensure the resulting maps are sufficiently cross-linked and to ensure that the maps still allow us to test the effect of scans at different elevations.

\begin{figure}[tp!]
\epsscale{0.8} 
\includegraphics[width=\columnwidth]{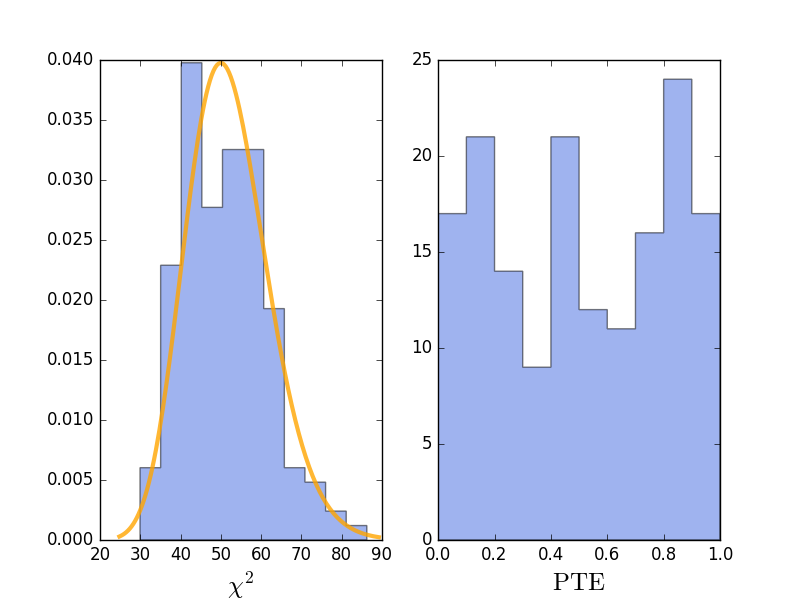}
\caption{{\it Left}: Histogram of $\chi^2$ from all 162 $\tau$, PWV, and elevation null tests is shown along with the expected distribution if the null tests were uncorrelated, although the correlations will have a negligible effect on this plot. The expectation is $\chi^2=n_{\ell,c}$ corresponding to the number of $\ell$ bins. {\it Right}: Distribution of the probability to exceed for each measure of $\chi^2$ on the left based on $n_{\ell,c}$ degrees of freedom.\\}
\label{fig:null_summary}
\end{figure}

The results of this null test are reported in 
Table~\ref{tab:null_results_el_pwv} and shown in Figure~\ref{fig:null_summary}. We see no evidence of an elevation-dependent effect.

\subsection{PWV}
\label{sec:pwv_test}
The atmosphere emits and absorbs at ACT frequencies, in large part due to the excitation of the vibrational and rotational modes of water molecules. Thus, the PWV is correlated with the level of optical loading on the detectors and to the level of atmospheric fluctuations. Both the increased loading and fluctuations could bias our maps. The low versus high PWV null test is designed to assess this possibility.

The median PWV is $W_v=0.88$\,mm with quartile breaks at 0.63 mm and 1.36 mm.
The PWV at which we split is different for each season and region. For D56 in s14, for example, TODs with $W_v<0.80\,$mm are ``low" and those with $W_v>0.80\,$mm are ``high." The dividing line ranges from $W_v=0.51\,$mm to $W_v=0.85\,$mm. In general, the TODs for null tests are split so that both subsets have similar noise levels, but for PWV splits there is an additional challenge.

\begin{figure}[tp!]
\epsscale{0.8} 
\includegraphics[width=\columnwidth]{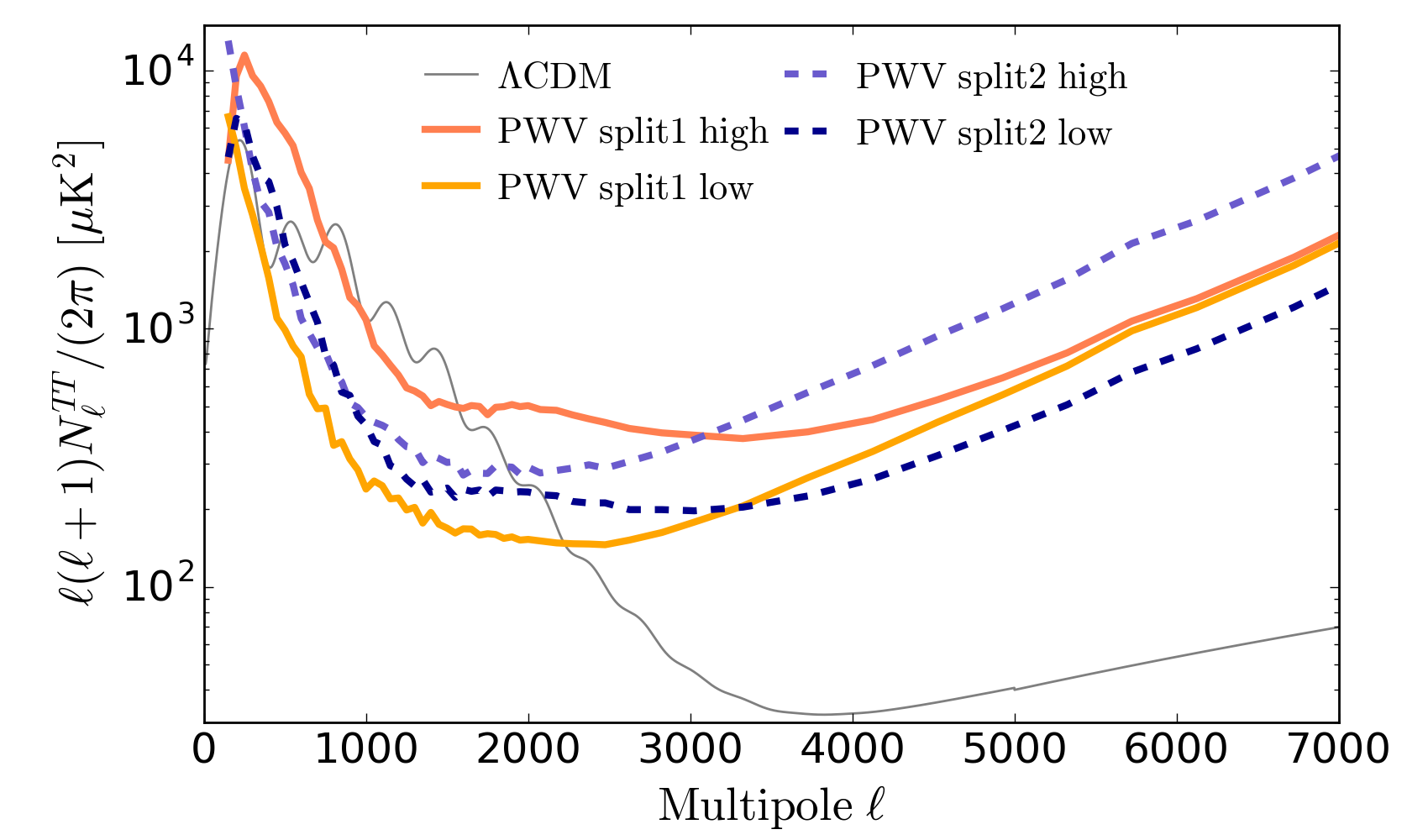}
\caption{Noise for different PWV splits for D56. Solid orange and yellow (dashed purple and blue) curves show the noise power spectra from data split into observations during high and low PWV respectively, keeping the small (large) angular scale noise similar in the two splits.}
\label{fig:pwv_null}
\end{figure}

 The noise spectra for the high-PWV and low-PWV splits are noticeably different. The $\ell\leq4000$ noise is due to the atmosphere and $\ell\geq4000$ noise is instrumental. So when the TODs are split such that the high-$\ell$ tails of the noise curves agree, as in Figure~\ref{fig:pwv_null}, the high-PWV split has more low-$\ell$ noise. Similarly, when the TODs are split such that the low-$\ell$ part of the curves agree, the high-$\ell$ parts do not. A few cases were tested and both PWV splits described above passed our null test. For the remaining tests the TODs were simply all split so that the high-$\ell$ tails of noise curves would agree.

We had one failure of this null test in the BN region 
at \freqa\,GHz for $\ell<600$. We are still investigating the source of this failure but as a precaution, we eliminated the \freqa\,GHz data for $\ell<600$ from the cosmological analysis. This was done before unblinding. After accounting for
this failure we see no evidence of a PWV-dependent effect. 
The results of this null test are reported in 
Table~\ref{tab:null_results_el_pwv} and Figure~\ref{fig:null_summary}.

\section{Additional systematic checks}
\label{sec:misc_syst}

As with any complex analysis, multiple decisions and checks of the data are made along the way. At a high level, all maps are visually inspected. Features that cannot be linked to ground pickup or regions of high noise are investigated. For example, an anomalous gain or time constant of a single detector in a single TOD is often visible in the maps. As another example, 
to identify contamination from beam side-lobes a different set of sky maps were made in Moon centered coordinates. Bright regions in these maps were identified, projected back into time-ordered form and subsequently flagged as cut (A20). Maps and spectra were made for data with the turn-around regions at the end of a scan excised. There was no effect and thus these segments were retained. A comprehensive assessment of these and other map-based investigations is given in A20.

For the spectra, we made ``waterfall plots" to look for outliers. To be specific, each row in the plot displayed pixels representing the $n_{\ell,c}$ elements of $\chi^2_{w,\ell}=(d_\ell-\bar{d_\ell})^2/\sigma_\ell^2$ where the $d_\ell$ are one of the spectra in Table~\ref{tab:num_ps}, $\bar{d_\ell}$ is the average spectrum of the region, and $\sigma_\ell^2$ is the uncertainty at each $\ell$ after excluding cosmic variance. Although different cut levels on 
$\chi^2_{w,\ell}$ were investigated, in the end no cut was made. All the spectra for this analysis come directly from the procedure in Section~\ref{sec:pipeline} with no additional cuts. 

\begin{table}[tp!]
\caption{Summary of Elevation and PWV null tests}
\vspace{-0.15in}
\begin{center}
\begin{adjustbox}{max width=\columnwidth}
\begin{tabular}{c c c}
\hline
\hline
Region & PWV $\chi^2/\nu$ (PTE) & Elevation $\chi^2/\nu$ (PTE)\\
\hline
D5 & - & 341/312 (0.12)\\
D6 & - & 289/312 (0.82)\\
D56 & 1152/1248 (0.98) & 1822/1872 (0.79) \\
BN & 1035/936 (0.013) & 1138/1248 (0.988)\\
\hline
\end{tabular}
\end{adjustbox}
\end{center}
\vspace{-0.1in}
{\small 
We report $\chi^2$/dof (PTE, probability to exceed) for the regions as shown. These values obtain after removing the \freqa\,GHz data that did not pass the null test.\\}
\label{tab:null_results_el_pwv}
\end{table}

\subsection{Unblinding}
\label{sec:unblinding}
All the tests described in the preceding sections were done before unblinding. After ``opening the box" to compare to models,
we noticed a lack in power in TT for $\ell<600$ relative to the {\sl WMAP} and {\sl Planck} data. Despite the large number of tests we did, we are still not certain of the source. We suspect it may be linked to our handling of the large low-$\ell$ noise from the atmosphere but have not yet found a definitive mechanism for such an effect. It is clear that the power at much lower $\ell$ is suppressed as shown in \citet{li/etal:2020}. As we show below, our cosmological results are broadly insensitive to using $\ell>600$, or $\ell>350$. However, as the TT spectrum for $\ell<600$ has been measured independently and with high accuracy by {\sl WMAP} and {\sl Planck}, for ACT we use TT at $\ell>600$ for our nominal data set. Since we do not know the source of the suppression, we use the pre-unblinding range for TE and EE, namely
$\ell>350$.

In addition, we found some features in the TE residuals that became more apparent when combining ACT with {\sl WMAP} in the parameter fits. This led to an assessment of potential sources of systematic error in ACT's TE spectrum. As a result we added a correction to the temperature to polarization leakage caused by the polarized sidelobes (L17) and a correction for the main beam temperature to polarization leakage (Section~\ref{sec:planet_mapping}). Both effects are described in more detail in A20. Neither effect was significant enough to fully ameliorate the tension between ACT's TE and the ACT plus {\sl WMAP} best fit model examined in A20. However, the corrections did reduce the residuals for ACT TE to $\Lambda$CDM (see Section~\ref{sec:leak_cor}).

\section{Diffuse Galactic Foreground Emission}
\label{sec:fg_diff}
\begin{table*}[tp!]
\caption{Dust power at 150 GHz in ($\mu$K)$^2$}
\vspace{-0.2in}
\begin{center}
\begin{tabular}{c c c c c c c }
\hline
Region & Window & TT & TE & EE & BB & CIB  \\
\hline\hline
AA & w0 & $39.7 \pm 0.8$ & $1.4 \pm 0.1$ & 
$0.5 \pm 0.0$ & $0.3 \pm 0.0$ & $10.1 \pm 0.4$ \\
AA & w1 & $17.4 \pm 1.2$ & $0.2 \pm 0.1$ & 
$0.2 \pm 0.1$ & $0.1 \pm 0.1$ & $6.3 \pm 0.7$ \\
AA & w2 & $2.8 \pm 0.8$ & $0.1\pm 0.1$ & 
$0.1\pm 0.0$ & $0.0 \pm 0.0$ & $8.1 \pm 0.5$ \\
AA & w3 & $3.0\pm 0.8$ & $0.1\pm 0.1$ & 
$0.1 \pm 0.0$ & $0.0 \pm 0.0$ & $8.4 \pm 0.5$ \\
AA & w4 & $6.9\pm 2.2$ & $0.2\pm 0.2$ & 
$0.1\pm 0.1$ & $0.1 \pm 0.1$ & $7.9 \pm 1.3$ \\
AA & w5 & $4.5\pm 3.4$ & $0.1\pm 0.3$ & 
$0.1\pm 0.1$ & $0.1\pm 0.1$ & $10.0\pm 1.9$ \\
BN &  & $4.5 \pm 0.3$ & $0.3\pm 0.1$ & 
$0.1\pm 0.0$ & $0.1\pm 0.0$ & $8.5\pm 0.2$ \\
D56 &  & $2.8\pm 0.5$ & $0.1\pm 0.1$ & 
$0.0\pm 0.1$ & $0.0\pm 0.1$ & $8.8\pm 0.5$ \\
D8 &  & $-0.2 \pm 0.6$ & $0.0 \pm 0.1$ & 
$0.1\pm 0.1$ & $0.1\pm 0.1$ & $8.8\pm 0.6$ \\
\hline
deep &  & $2.8\pm 0.5$ & $0.1\pm 0.1$ & 
$0.0\pm 0.1$ & $-0.0\pm 0.1$ & $8.8\pm 0.5$ \\
wide  & & $8.8\pm 0.3$ & $0.4\pm 0.0$ & $0.1\pm0.0$ & $0.1\pm0.0$ & $8.5\pm0.2$ \\
\hline
\end{tabular}
\end{center}
\vspace{-0.1in}
{\small The level of dust and CIB emission in the ${\cal D}_\ell$ spectrum. The entries correspond to $a_{\rm dust} {\cal{F}}^\mathrm{d}_{\rm{150}}g_1({\rm{150}})^2$ and
$a_{\rm CIB}{\cal F}_{150}^{\rm CIB}g_1^2(150)$ in the right-most column. All $a_{\rm dust}$ ($a_{\rm CIB}$) values are relative to the CMB and for a pivot scale of $\ell=500$ (3000). The errors are statistical only, do not include systematic uncertainty, and are rounded to the nearest tenth. Regions D1, D5, and D6 are too small for a robust fit thus ``deep" matches D56.}
\label{tab:dust_per_region}
\end{table*}

For all of our maps, we first apply the {\sl Planck} cosmology mask. In the remaining area, the diffuse Galactic foreground emission is on the order of 1\% the CMB in power for TT, TE and EE. This is one reason we pass the consistency tests between \freqa and \freqb\,GHz without accounting for it. For BB the dust emission is comparable to the lensing signal in some multipole ranges in some regions and so more care is required. In this section we compute the level of Galactic foreground emission using the {\sl Planck} 353\,GHz, and {\sl WMAP} K-band (at 22.4~\,GHz) maps as templates for dust and synchrotron emission. We fit the components separately. Since synchrotron emission is below our detection threshold we do not consider the correlations between the two components. In the next section we use results from the fit as priors in the cosmological likelihood. In all cases, our treatment of the foregrounds in regards to their effects on the cosmological parameters is done in the power spectra.
As part of this analysis we do not produce foreground-subtracted data products. However, \citet{madhavacheril/etal:2019} do produce component-separated maps but those were not part of this analysis.

In CMB temperature units, the power spectrum in our frequency range from diffuse Galactic sources is modeled as the CMB plus two foreground components:
\begin{equation} 
\mathcal{D}_\ell = \mathcal{D}_\ell^{\rm CMB} + \mathcal{D}_{\ell,\nu}^{\rm dust} + \mathcal{D}_{\ell,\nu}^{\rm sync.}.
\end{equation}
We model the dust power spectrum as 
\begin{equation} 
\mathcal{D}_{\ell,\nu}^{\rm dust} = a_{\rm dust}(\ell/500)^{\alpha_d+2}{\cal{F}}_{\nu}^d g_1(\nu)^2,
\label{eq:dust_model}
\end{equation}
where $a_{\rm dust}$ is the dust power in antenna temperature units,
$\alpha_d$ is $-2.42/-2.54$ for EE/BB polarization \citep{planck_poldust:2018}, $-2.4$ for TE, and $-2.6$ for TT \citep{planck_spectra:2019}.
The antenna to CMB temperature conversion factor is
$g_1(\nu)$,\footnote{$g_1(\nu)= (x^2 e^x/(e^x-1)^2)^{-1}$ with $x=h\nu/k_B T_{\rm CMB}$.} and ${\cal{F}}_{\nu}^d$ is the {\sl Planck} modified blackbody dust model in antenna temperature
\begin{equation}
{\cal{F}}^\mathrm{d}_{\nu} =
\big((\nu/\nu_\mathrm{353})^{\beta_\mathrm{d}-2} B_\nu(T_d)/B_{\nu=\nu_\mathrm{353}}(T_d)\big)^\mathrm{2},
\label{eq:dust_model_1}
\end{equation}
with $\mathrm{\beta_d} = 1.5$, $T_d = 19.6$\,K, and $\nu_{353} = 364.2$\,GHz. The effective frequency was found iteratively using the {\sl Planck} color corrections \citep{planck_color:2014,choi/2015}. We report all results for dust emission scaled to 150\,GHz.

Similarly, we model the synchrotron spectrum as
\begin{equation} 
\mathcal{D}_{\ell,\nu}^{\rm sync} = a_{\rm sync}(\ell/500)^{\alpha_s+2}(\nu/\nu_{22})^{2\beta_s}g_1(\nu)^2,
\end{equation}
where $a_{\rm sync}$ is the synchroton power in antenna temperature units,
$\alpha_s$ is approximated as $-2.8$ for EE/BB polarization \citep{planck_int_dust} and $-2.7$ for TE and TT. We use $\nu_{22}=22.4$\,GHz and $\beta_s=-2.7$ to be conservative. Although $\alpha_s$ is different for TT and EE, the effect is negligible for our purposes.

We compute TT, TE, EE, BB auto- and cross-frequency power spectra with ACT, {\sl Planck}, and {\sl WMAP}. For dust we use $\mathcal{D}_\ell^{150}$, $\mathcal{D}_\ell^{150\times353}$, and $\mathcal{D}_\ell^{353}$,
and for synchrotron we use $\mathcal{D}_\ell^{22}$, $\mathcal{D}_\ell^{22\times98}$, and $\mathcal{D}_\ell^{98}$. The maps are weighted with the ACT spatial window functions (Section~\ref{sec:spatial_window}) to compute $\mathcal{D}_\ell^{150}$ and $\mathcal{D}_\ell^{150\times353}$ and with the {\sl Planck} spatial window functions to compute $\mathcal{D}_\ell^{353}$ (similarly for power spectra with {\sl WMAP}). 

For the uncertainties of the {\sl Planck} and {\sl WMAP} foreground emission, we analytically estimate the covariance matrices (diagonal and pseudo-diagonal elements) but correct the effective number of modes with factors derived from comparing ACT simulation error bars to analytic error bars. The simulations are typically 1.1 to 1.3 higher near $\ell=300$ dropping to unity by $\ell=2000$. Thus we multiply the  {\sl Planck} and {\sl WMAP} error bars by these factors.

Because {\sl Planck}'s 353\,GHz map contains emission from both diffuse dust in our galaxy and from the cosmic infrared background (CIB), fitting for the dust model in Equation~\ref{eq:dust_model} alone is not sufficient. In order to extract only the Galactic term, we fit to the above plus
\begin{equation}
    a_{\rm CIB}{\cal F}_\nu^{\rm CIB}g_1(\nu)^2
\end{equation}
where ${\cal F}_\nu^{\rm CIB}$ is a template based on the third and fourth terms on the right hand side of Equation~\ref{eq:sec_model}, that sums the clustered and Poisson CIB components. An estimate of three coefficients associated with those terms, $A_{\rm d}$, $A_{\rm c}$ and $\beta_{\rm c}$, is needed to get the relative weight between the CIB clustered and Poisson components and to re-scale the template between frequencies. These values are found with an initial run of the multi-frequency likelihood code (described in the next section) with no priors imposed on the level of Galactic dust, that is, with freely varying dust amplitudes. The free dust parameter has no impact on the CIB estimates from the full likelihood because much of the support for the CIB model comes from $\ell>2000$. The value of $a_{\rm dust}$ in the Galactic fit, on the other hand, is sensitive to the choice of the CIB model because the CIB and dust are degenerate for $600<\ell<3000$ for constraining the total dust-like emission.

To evaluate the dust power, we form the map difference power spectrum to remove $\mathcal{D}_\ell^{\rm CMB}$. In the limit that the maps contain only the CMB, Galactic dust, and the CIB, and that the Galactic and CIB emission are uncorrelated, the spectrum of the residual in the map difference is given by:
\begin{eqnarray} 
\label{eq:dust_null}
\mathcal{D}_\ell^{A} + \mathcal{D}_\ell^{B} - 2\mathcal{D}_\ell^{A\times B} &= & a_{\rm dust}(\ell/500)^{\alpha_d+2}\\ \times  \big({\cal{F}}^d_{\rm{A}}g_1(\nu_{\rm A})^2 + {\cal{F}}^d_{\rm{B}}&g_1&(\nu_{\rm B})\rm{^2} - 2\sqrt{{\cal{F}}^d_{\rm{A}}{\cal{F}}^d_{\rm{B}}}g_1(\nu_{\rm A})g(\nu_{\rm B})\big) \notag \\
+ a_{\rm CIB}T_\ell\big({\cal{F}}^{\rm CIB}_{\rm{A}}&g_1&(\nu_{\rm A})^2 
+ {\cal{F}}^{\rm CIB}_{\rm{B}}g_1(\nu_{\rm B})\rm{^2} \notag \\
 &-& 2\sqrt{{\cal{F}}^{\rm CIB}_{\rm{A}}{\cal{F}}^{\rm CIB}_{\rm{B}}}g_1(\nu_{\rm A})g(\nu_{\rm B})\big) \notag
,
\notag
\end{eqnarray}
where $A$ is ACT's \freqb\,GHz map $B$ is {\sl Planck}'s 353 GHz map.
We then solve for $a_{\rm dust}$ and $a_{\rm CIB}$ with a linear least squares fit taking the left hand side of the expression as the data and the right side as the model. The uncertainty is found with multiple simulations drawn from the analytic covariance matrix.  
The associated covariance matrix and details are given in Appendix~\ref{appen:fgcalcs}. Although our fitted values for $a_{\rm CIB}$ are consistent with those found by {\sl Planck}, we do not use them in the cosmological analysis. 

\begin{figure}[!tp]
\epsscale{1.1}
\includegraphics[width=\columnwidth]{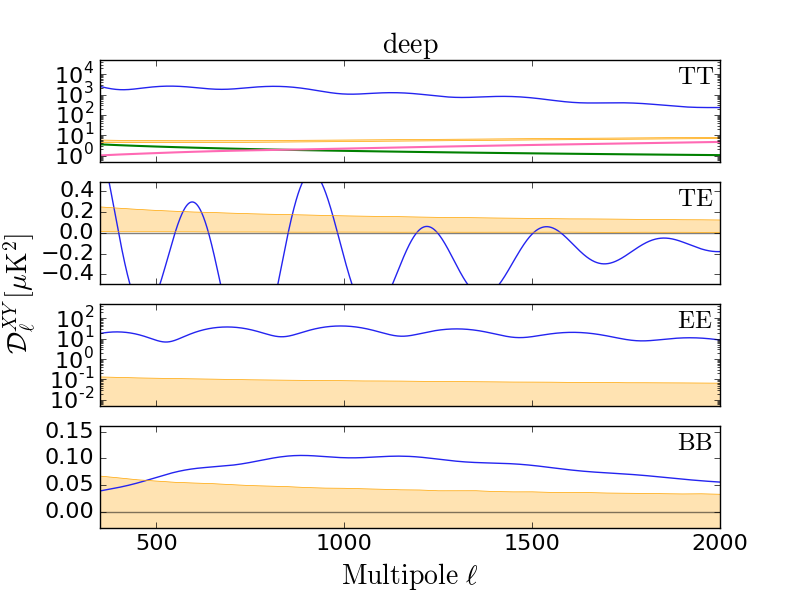}
\includegraphics[width=\columnwidth]{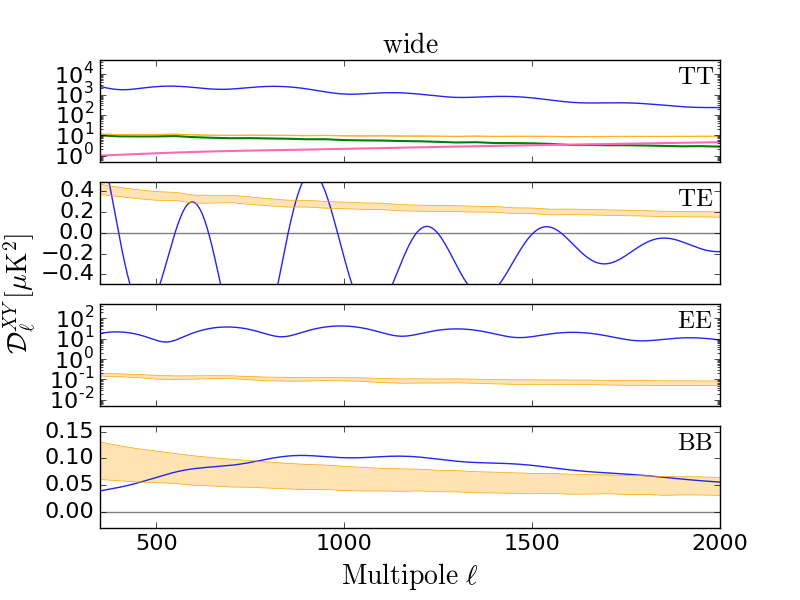}
\caption{The diffuse dust emission at \freqb\,GHz for the deep (top, 15 mJy source cut) and wide (bottom, 100 mJy source cut) Galactic dust levels with the statistical 1$\sigma$ error band in yellow and $\Lambda$CDM theory spectra in blue. For TE, the theory spectrum was multiplied by 0.01. The falling (rising) solid line in green (pink) in the TT panel is the Galactic dust (CIB).}
\label{fig:diff_dust}
\end{figure}

The solution for $a_{\rm dust}$ depends on region as shown in Table~\ref{tab:dust_per_region}. For use in the likelihood, we also fit for the net amount of dust in the deep and wide regions.
These are plotted in Figure~\ref{fig:diff_dust}. To check the results we can also simply scale the {\sl Planck} 353\,GHz spectra to ACT frequencies after accounting for the CMB. In all cases the extrapolated results are consistent with the fits.

\begin{figure}[tp!]
\epsscale{0.8} 
\includegraphics[width=\columnwidth]{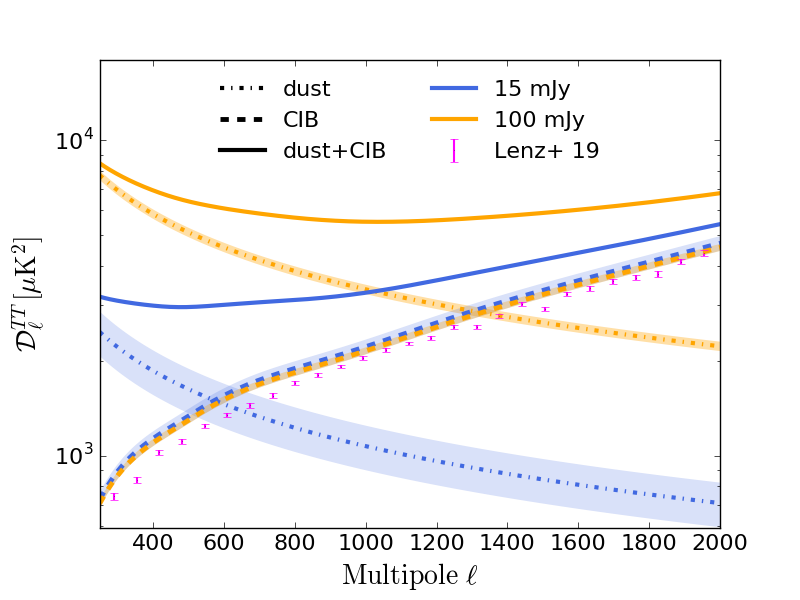}
\caption{The composition of the 353 GHz power spectrum from Planck. After subtracting the CMB power, one is left with a combination of Galactic dust and CIB. The (falling) dash-dot lines show the dust spectra in the wide (100 mJy threshold) and deep (15 mJy threshold) regions. The (rising) dashed lines similarly show the CIB spectra. The solid lines show the total. Because of the shape of the dust and CIB spectra in this region, the two components can play off each other to combine for the same total. For our template, $a_{\rm dust}$ and $a_{\rm CIB}$ have a correlation coefficient of $-0.9$
for $350<\ell<2000$. A different CIB template can also fit if the dust level is adjusted accordingly. 
However, the agreement of the \citet{lenz/etal:2019} CIB model with the CIB level we deduce independently from the wide and deep regions lends confidence to our determination of $a_{\rm dust}$.}
\label{fig:cib_dust}
\end{figure}

As a check of the method, we compare to the \citet{lenz/etal:2019} CIB power spectrum. In contrast to our model, theirs was fitted for $70<\ell<1500$. To clean Galactic contamination from the {\sl Planck} maps they used HI4PI neutral hydrogen maps \citep{h14pi:2016} in the low dust regions. Although we cannot compare their HI derived dust levels to ours, because they depend on region, we can compare the deduced isotropic CIB component. The result is shown in Figure~\ref{fig:cib_dust}. It is reassuring that the fitted level for the CIB in both deep and wide regions is similar even though the dust levels are quite different. An approximation of the systematic error in $a_{\rm dust}$ is the difference between the \cite{lenz/etal:2019} spectrum and our fitted level for the template at $\ell=500$, or roughly 20\%. 

One of the limitations of the current treatment of foreground emission is that we characterize the level of dust emission in the wide region with a single parameter. As 
Table~\ref{tab:dust_per_region} shows, there is a large variation in dust amplitude for w0 to w5. Future analyses will address a more spatially fine-grained model. 

The solution for $a_{\rm sync}$ is similar but with the model given by  
\begin{eqnarray} 
\label{eq:sync_null}
\mathcal{D}_\ell^{A} + \mathcal{D}_\ell^{B} - 2\mathcal{D}_\ell^{A\times B} &= & a_{\rm sync}\big(\frac{\ell}{500}\big)^{\alpha_s+2}\notag\\ \times  \bigg[\bigg(\frac{\nu_A}{\nu_{22}}\bigg)^{2\beta_s}g_1(\nu_{\rm A})^2 &+& \bigg(\frac{\nu_B}{\nu_{22}}\bigg)^{2\beta_s}g_1(\nu_{\rm B})\rm{^2}\\ 
&-& 2\bigg(\frac{\sqrt{\nu_A\nu_B}}{\nu_{22}}\bigg)^{2\beta_s} g_1(\nu_{\rm A})g_1(\nu_{\rm B}))\bigg].
\notag
\end{eqnarray}

In this case, all fits are consistent with zero as shown in Appendix~\ref{appen:fgcalcs}. If instead of fitting we simply extrapolate the {\sl WMAP} spectrum for our region, there are indications that at \freqa\,GHz there may be a contribution in BN in TT at a third or less the dust level at \freqb\,GHz.  

For both the Galactic dust and synchrotron emission, our model should be interpreted as a baseline upon which future investigations will improve. For example, we do not yet include a possible CO contribution. And, as noted, we quantify all the emission from deep and wide regions with one number each. The primary change in cosmological parameters between including a Galactic dust correction or not is an upward shift of $N_{\rm eff}$ by about $0.5\sigma$ (see A20). We would expect future potential shifts from improvements on the Galactic model to be less than this. 

\section{Parameter Likelihood}
\label{sec:likelihood}

Following the methodology described in \cite{dunkley/etal:2013}, we use the set of 10 coadded TT/TE/EE spectra, described in Sections~\ref{sec:data_select} and \ref{sec:pipeline}, and their covariance matrices to build a multi-frequency Gaussian likelihood that models known millimeter-wave emission present in the data: 
\begin{eqnarray}
-2\rm{ln}\mathcal{L} = (C_b^{\rm th}-C_b^{\rm data})^T {\bf \Sigma}^{-1}
(C_{b^\prime}^{\rm th}-C_{b^\prime}^{\rm data}) \,,
\end{eqnarray} 
where $C_b^{\rm th} = C_b^{\rm CMB} + C_b^{\rm FG}$ is the binned theoretical model including CMB, Galactic and extragalactic foreground emission (FG), $C_b^{\rm data}$ is the data vector and $\Sigma$ the band power covariance matrix described in Section~\ref{sec:act_covmat}. 
The deep and wide regions are then combined at the likelihood level, 
\begin{eqnarray}
 -2\rm{ln}\mathcal{L}_{\rm ACT}=-2\rm{ln}\mathcal{L}_{\rm ACT, deep}-2\rm{ln}\mathcal{L}_{\rm ACT, wide}. 
\end{eqnarray}
As baseline we consider $600.5\leq\ell\leq7525.5$ for all TT spectra and $350.5\leq\ell\leq7525.5$ for all TE/EE spectra. This results in 47 $\ell$ bins in each TT spectrum and 52 in the TE/EE spectra.
For some aspects of the analysis, noted below, we include data from MBAC in DR2 by adding $-2\ln\mathcal{L}_{\rm MBAC}$ to the likelihood.
Its 217\,GHz channel is helpful in constraining foreground emission. 

\subsection{Foreground model for TT, TE, and EE}
\label{sec:likelihood_foregrounds}
In temperature the foreground model is the same as in \cite{dunkley/etal:2013}. A similar model is used to fit the {\sl Planck} power spectra \citep{planck_spectra:2019}. The equation for the model is given in Appendix~\ref{appen:fgcalcs} and includes:

1)\,The thermal and kinetic Sunyaev-Zel'dovich effects \citep{sunyaev/zeldovich:1972}. These are characterized with amplitude parameters $A_{\rm tSZ}$ and $A_{\rm kSZ}$ that scale the template spectrum (\citet{battaglia/etal:2012} for tSZ and \citet{battaglia/etal:2010} for kSZ) to $\ell=3000$ at 150 GHz.

2)\,Both Poisson and clustered terms for the CIB. These are parametrized with an amplitude parameter, $A_{\rm d}$, that scales a shot noise spectrum for the Poisson term; a separate amplitude parameter, $A_{\rm c}$, that scales a hybrid template ({\sl Planck} CIB + $\ell^{0.8}$; see Appendix~\ref{appen:fgcalcs}) for the clustered component; and a third parameter, $\beta_{\rm c}$, that scales the frequency spectrum. Because of the small CIB clustered contribution at ACT frequencies we impose a Gaussian prior on $A_{\rm c}$ of $4.9\pm0.9$ from ACT MBAC constraints~\citep{dunkley/etal:2013}. In fits that involve 
$\cal{L}_{\rm MBAC}$ we remove this prior.

3)\,A cross-correlation signal arising from ${\rm tSZ}$ and the clustered CIB, scaled with a correlation parameter, $\xi$ \citep{addison/dunkley/spergel:2012}.

4)\,Different levels of power from unresolved radio sources for the deep and wide regions as characterized by $A_{\rm s, d}$ and $A_{\rm s, w}$ respectively.
Both are referenced to $\ell=3000$ at 150 GHz. Similarly to \cite{dunkley/etal:2013} we impose a Gaussian prior on these amplitudes. For the deep region, $A_{\rm s, d}=3.1\pm0.4$ as in previous ACT analyses because it has the same 15\,mJ flux cut. For the wide region, 
$A_{\rm s, w}=22.5\pm3.0$ based on a prediction of the residual level for a 100\,mJ flux cut and a similar 15\% uncertainty. In practice the prior on $A_{\rm s, w}$ has no effect because the data constrain the parameter better than the prior.

5)\,Diffuse Galactic emission. For thermal dust we use the assessment in Section~\ref{sec:fg_diff} to impose a prior on the amplitude at $\ell=500$ at 150 GHz separately in both the wide and deep regions, $A_{\rm dust, d}^{\rm TT}$ and $A_{\rm dust, w}^{\rm TT}$, scaling a power law template ($C_\ell\propto\ell^{-0.6}$). We found that Galactic synchrotron emission is negligible in our regions and therefore it is not included in the baseline likelihood model.

For polarization we include:

6)\,Poisson sources in both TE and EE with a free amplitude at $\ell=3000$ at 150 GHz, $A_{\rm ps}^{\rm TE}$ and $A_{\rm ps}^{\rm EE}$. We impose a positive prior on $A_{\rm ps}^{\rm EE}$ but allow $A_{\rm ps}^{\rm TE}$ to take negative values;

7)\,Diffuse polarized Galactic dust emission. We vary the amplitude of thermal dust emission, with Gaussian priors, at $\ell=500$ at \freqb\,GHz in each of deep and wide TE and EE, $A_{\rm dust, d}^{\rm TE}$, $A_{\rm dust, w}^{\rm TE}$, $A_{\rm dust, d}^{\rm EE}$ and $A_{\rm dust, w}^{\rm EE}$. Following {\sl Planck}, the dust emission is modeled as a power law $C_\ell\propto\ell^{-0.4}$. As for temperature, we find negligible levels of synchrotron emission in both regions. 

All these components have color corrections as shown in Appendix~\ref{appen:fgcalcs} and use updated effective frequencies for the ACTPol receiver.

\subsection{Polarization efficiencies}
While an overall calibration is applied to the spectra before coadding them into the wide and deep regions, and the uncertainty is propagated to the covariance matrix, polarization efficiencies are left as free parameters in the multi-frequency likelihood. We compare the data with a corrected model as 
\begin{eqnarray}
{C_{\ell, ij}^{\rm d,TE}} &=& y^{\rm P}_{j}{C_{\ell,ij}^{\rm m,TE}} \\
{C_{\ell,ij}^{\rm d,EE}} &= &y^{\rm P}_i y^{\rm P}_j{C_{\ell,ij}^{\rm m,EE}},
\end{eqnarray}
where for TE, to account for $T_iE_j \neq T_jE_i$, we have:
\begin{eqnarray}
{C_{\ell,98\times150}^{\rm d,TE}} &=& y^{\rm P}_{150} {C_{\ell,98\times150}^{\rm m,TE}}\,,  \\
{C_{\ell,150\times98}^{\rm d,TE}} &=& y^{\rm P}_{98} {C_{\ell,150\times98}^{\rm m,TE}} \,.
\end{eqnarray}
The $i$ and $j$ subscripts denote different spectra, the $d$ superscript denotes data and $m$ denotes the model leading to the expectation that $y^P<1$. 
The $y^{\rm P}$ parameters are varied with uniform prior between 0.9 and 1.1. \\

To summarize the preceding two sections, there are eighteen active foreground and nuisance parameters ($A_{\rm tSZ}$, $A_{\rm kSZ}$, $\xi$, $A_{\rm d}$, $A_{\rm c}$, $\beta_{\rm c}$, $A_{\rm s, d}$, $A_{\rm s, w}$, $A_{\rm dust, d}^{\rm TT}$, $A_{\rm dust, w}^{\rm TT}$, $A_{\rm dust, d}^{\rm TE}$, $A_{\rm dust, w}^{\rm TE}$, $A_{\rm dust, d}^{\rm EE}$, $A_{\rm dust, w}^{\rm EE}$, $A_{\rm ps}^{\rm TE}$, $A_{\rm ps}^{\rm EE}$, $y^{\rm P}_{\rm 98}$, $y^{\rm P}_{\rm 150}$).

\subsection{Leakage Corrections}
\label{sec:leak_cor}
The main beam leakage and residual leakage from the polarized sidelobes terms (see A20) are included in the likelihood with a leakage-corrected model for TE and EE at each call:
\begin{eqnarray}
T_iE_j^{'} &=& T_iE_j+T_iT_{j}\gamma_{j} \\
E_iE_j^{'} &=& E_iE_{j}+T_iE_j\gamma_{i}+T_jE_{i}\gamma_{j}+T_iT_{j}\gamma_{i}\gamma_{j} \,.
\end{eqnarray}

The $\gamma$ factors encode the computed leakage and its $\ell$ dependence. At \freqa\,GHz, the maximum of $\gamma_{98}=0.038$; at \freqb\,GHz, $|\gamma_{150}|<0.0035$.
In the baseline case, the $\gamma$ factors are fixed to their nominal amplitudes because their uncertainties have already been incorporated in the data covariance matrix. We fit two scaling factors (one for \freqa\,GHz and one for \freqb\,GHz) multiplying the nominal values for the gamma factors and find that the data support the baseline model (the scaling factors are unity). The main impact of these leakage corrections, as opposed to no leakage correction ($\gamma=0$), is to reduce the TE residuals to $\Lambda$CDM.  The residuals are reduced to different degrees depending on the region and $\ell$ range. For the deep region, where the effect is most pronounced, 
the leakage at \freqb\,GHz is larger than at \freqa\,GHz whereas the reverse is true for the wide region.

\subsection{Validation on simulations}
We validate the full power spectrum and likelihood pipeline on a suite of 100 CMB plus foreground simulations following the prescription in Section~\ref{sec:sims}. As done for the data, we combine at the likelihood level 100 simulations for the deep region and 100 for the wide region. 

We run cosmological estimates from the simulations using the public CosmoMC package \citep{2002PhRvD..66j3511L} and consider the basic six $\Lambda$CDM cosmological parameters: $\Omega_b h^2$, $\Omega_c h^2$, $\theta$, $\tau$, $A_s$ and $n_s$, with a pivot scale $k_0 = 0.05$ Mpc$^{-1}$, and fix the total mass of neutrino particles to 0.06\,eV. We also vary all foreground parameters in various combinations. 
We recover unbiased results for all parameters and in particular the cosmological parameters are all within $0.2\sigma$ of the input value. To guarantee statistically unbiased results for the simulations, we varied all parameters in symmetric ranges, even allowing amplitudes that for physical reasons must be positive (e.g., the amplitude of the term for Poisson sources) to take negative values. These unphysical values were not included when fitting the data.

\subsection{Inputs to the likelihood}
\begin{figure*}[!h]
\centering
\includegraphics[width=5.5truein]{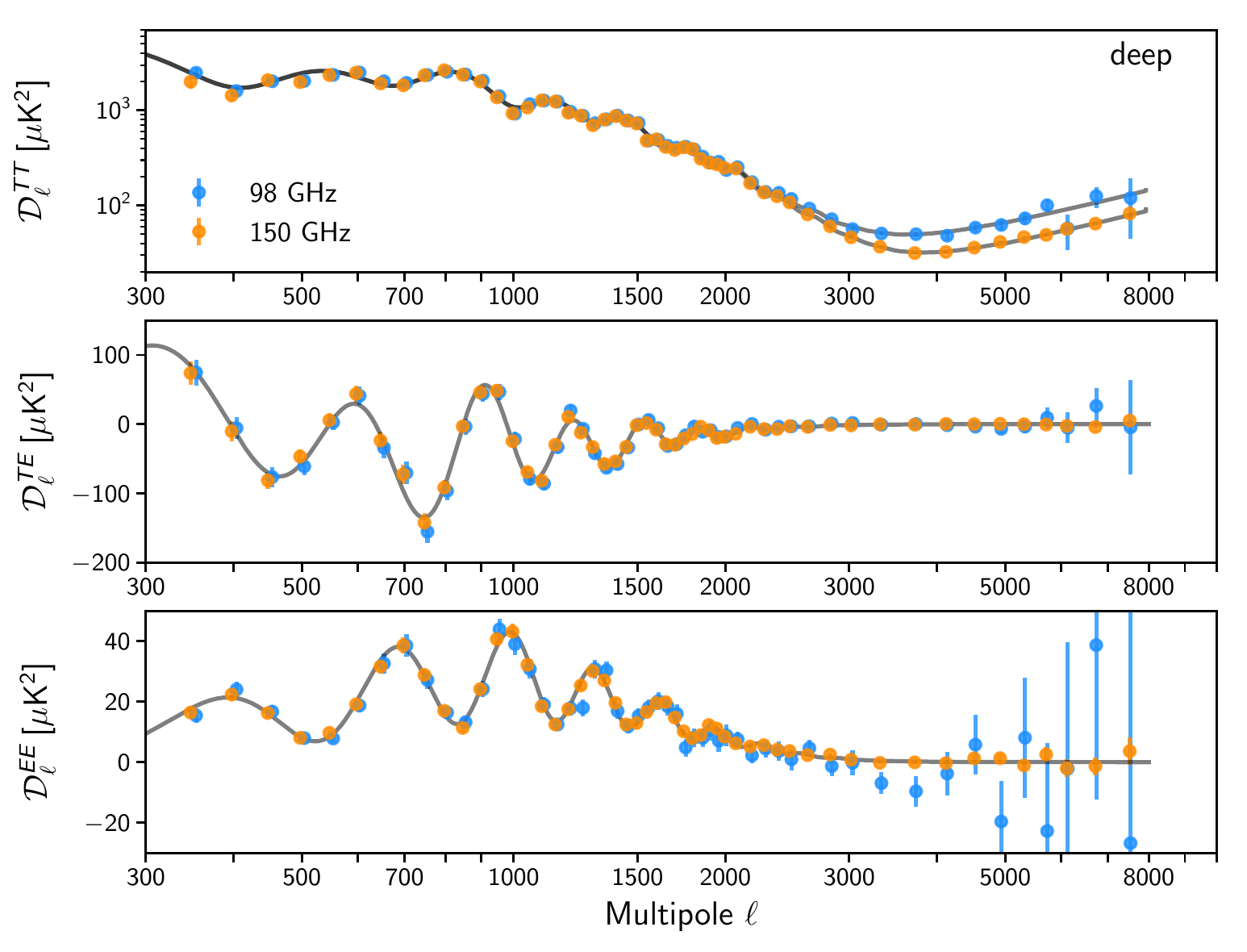}
\includegraphics[width=5.5truein]{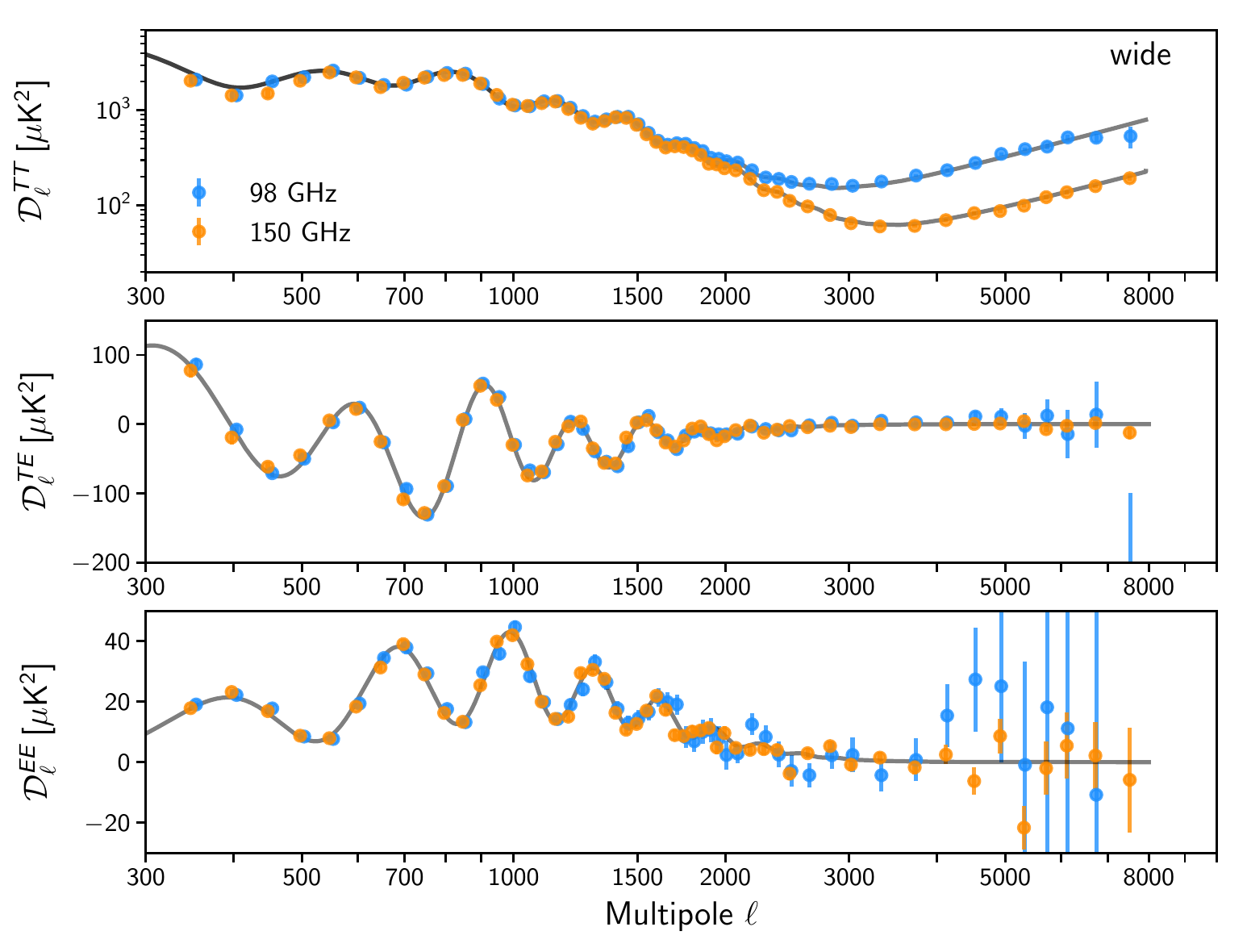}
\caption{The deep (top) and wide (bottom) spectra used in the likelihood. The gray line is the best fit $\Lambda$CDM plus foreground model for ACT only. The \freqa$\times$\freqb\,GHz spectra are omitted for clarity. The full coadd of these spectra plus the frequency cross spectra is given in Table~\ref{tab:specs}.}
\label{fig:likelihood_input}
\end{figure*}

Figure~\ref{fig:likelihood_input} shows the raw \freqa and \freqb\,GHz spectra that are inputs for the likelihood. A number of features are evident. 
$a$)\,There is a power deficit in TT compared to the best-fitting model for $\ell<600$ especially in the wide region. $b$)\,The agreement between \freqa\,GHz and \freqb\,GHz is clear. $c$)\,The Poisson point-source tail has a higher amplitude in the wide region. $d$)\,The error bars are smaller at high $\ell$ in the deep region when compared to the wide region, but at low $\ell$ the error bars for the wide region are comparatively smaller. The latter is due to cosmic variance.

\subsection{Likelihood results}
The multi-frequency likelihood has two primary applications: we use it directly in joint cosmological and foreground/nuisance parameters fits as done for the simulations, or we use it to derive and separate the CMB-only power spectrum and the foreground levels as in
\citet{dunkley/etal:2013}. These two approaches lead to consistent cosmological results.

\subsubsection{ACTPol joint $\Lambda$CDM and foreground fit}
\label{subsec:lcdm_fit}

\begin{table}[tp!]
\caption{Standard $\Lambda$CDM parameters and foreground parameters}
\vspace{-0.2in}
\begin{center}
\begin{tabular}{lcc}
\hline\hline
 Parameter & Prior & ACTPol\\
\hline
$100\Omega_b h^2$ & & $2.145 \pm 0.031$\\
$100\Omega_c h^2$ & & $11.84 \pm 0.38$ \\
$10^4 \theta_A$ & & $104.221 \pm 0.071$\\
$\tau$ & $0.065 \pm 0.015$ & $0.063 \pm 0.014$\\
$n_s$ & & $1.006 \pm 0.015$\\
ln$(10^{10} A_s)$ & & $3.046 \pm 0.030$\\
\hline
$H_0$ & & $ 67.6 \pm 1.5$\\
$\sigma_8$ & & $ 0.825 \pm 0.016 $\\
\hline
$A_{\rm tSZ}$ &\gt 0& $5.29 \pm 0.66$\\
$A_{\rm kSZ}$ &\gt 0& $\lt 1.8$ \\
$\xi$ & [0,0.2] & $\gt 0.047$ \\
$A_{\rm d}$ & & $6.58 \pm 0.37$ \\
$A_{\rm c}$ & $4.9 \pm 0.9$ & $3.15 \pm 0.72$ \\
$\beta_c$ & & $2.87^{+0.34}_{-0.54}$\\
$A_{\rm s, d}$& $3.1 \pm 0.4$ & $3.74 \pm 0.24$\\
$A_{\rm s, w}$ & & $22.56 \pm 0.33$\\
$A^{\rm TT}_{\rm dust, d}$&  $2.79 \pm 0.45$ & $2.79 \pm 0.44$\\
$A^{\rm TT}_{\rm dust, w}$ & $8.77 \pm 0.30$ & $8.70 \pm 0.30$\\
$A^{\rm TE}_{\rm dust, d}$& \gt 0, $0.11 \pm 0.10$ & $\lt 0.27$\\
$A^{\rm TE}_{\rm dust, w}$ & $0.36 \pm 0.04$ & $0.355 \pm 0.040$\\
$A^{\rm EE}_{\rm dust, d}$ & \gt 0, $0.04 \pm 0.08$ & $\lt 0.17$\\
$A^{\rm EE}_{\rm dust, w}$ & $0.13 \pm 0.03$ & $0.130 \pm 0.030$\\
$A^{\rm TE}_{\rm ps}$ & [$-1$,1]& $0.042\pm0.055$\\
$A^{\rm EE}_{\rm ps}$ &\gt 0 & $\lt 0.064$\\
$y^{\rm P}_{98}$ && $0.9853 \pm 0.0054$\\
$y^{\rm P}_{150}$ && $0.9705 \pm 0.0045$\\
\hline
\end{tabular}
\end{center}
\vspace{-0.1in}
\small{Uncertainties are 68\% confidence level or 95\% upper/lower limits. Ranges of variation and priors are also reported. The TT $\ell<600$ data have been cut for the above. \\}
\label{table:lcdmfg}
\end{table}

Our baseline results for standard $\Lambda$CDM and eighteen foreground and nuisance parameters are shown in Table~\ref{table:lcdmfg}. Because we do not measure the polarization at low $\ell$ we adopt a prior of $\tau=0.065\pm 0.015$. It encompasses expectations from {\sl Planck} \citep{planck_spectra:2019} and {\sl WMAP} \citep{weiland/etal:2018} as noted in~\cite{Natale:2020owc}. An interpretation of the results is given in A20. 

 The fit has a $\chi^2=1061$ (500 for deep and 560 for wide, rounded to the first digit) for 1010 band powers. Accounting for the 24 parameters and nine priors results in 995 degrees of freedom, corresponding to a reduced $\chi^2=1.07$ and PTE of 0.07.  When the total $\chi^2$ is broken down by spectrum, we find
$\chi^2=311$ for TT (148 for deep and 163 for wide), $\chi^2=402$ for TE (201 for deep and 201 for wide), and $\chi^2=322$ for EE (126 for deep and 196 for wide). To put these contributions in context, there are 282 TT, 416 TE, and 312 EE band powers.
The $\chi^2$ for deep and wide regions may be added because the regions are uncorrelated and separate in the likelihood. In contrast, the $\chi^2$ for the separate spectra (TT, TE, and EE) are combined with a covariance matrix and therefore the total is not the sum of the individual $\chi^2$s. 
Considering only $\ell<2000$ for this fit we find $\chi^2=639$ (313 for deep and 326 for wide, 630 total band powers), suggesting that part of the source of the $\Delta\chi^2=60$ between deep and wide in the full fit is due to foreground and secondary source modeling (see also Section~\ref{sec:Fg_levels}).

\begin{figure}[tp!]
    \centering
    \includegraphics[width=\columnwidth]{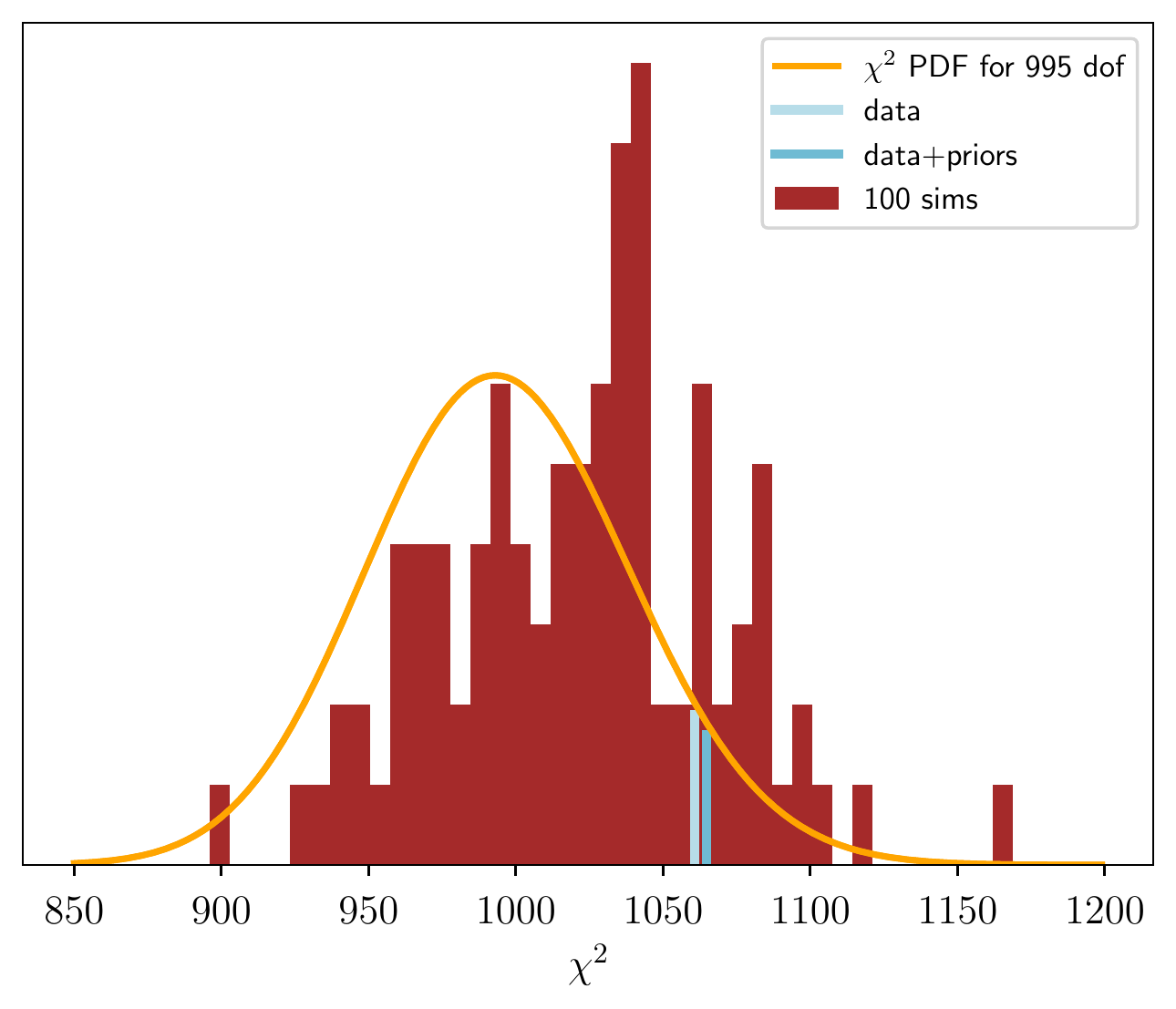}
    \caption{Best-fit $\Lambda$CDM $\chi^2$ from the data (blue), 100 simulations (dark red) and the expected Gaussian PDF from 995 degrees of freedom (computed from 1010 data points, 24 varied parameters and 9 priors). To include the impact of marginalizing over a dust prior in simulations having no dust levels, we impose the same prior width as for the data but center the prior on zero for all dust parameters. The simulations do not include leakage corrections in the modeling. The height of the blue lines is normalized to the distribution of the analytic prediction at the $\chi^2$ for the data. Different shades of blue show the $\chi^2$ contribution coming from the data likelihood only (light blue) or when also including the term coming from the priors (dark blue). The comparison between the simple analytic Gaussian distribution and the data/simulations highlights the complexity of our covariance matrix which includes multiple non-Gaussian terms. The PTE for the best fit model against simulations is higher than the 0.07 quoted for the analytic expression and in either case the best fit is well within expectations for $\Lambda$CDM.}
    \label{fig:chi2datasims}
\end{figure}

\begin{figure}[tp!]
    \centering
    \includegraphics[width=\columnwidth]{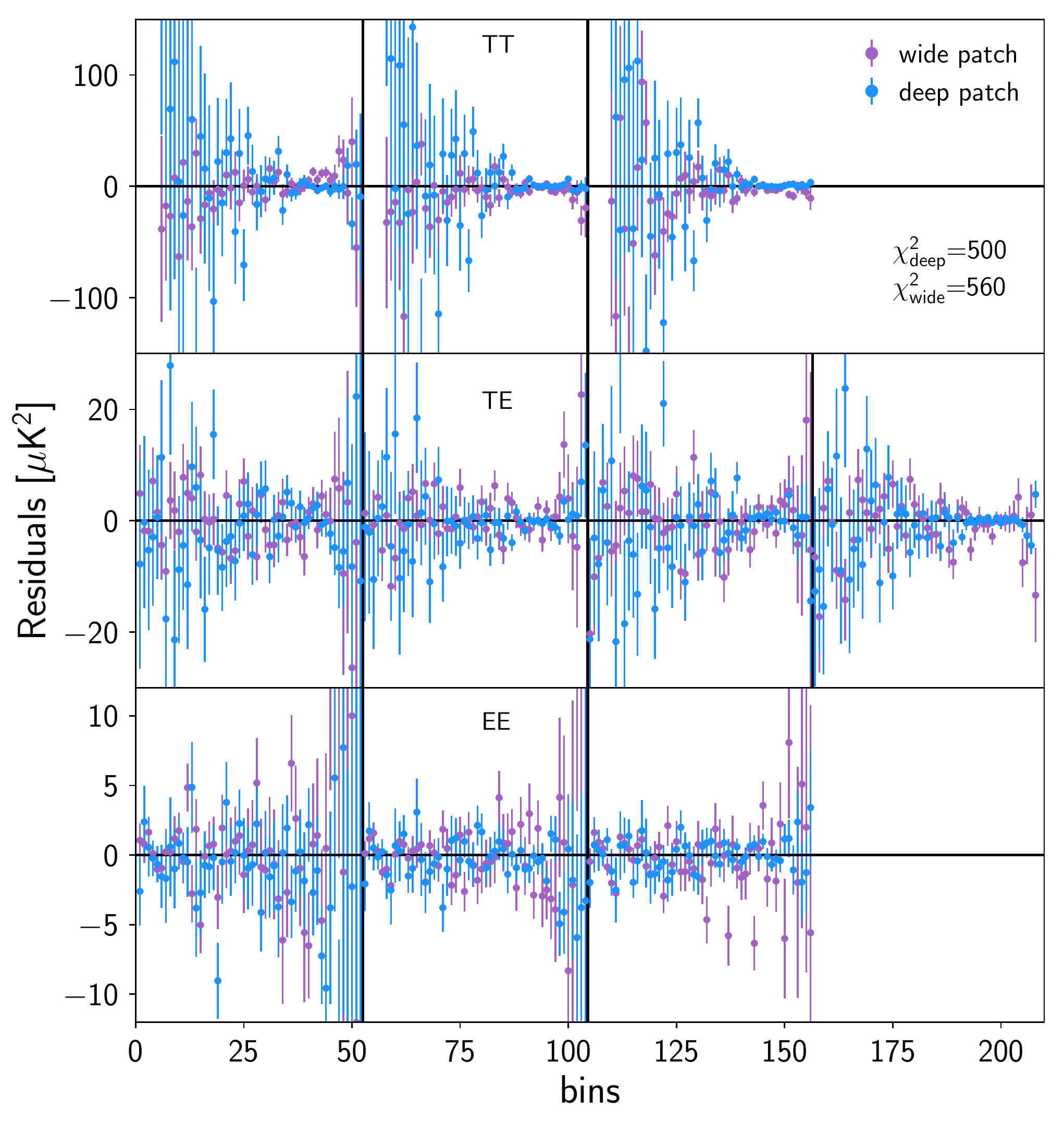}
    \caption{Residuals between the ACTPol deep (blue) and wide (purple) data and the best-fit $\Lambda$CDM plus foreground model. The panels span: TT \freqa $\times$\freqa, \freqa $\times$\freqb, \freqb $\times$\freqb (top), TE \freqa $\times$\freqa, \freqa $\times$\freqb, \freqb $\times$\freqa, \freqb $\times$\freqb (middle), and EE \freqa $\times$\freqa, \freqa $\times$\freqb, \freqb $\times$\freqb (bottom). Bins 1, 53, 105, and 157 corresponds to $\ell=350.5$, and bins 52, 104, 156, and 208 to $\ell=7525.5$.  The $\chi^2$ values are for the joint deep plus wide $\Lambda$CDM best-fit, for which $\chi^2=1061$, but broken down by region. For each region we fit 505 band powers with 16 common cosmological and foreground parameters and 8 separate foreground parameters thus we do not compute separate degrees of freedom for each 
    region. As described in Section~\ref{subsec:lcdm_fit}, when broken down by spectrum type, $\chi^2=312$ for TT only, $\chi^2=402$ for TT only, and $\chi^2=322$ for EE only.}
    \label{fig:resalllcdm}
\end{figure}

\begin{figure}[tp!]
    \centering
    \includegraphics[width=\columnwidth]{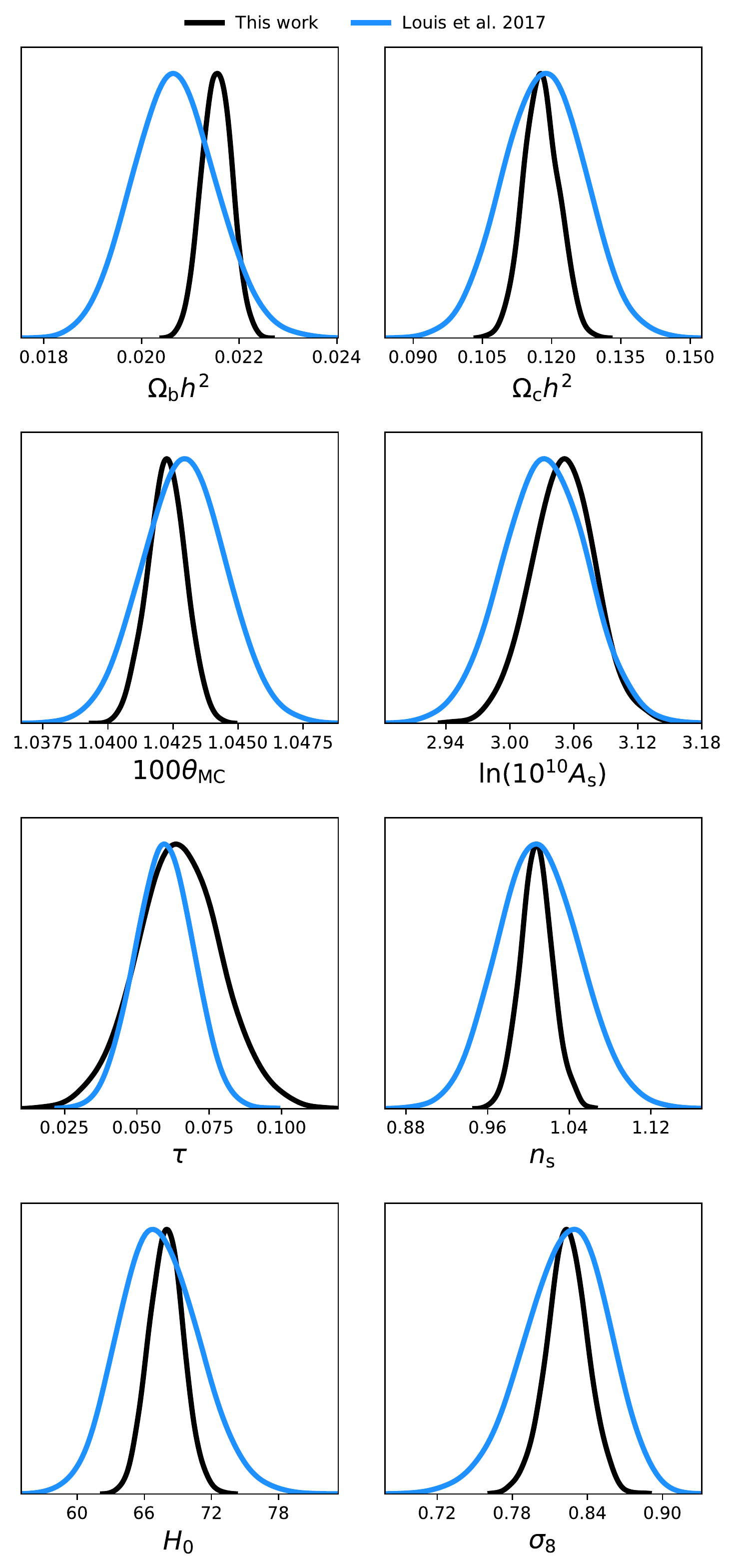}
    \caption{Constraints on the basic $\Lambda$CDM parameters as obtained in this work and the previous results reported in \cite{louis/2017}. We note that the difference in amplitude and optical depth are due to the different $\tau$ prior used in the two analyses.}
    \label{fig:CLlcdm}
\end{figure}

Given the structure of our covariance matrix, a better assessment of the goodness of the fit comes from comparing the result with simulations as shown in Figure~\ref{fig:chi2datasims}. The agreement with $\Lambda$CDM is well within expectations. Residuals for the deep and wide spectra with respect to this best-fit model are shown in Figure~\ref{fig:resalllcdm}. Figure~\ref{fig:CLlcdm} also shows good agreement between the basic $\Lambda$CDM parameters from L17 and this work. The $\Lambda$CDM model continues to describe the ACT data.

\begin{figure*}[tp!]
    \centering
    \includegraphics[width=\textwidth]{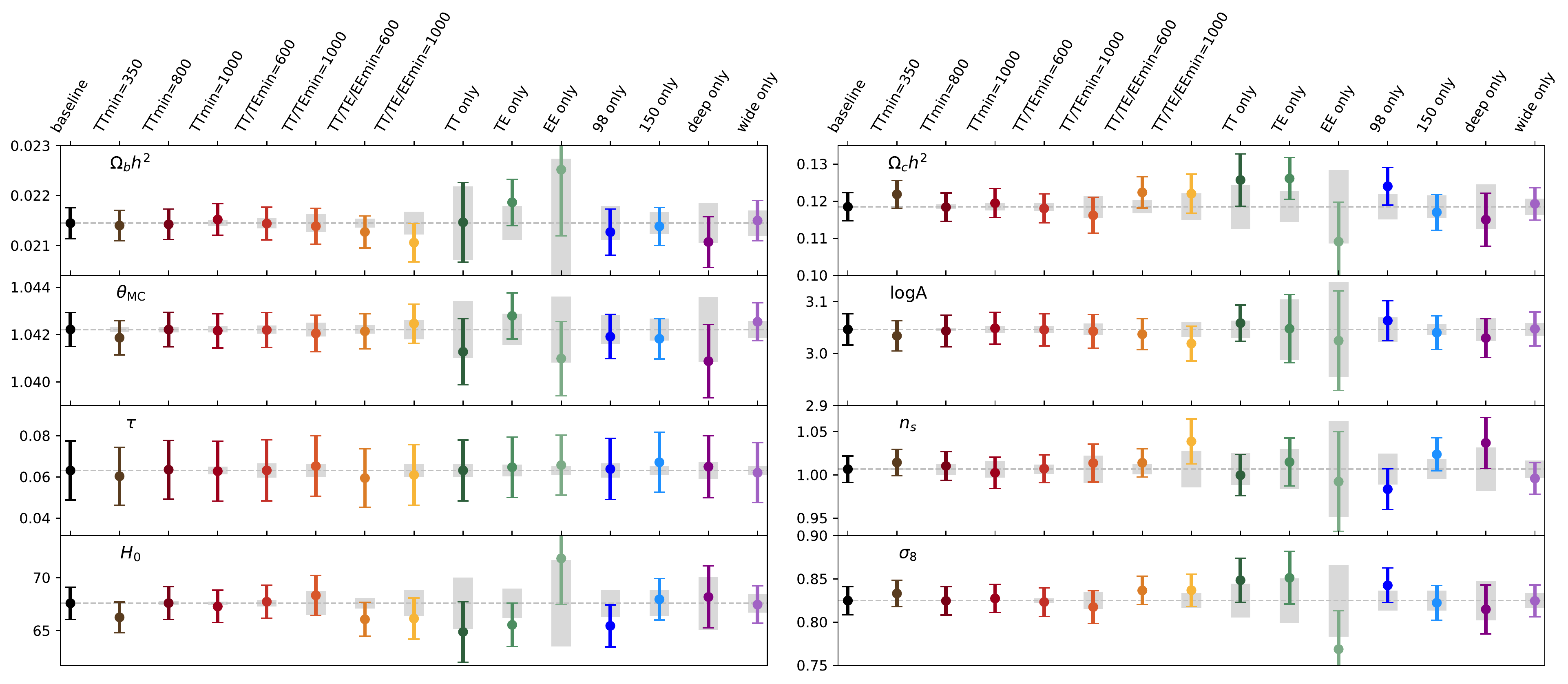}
    \caption{Tests of robustness on the cosmological results: the ACTPol full dataset (black baseline results) is compared to constraints from a subsets of the data (colored points). The expected statistical fluctuation is shown with the gray bands following \citet{2019arXiv191107754G}.}
    \label{fig:wisk}
\end{figure*}

We ran a number of tests of robustness on our cosmological results as summarized in Figure~\ref{fig:wisk} and described below. When we compare results from a subset of data to the baseline results we also report the expected statistical shift following the procedure of \cite{2019arXiv191107754G}. 
The tests are:

1)\,We explore variations in cosmological parameters for different choices of minimum starting multipole in TT, or TT/TE or TT/TE/EE.

2)\,We compare the cosmological parameters deduced from TT, TE, and EE separately. 

3)\,We compare parameters from \freqa\,GHz-only and \freqb\,GHz-only to results from the combination of both.

4)\,We compare parameters separately from the deep and wide regions. 

We conclude that our results are robust against all of these different selections of subsets of the data.

Of particular note are the results on $H_0$. As discussed above, the primary motivation for the blinding was to avoid bias on its value. However, it turns out to be one of the more robustly determined parameters. When keeping the original $350.5\leq\ell\leq7525.5$ range for TT, $H_0=66.2\pm1.4$ km/s/Mpc. After selecting the baseline range for TT of $600.5\leq\ell\leq7525.5$, $H_0=67.6\pm1.5$ km/s/Mpc; 
without correcting for the temperature to polarization leakage, $H_0=67.4\pm1.5$ km/s/Mpc; when including MBAC data (see next subsection and A20)  $H_0=67.9\pm1.5$ km/s/Mpc. The robustness to $H_0$ is due to our measurement of the amplitudes of the TT/TE/EE acoustic peaks as discussed in A20. In summary, the value for $H_0$ for ACT alone is consistent with {\sl Planck}'s CMB-based measurement and inconsistent with the $z<1$ SH$_0$ES Cepheids/supernovae based measurement \citep{riess/etal:2019} at $>3\sigma$. 

In the ACT-only analysis, the best fit $n_s$ is somewhat higher than expectations from {\sl WMAP} and {\sl Planck} and the best fit $\Omega_b h^2$ is somewhat lower. As demonstrated in A20, this appears in part due to the lack of low-$\ell$ information in ACT. We also found, as shown in A20, that changing the amplitude of TE alone moves one up (higher TE, higher $n_s$) and down (lower TE, lower $n_s$) the $\Omega_bh^2-n_s$ degeneracy line. However, at this point we have no reason to believe our reported TE spectrum is inaccurate.

\subsubsection{CMB-only TT,TE,EE power spectra}
\begin{figure}[tp!]
    \centering
    \includegraphics[width=\columnwidth]{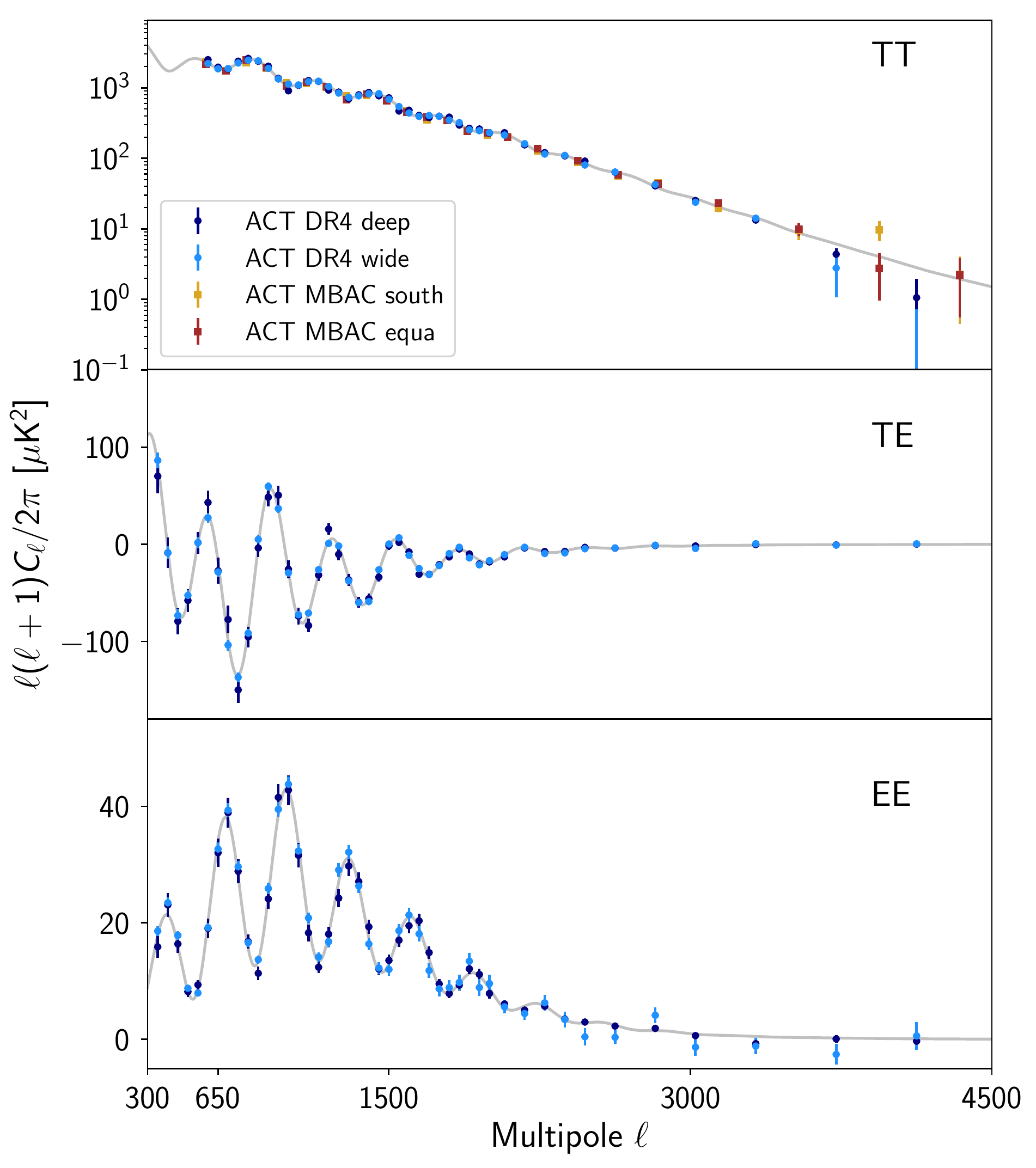}
    \caption{CMB-only band powers obtained after marginalizing over foreground emission plotted with the ACT DR4 $\Lambda$CDM best fit theory. }
    \label{fig:cmbspec}
\end{figure}

For the second application of the multi-frequency likelihood, we implement the Gibbs sampling technique introduced in~\cite{dunkley/etal:2013} to extract the CMB-only power present in the data and generate a foreground-marginalized spectrum and covariance matrix.  We extend the formalism to also marginalize over the leakage corrections\footnote{We modify the Gibbs sampling process by first expanding the matrix mapping the CMB vector into the multi-frequency data to include the mixing introduced by the leakage and then correcting the full foreground and CMB realization with the leakage terms when comparing with the measurements.} during the extraction so that no leakage treatment is needed when using the CMB-only products. To improve the estimation of the CMB part, we add intensity power spectra estimated at both 150 and 220 GHz by the previous ACT receiver, MBAC \citep{swetz/etal:2011, dunner/etal:2013}. The spectra are shown in Figure~\ref{fig:cmbspec}. 

The compressed Gaussian likelihood using the DR4 spectra and covariance obtained here is used in A20 for our extended cosmological analysis.\footnote{The full multi-frequency likelihood, \texttt{actpolfull\_dr4}, and this CMB-only likelihood, \texttt{actpollite\_dr4}, are both available on LAMBDA.} Because we ignore correlations between the new DR4 spectra and the MBAC spectra in the CMB extraction, we do not retain the MBAC CMB component for cosmological analysis, i.e., we only use MBAC to better constrain the high-frequency foreground emission as shown in the next subsection and this in turn improves the DR4 CMB spectra extraction. 
This indirect contribution from combining with MBAC spectra impacts some cosmological results, with tighter constraints obtained on beyond $\Lambda$CDM parameters (see A20) than from the no-MBAC multi-frequency likelihood.
The CMB-only TT, TE, and EE only spectra are given in Table~\ref{tab:specs}. 

\subsubsection{Foreground levels}\label{sec:Fg_levels}
\begin{figure*}[tp!]
    \centering
    \includegraphics[width=\textwidth]{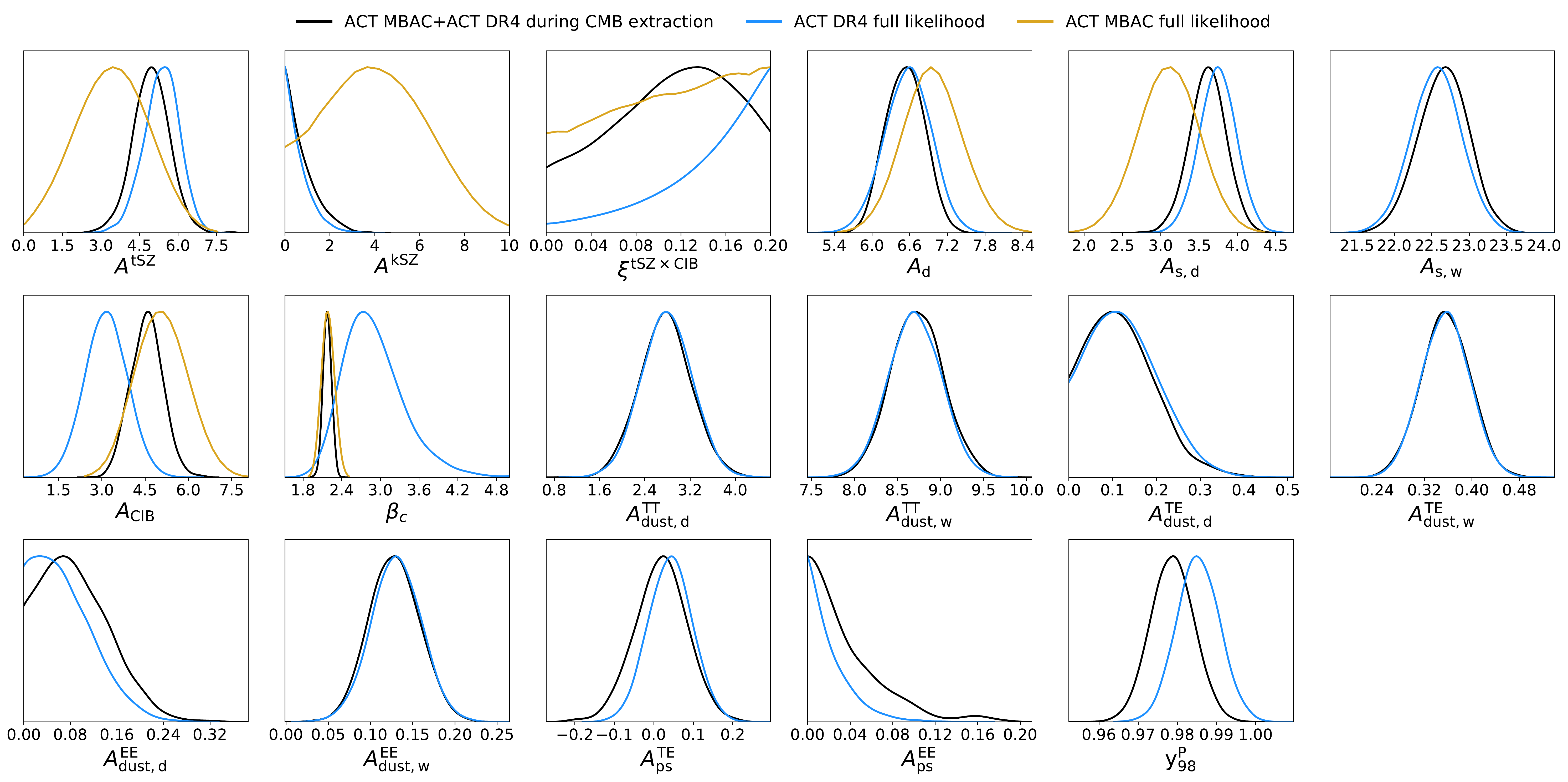}
    \caption{A posteriori distributions  of the foreground parameters using the full multi-frequency likelihood in a $\Lambda$CDM fit (this paper), when extracting the CMB-only spectrum with MBAC and DR4, and for $\Lambda$CDM from ACT MBAC only. More analysis is needed than presented here to interpret some of the results, for
    example the limit on the kSZ amplitude depends strongly on the assumptions made about CIB.}  
    \label{fig:fg}
\end{figure*}

Results on the foreground parameters are shown in Figure~\ref{fig:fg}. We report constraints from three implementations of the likelihood: 
the full multi-frequency run in a $\Lambda$CDM fit without MBAC (the baseline in this paper); a run for extracting the CMB-only spectrum with MBAC and DR4; and a run for $\Lambda$CDM from ACT MBAC only~ \citep{dunkley/etal:2013}. We find good agreement between the DR4 and ACT's MBAC measurements of these components.

We tested some assumptions of the foreground model by, for example, choosing different effective frequencies, using scale-dependent color corrections that take into account the full bandpass information  \citep{madhavacheril/etal:2019}, varying parameters in broader ranges, and imposing different dust priors. None of these had an impact on the cosmological results. We note that this is the minimal foreground model we need to fit our data, and we report the constraints on foreground parameters for this baseline case. However, a more thorough analysis is needed to fully interpret the foreground results; for example the limit on the kSZ amplitude depends strongly on the assumptions made about CIB. 

\begin{figure}[tp!]
    \centering
    \includegraphics[width=\columnwidth]{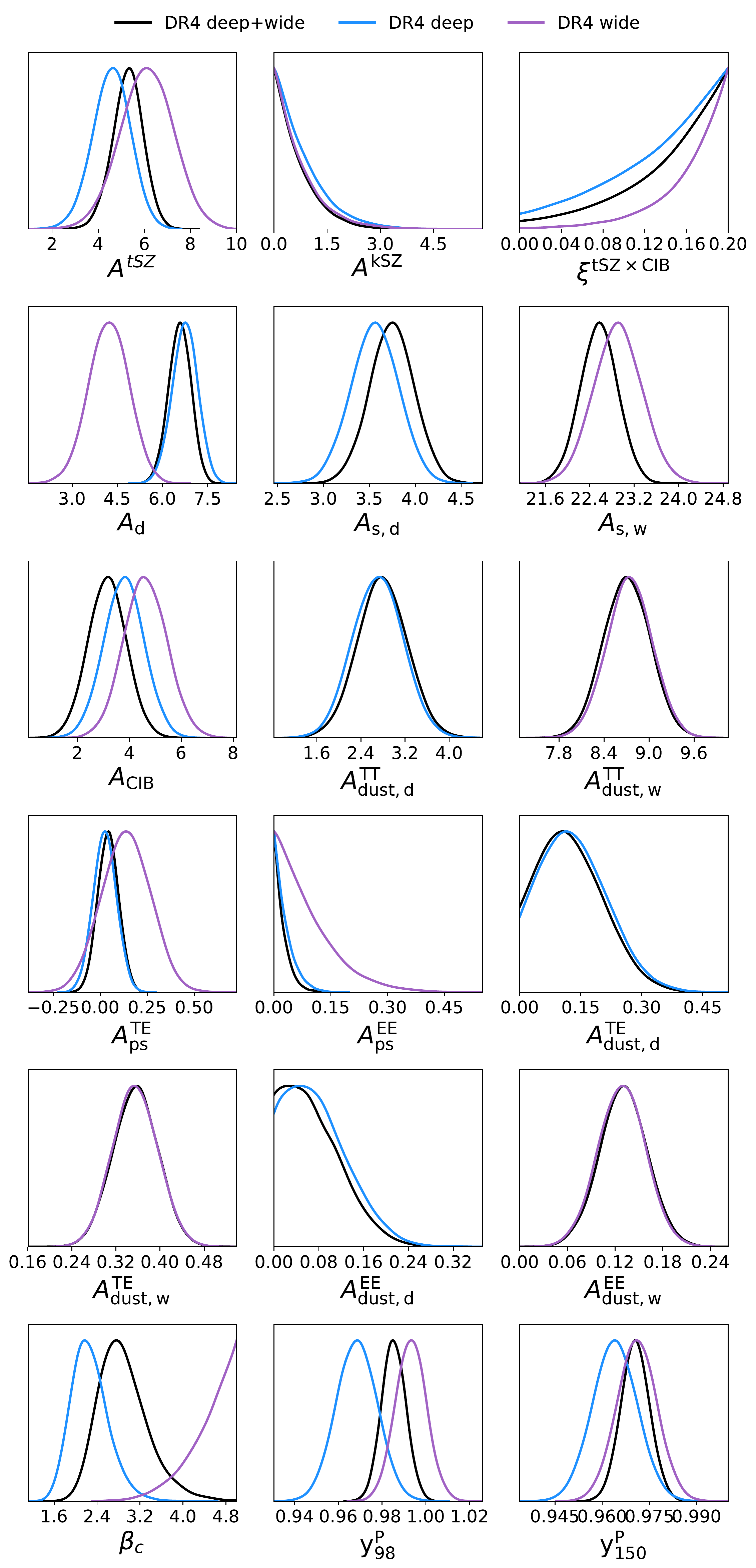}
    \caption{The foreground parameters for the deep and wide regions separately and combined.}
    \label{fig:fgdw}
\end{figure}

Figure~\ref{fig:fgdw} shows that there is some difference between the deep- and wide-only parameters. In particular the wide region shows a preference for a high $\beta_c$, which scales the CIB's frequency spectrum, that is ruled out by the addition of the 220 GHz MBAC data. One source of the discrepancy may be that we characterized the foreground emission in the wide region with a single parameter even though there is a lot of spatial variation within the region  (see Section~\ref{sec:fg_diff}). We note though that the high $\beta_c$ has no impact on the cosmology; the wide-only results with TT spectra cut at $\ell_{\rm max}=2000$ give the same cosmological parameters.

\section{Additional results}
\label{sec:results}

\subsection{TB, EB, and lensing B-modes}
In $\Lambda$CDM, there should be no signal in the TB and EB spectra. We test this by coadding all the spectra first by frequency and then by combining frequencies. For each we compute the $\chi^2$ for the null hypothesis for the $n_{\ell,c}$ elements of the spectrum. These are reported in Table~\ref{tab:tbeb_null}. We did not subtract dust emission because at the levels found above it is not a significant contaminant for this test. To perform the coaddition and determine the $\chi^2$, the full covariance matrix was used. We conclude there is no significant signal in either the TB or EB spectra, consistent with $\Lambda$CDM.

\begin{table}[!tp]
\caption{Summary of the $\chi^2$ for various frequency combinations of null spectra for TB and EB. }
\vspace{-0.15in} 
\begin{center}
\begin{tabular}{c c c}
\hline
 & TB ($\chi^2$, PTE) & EB ($\chi^2$, PTE)\\
\hline
\freqa\,GHz & 68.5 (0.06) &  53.9 (0.40) \\
\freqa $\times$ \freqb\,GHz & 44.0 (0.78) &  74.4 (0.02) \\
\freqb $\times$ \freqa\,GHz & 58.7 (0.24) &  $\cdots$ \\
\freqb\,GHz & 45.1 (0.74) &  57.0 (0.30) \\
Combined & 60.3 (0.20) &  68.4 (0.06) \\
\hline
$\sigma_{{\cal D}_\ell}$, $\ell=500.5$ & 2.4 ($\mu$K)$^2$& 0.16 ($\mu$K)$^2$\\
$\sigma_{{\cal D}_\ell}$, $\ell=2000.5$ &1.3 ($\mu$K)$^2$ & 0.42 ($\mu$K)$^2$\\
$\sigma_{{\cal D}_\ell}$, $\ell=4125.5$ & 0.50 ($\mu$K)$^2$ & 0.52 ($\mu$K)$^2$\\
\hline

\hline
\end{tabular}
\end{center}
\vspace{-0.1in} 
\small{The probability to exceed (PTE) is computed from $\chi^2$ with $n_{\ell,c}$ degrees of freedom. We give the uncertainties at three values of $\ell$. The first two have
$\Delta\ell=50$; the third has $\Delta\ell=400$. These may be compared to the TE and EE uncertainties in Table~\ref{tab:specs}.  }
\label{tab:tbeb_null}
\end{table}

\begin{figure}[tp!]
    \centering
    \includegraphics[width=\columnwidth]{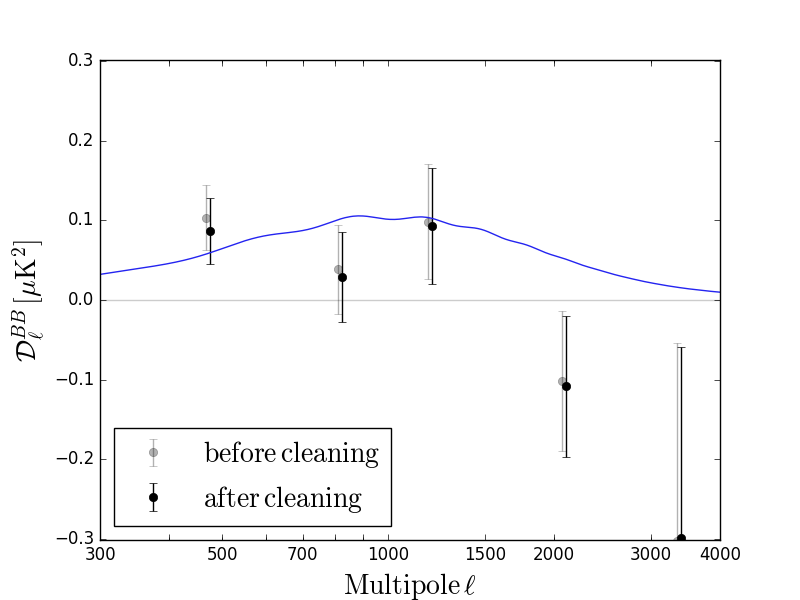}
    \caption{The lensing B-mode spectrum. The parameter $a_{\rm Blens}$ scales the amplitude of  the theory curve in blue. The values from before foreground subtraction are shown in gray.}
    \label{fig:bmodes}
\end{figure}

\citet{zaldarriaga/seljak:1998} showed that the gravitational lensing of an E-mode signal produces lensing B-modes. Initial observations of the effect have been reported \citep[e.g.,][]{hanson/etal:2013,polarbear:2014,ade/etal:2015}. Given the current level of sensitivity, visualizing the effect requires both coadding the spectra and combining $\ell$ bins. Before combining the wide and deep regions, we subtract contamination from the dust B-modes (including error propagation) shown in Figure~\ref{fig:diff_dust} in the wide region. After subtraction, we use the covariance matrix to combine the resulting spectrum into the five bins given in Table~\ref{tab:bb_data}. These are plotted in Figures~\ref{fig:bmodes} and \ref{fig:recent_cmb}. 
We simulated the leakage of E-modes into B-modes due to the unevenly weighted and cut polarization maps \citep{bunn/etal:2003} and found the effect negligible for our coverage and statistical weight.

\begin{table}[!tp]
\caption{The BB spectrum}
\vspace{-0.15in} 
\begin{center}
\begin{tabular}{c c c c}
\hline
\hline
 $\ell$ & $\ell_{min}$& $\ell_{max}$ & ${\cal D}_{BB}$ ($\mu$K)$^2$\\
\hline
475 & 300 & 650 & $0.090\pm 0.043$ \\
825.5 & 651 & 1000 & $0.029\pm 0.057$ \\
1200.5 & 1001 & 1400 & $0.094\pm 0.073$ \\
2100.5 & 1401 & 2800 & $-0.113\pm 0.092$ \\
3400.5 & 2801 & 4000 & $-0.30\pm 0.24$ \\
\hline
\end{tabular}
\end{center}
\label{tab:bb_data}
\end{table}

To assess the lensing signal, we parameterize the $\Lambda$CDM prediction with a single scaling parameter, $a_{\rm Blens}$. We then compute the likelihood of $a_{\rm Blens}$ given our data.  We find that for $300<\ell<1400$, $a_{\rm Blens}=0.79\pm0.38$, where $a_{\rm Blens}=1$ corresponds to $\Lambda$CDM. If the upper range is extended to 2800 or 4000 then $a_{\rm Blens}=0.60\pm0.36$. Thus, the data are consistent with $\Lambda$CDM but the test is not powerful. In L17 we reported $a_{\rm Blens}=2.03\pm 1.01$. Even though we have added much more data and reduced the uncertainty on the spectrum by a factor of 2.8, the combination of noise fluctuations and the new analysis techniques have resulted in a measurement of similar significance. 

\subsection{Noise power spectra}

\begin{figure}[!h]
\begin{center}
\vspace{-0.1in}
\includegraphics[width=\columnwidth]{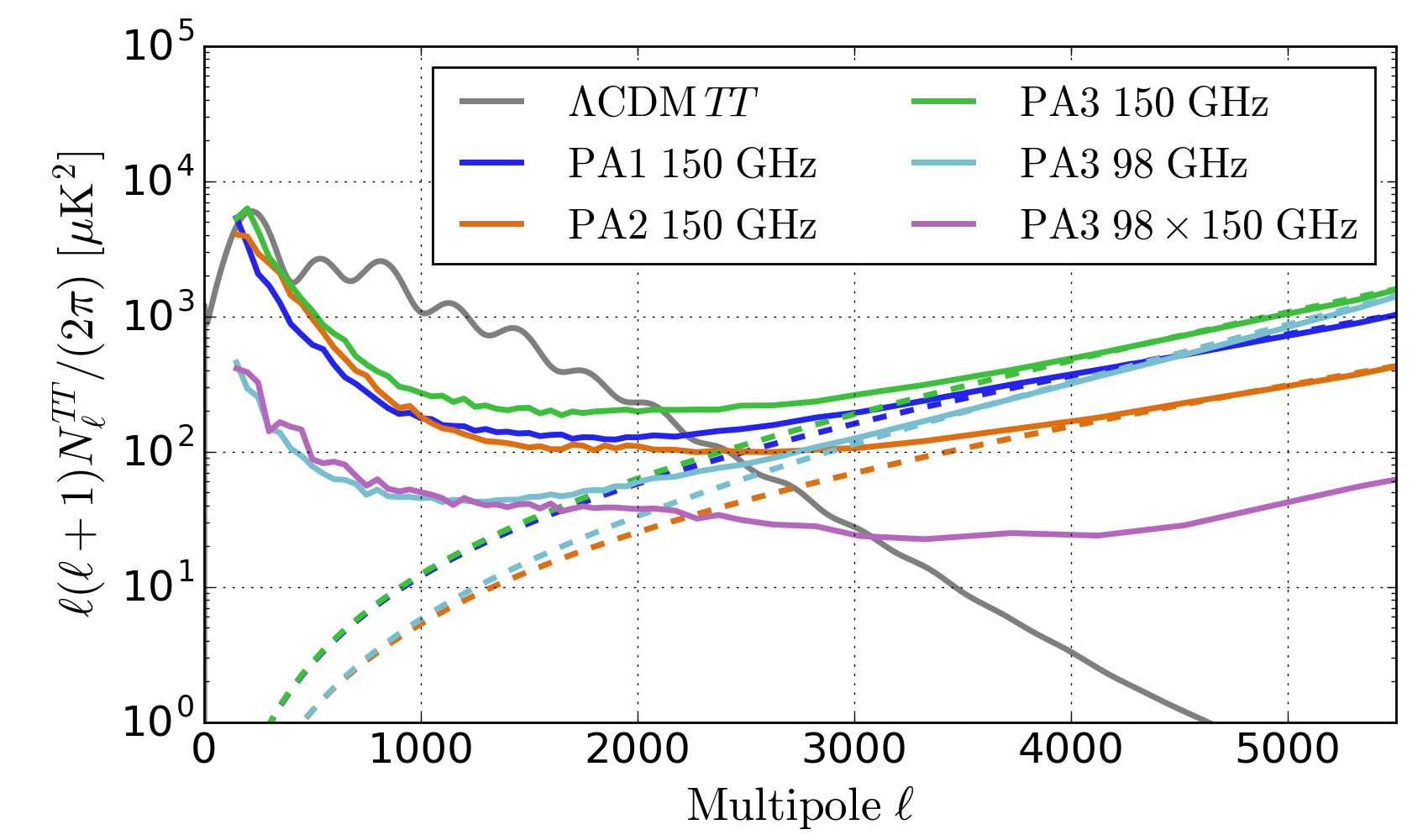}
\includegraphics[width=\columnwidth]{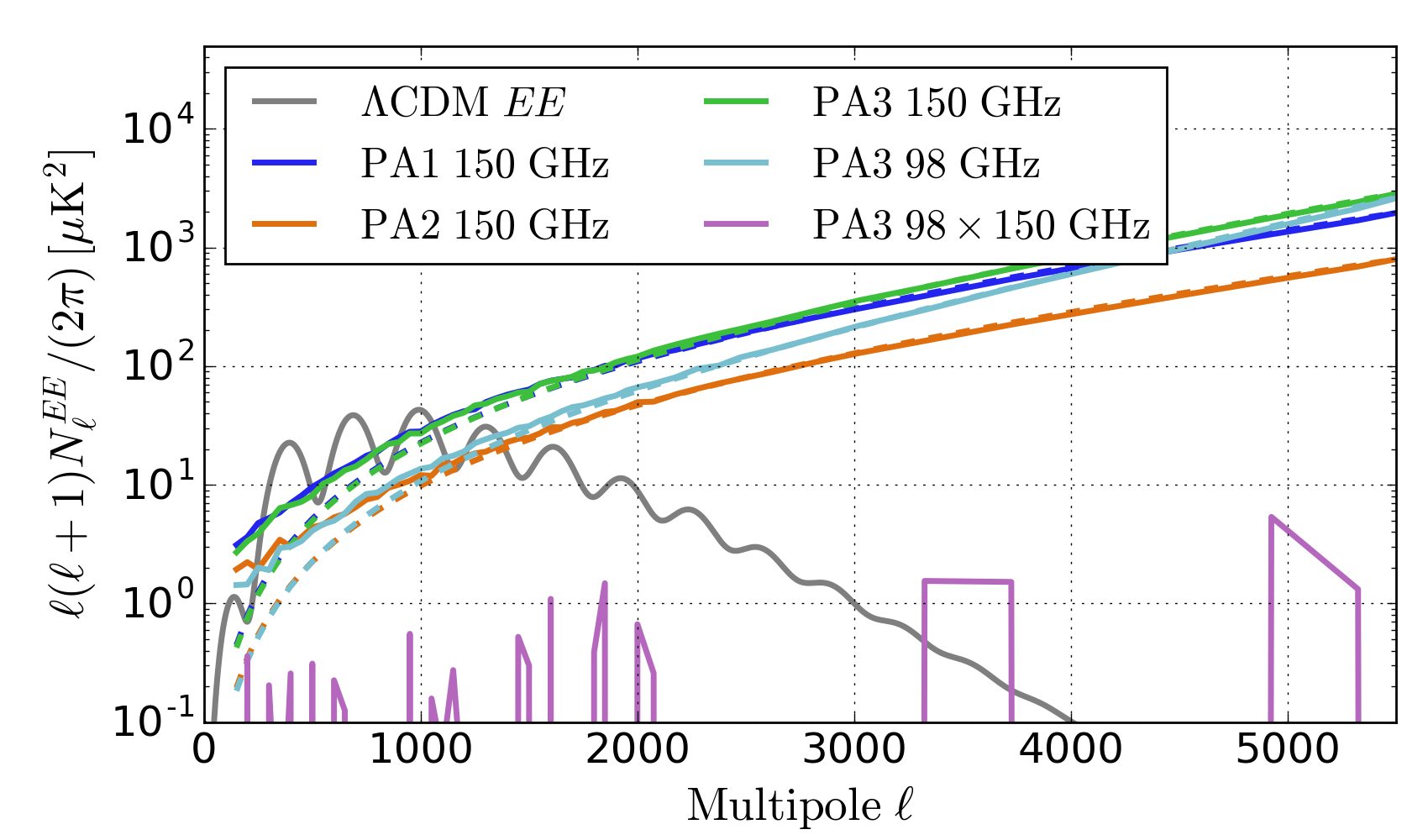}
\vspace{-0.2in}
\caption{The TT (top) and EE (bottom) noise power spectrum of different arrays are shown for s15/D56 in solid lines. The white noise levels are shown in dashed lines. The TT $1/\ell$ spectrum is dominated by atmospheric fluctuations at $\ell<1500$, which is smaller at \freqa GHz (PA3) than at \freqb GHz. PA2 is the most sensitive array, shown by the overall white noise level. Below this noise level, there is a small noise correlation seen in TT between \freqa and \freqb GHz channels for the dichroic array (PA3). This correlation is not significant in polarization. }
\label{fig:noise_ps}
\end{center}
\end{figure}

The \freqa and \freqb\,GHz TT/EE noise power spectra for D56 are shown in Figure~\ref{fig:noise_ps}. Atmospheric fluctuation power is much lower at \freqa\,GHz as seen by its low $\ell$ behavior. We also note the improvement of the noise in the $1000<\ell<2500$ regime relative to the noise level of the same data set analyzed in L17. This improvement results from optimization of the mapmaking pipeline as described in A20. In particular, Fourier transforms with the gaps in the TODs simply filled with lines were previously causing excess power in the estimation of the noise covariance matrix in the maximum likelihood mapmaking method. These gaps are now better handled with interpolations based on uncut samples and the knowledge of detector correlations prior to the estimation of the noise model, leading to a suppression of this excess noise.

Because atmospheric emission is largely unpolarized, the polarization noise spectra do not have the $\ell<1500$ upturn the temperature spectra have. This is independent of frequency unlike in temperature, where \freqa\,GHz has a lower $\ell$ upturn than does \freqb\,GHz. Preliminary studies indicate the bath temperature fluctuations coupled with differences in conductances for detectors of orthogonal polarization can lead to an excess low-$\ell$ noise power in polarization. 

Lastly, there is a noise correlation measured between the two frequency channels for the dichroic array (PA3) because the detectors for the two frequency channels share the antenna in each pixel. Hence the same atmosphere fluctuation will be seen by both. We measure the correlation in temperature but not in polarization. This correlation in temperature was included in generating noise simulations in Section~\ref{sec:noise_sim}. 

\begin{figure}[tp!]
\centering
\includegraphics[width=\columnwidth]{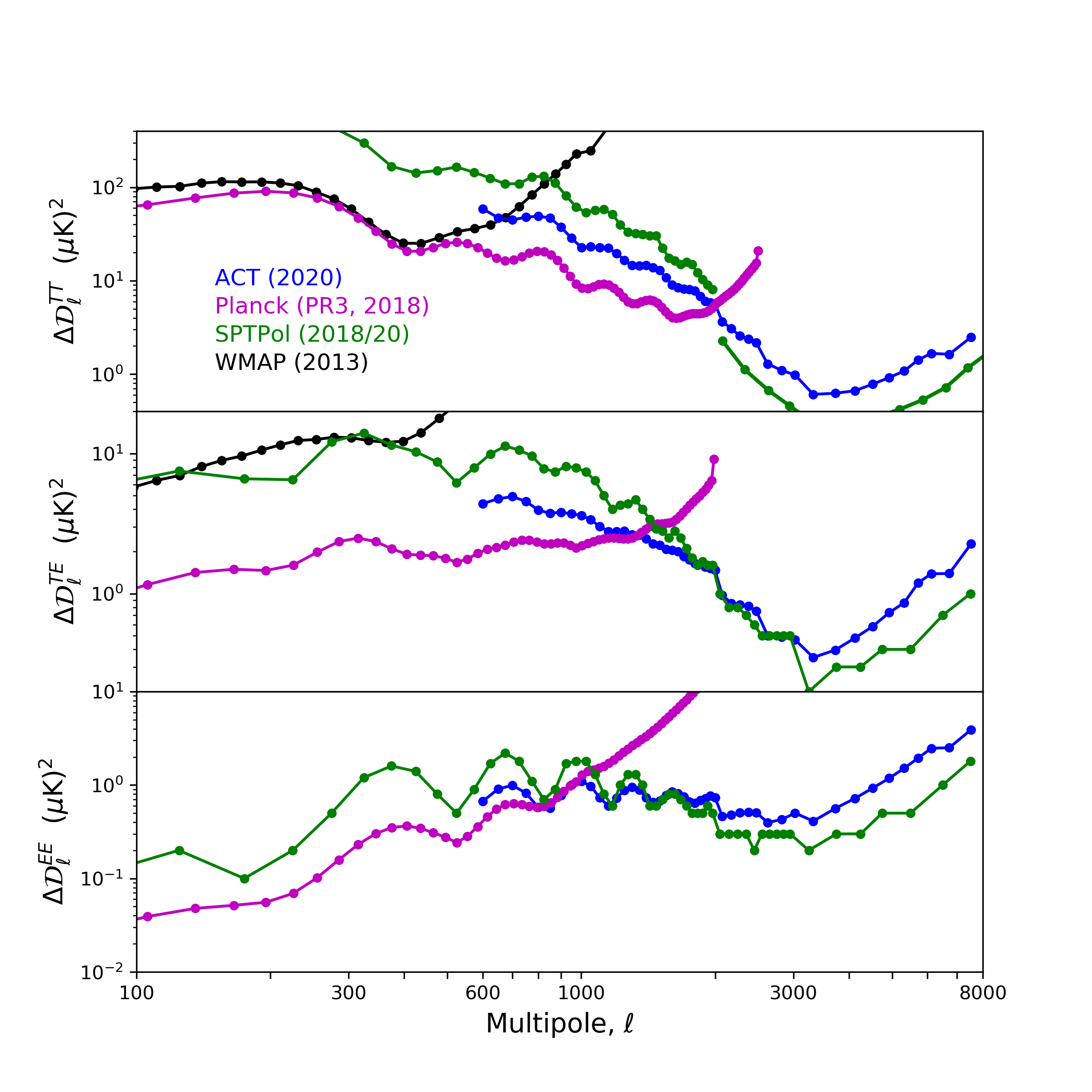}
\caption{The power spectrum error bars from this work, \citet{planck_spectra:2019}, \citet{reichardt/etal:2020} (SPT 150 GHz TT $\ell>2000$), \citet{henning/etal:2018} (SPT 150 GHz TT/TE/EE), and \citet{bennett/etal:2013} ({\sl WMAP}).
The dots show the $\ell$ bin center. The curves should not be overinterpreted: the error bar can be reduced by combining $\ell$ bins. Also, we show only 150 GHz for SPT whereas others show the best CMB spectrum. The bumps in the curves show where the measurements are cosmic-variance limited.}
\label{fig:recent_cmb_noise}
\end{figure}

The noise power spectra show the $N_b$ term in Equation~\ref{eq:1}. The error bar that enters the power spectrum also has the cosmic variance term and accounts for the number of modes as in Equation~\ref{eq:1}.
Figure~\ref{fig:recent_cmb_noise} compares the power spectrum error bars
from {\sl WMAP}, {\sl Planck}, SPT and ACT. The curves give an indication of how different data sets complement each other. 

\subsection{Direct comparison to {\sl Planck}}
\label{sec:planck_compare}

{\sl Planck} overlaps all ACT regions and has sufficiently low noise that a direct map-level comparison provides an important check for systematic error in both measurements. After unblinding and correcting for the TE leakage (Section~\ref{sec:misc_syst}), ACT TT, TE, EE power spectra are compared to the ACT$\times${\sl Planck} (AP),  {\sl Planck}$\times${\sl Planck} (PP), and, for TE, the {\sl Planck}$\times$ACT (PA) spectra in the same region. 
We use spatial window functions with ACT ({\sl Planck}) inverse noise variance maps (Section~\ref{sec:spatial_window}) to compute the AA and AP (PP) spectra. We limit the comparison to D56 and BN because they constitute the majority of the DR4 sensitivity. 

To assess the comparison we use simulations. For the signal, we use the same simulations for {\sl Planck} and ACT (Section~\ref{subsec:sig_sim}). For the {\sl Planck} noise, we download {\bf FFP10}\footnote{ At \url{http://pla.esac.esa.int/} 300 simulations are available for each of half-mission 1 and 2.} simulations, rotate them to equatorial coordinates, and project them into the ACT CAR pixelization. The uncertainties on the ACT and {\sl Planck} null power spectra are computed from the dispersion of the null power spectra of the simulations. Because we have 300 {\sl Planck} simulations we use 300 ACT simulations to match them.

We multiply the {\sl Planck} polarization maps by 0.9829 to account for polarization efficiency fitted in the {\sl Planck} likelihood \citep{planck_cosmo:2019} but we do not yet include the {\sl Planck} leakage beam for these regions. Also, we use only the diagonal terms of {\sl Planck}'s covariance matrix. To ensure a similar comparison to ACT, we use only the diagonal uncertainties for ACT as well. Lastly, we do not include a detailed accounting of foreground emission. For these reasons, this analysis should be viewed as a first pass that will improve with further investigation. 

\begin{figure}[tp!]
\centering
\includegraphics[width=0.496\textwidth]{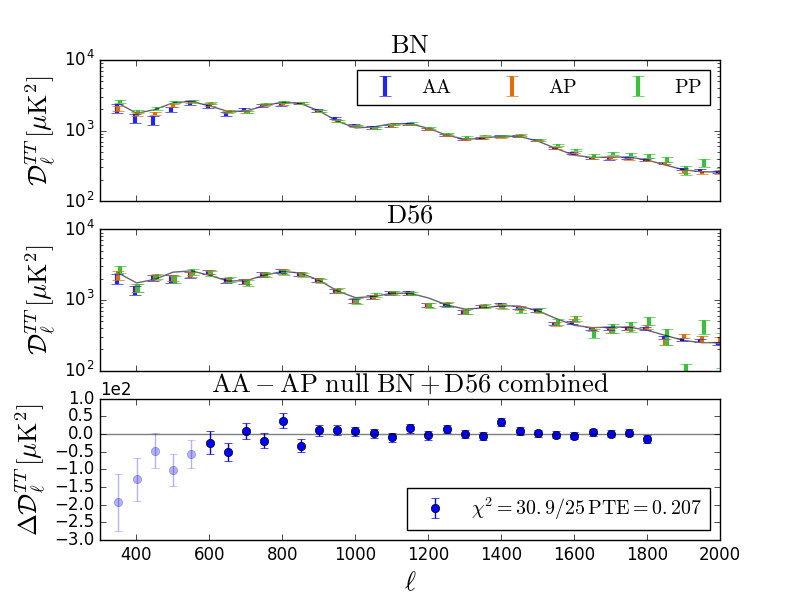}
\includegraphics[width=0.496\textwidth]{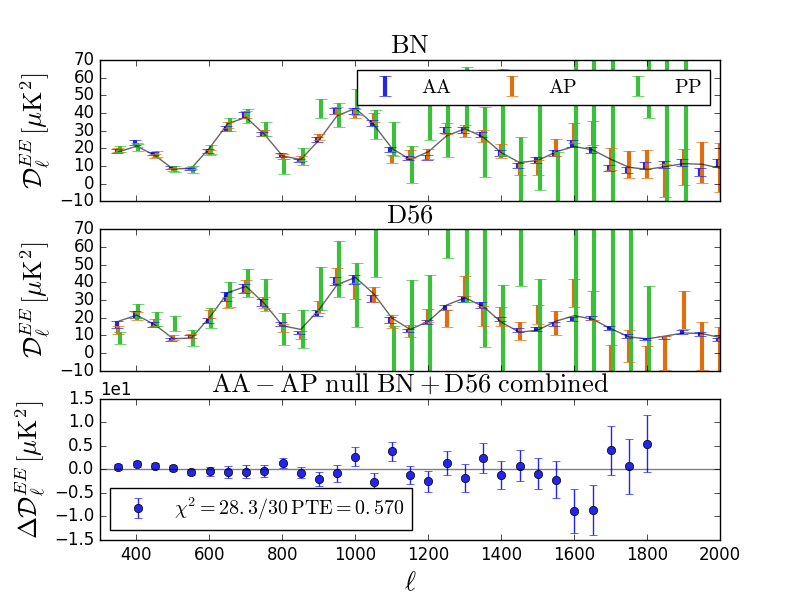}
\caption{A comparison of the ACT (AA), ACT$\times${\sl Planck} (AP), and {\sl Planck} (PP) spectra in the BN and D56 regions for the TT and EE spectra. These spectra include sample variance. Both bottom panels show the weighted sum of AA$-$AP of BN and D56 along with the $\chi^2$ to the null spectrum for the 25 $\ell$ bins for TT and the 30 $\ell$ bins for EE. For TT, the first 5 bins (shown in faint color) are not part of the cosmological analysis. By construction these spectra minimize sample variance.}
\label{fig:AP_TT_EE}
\end{figure}
\smallskip

Figure~\ref{fig:AP_TT_EE} shows the comparison for TT (top triplet) and EE (lower triplet). The top panel in the TT triplet shows the three spectra for BN. The middle panel shows the same for D56. The low AA values in the first few bins show why $\ell=600$ was chosen as the first bin of the TT spectrum (see also TT panel in Figure~\ref{fig:likelihood_input} and Section~\ref{sec:unblinding}). The bottom panel shows the weighted sum of the difference between AA and AP for the BN and D56 spectra. Because there is reduced sample variance in this difference, the statistical noise may be examined. The $\chi^2$ to the difference spectrum is computed for $600\leq\ell\leq1800$. This test shows the consistency between ACT and {\sl Planck} well below the sample variance limit, as can be assessed by comparing the error bars to those in Table~\ref{tab:specs}. At the current maturity of the analysis, the TT spectra are not straightforward to compare because {\sl Planck}'s point source mask and cut level are different than ACT's. Table~\ref{tab:act_planck_chi2_pte} in Appendix~\ref{appen:planck_compare} shows more consistency checks. While there are a number of low PTE values, we find the overall agreement acceptable.

The bottom triplet of panels in Figure~\ref{fig:AP_TT_EE}  show a similar set of spectra but for EE. It is clear that ACT is more sensitive than {\sl Planck} in these regions but the similarity between AA and AP is promising. Here the $\chi^2$ to the null spectrum is computed for $350\leq\ell\leq1800$. Again, Table~\ref{tab:act_planck_chi2_pte} in Appendix~\ref{appen:planck_compare} shows more consistency checks. In EE, there is good agreement between ACT and {\sl Planck}.

Table~\ref{tab:act_planck_chi2_pte} also shows the consistency with $\Lambda$CDM as assessed with $\chi^2$ for the best fit ACT-only model (Section~\ref{subsec:lcdm_fit}) and the {\sl Planck} model \citep{planck_cosmo:2019}. There is generally good agreement with the $\Lambda$CDM model in light of the caveats about TT.

Table~\ref{tab:act_planck_ee_pte_matrix} shows the differences between various spectral combinations within BN and D56 for different frequency combinations for EE. For {\sl Planck} we use only the 143 GHz maps. The $\chi^2$ are computed for $350\leq\ell\leq1800$ using the diagonal elements of the covariance matrix. The PTEs are shown graphically in Figure~\ref{fig:TE_pte_hist}.  Again we conclude that the ACT and {\sl Planck} agree well.

\begin{figure}[tp!]
\centering
\includegraphics[width=\columnwidth]{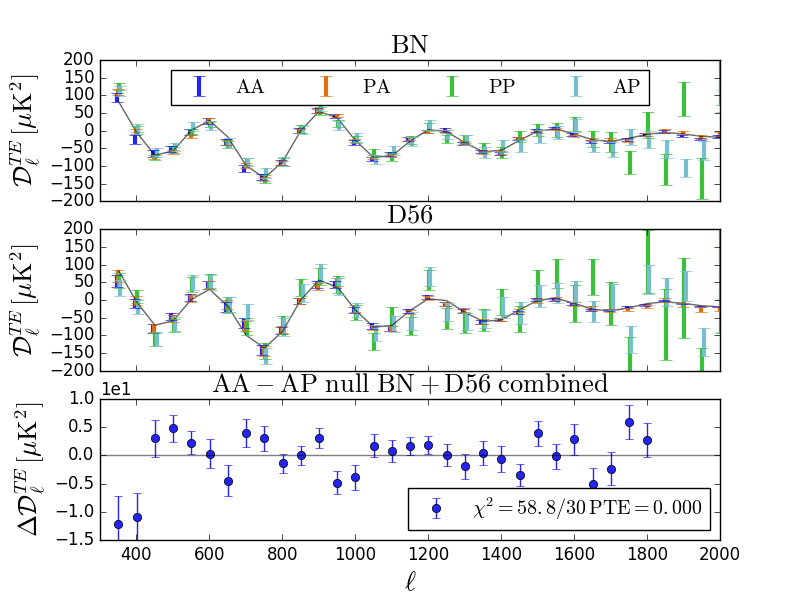}
\caption{ The TE spectra for the BN and D56 regions in the top two panels along with the weighted sum of AA$-$AP of BN and D56 in the bottom. Sample variance is included in the top two but is minimized by construction in the bottom panel. The figure highlights the failure of a null test. Table~\ref{tab:act_planck_te_pte_matrix} shows that of the 72 similar null tests, just three had larger $\chi^2$. However, Table~\ref{tab:act_planck_te_pte_matrix} also shows that AA is internally consistent, Table~\ref{tab:act_planck_chi2_pte} shows that AA is in good agreement with $\Lambda$CDM, and even for {\sl Planck} alone there is considerable spread in the first two bins.}
\label{fig:AP_TE}
\end{figure}

\begin{figure}[tp!]
\centering
\includegraphics[width=\columnwidth]{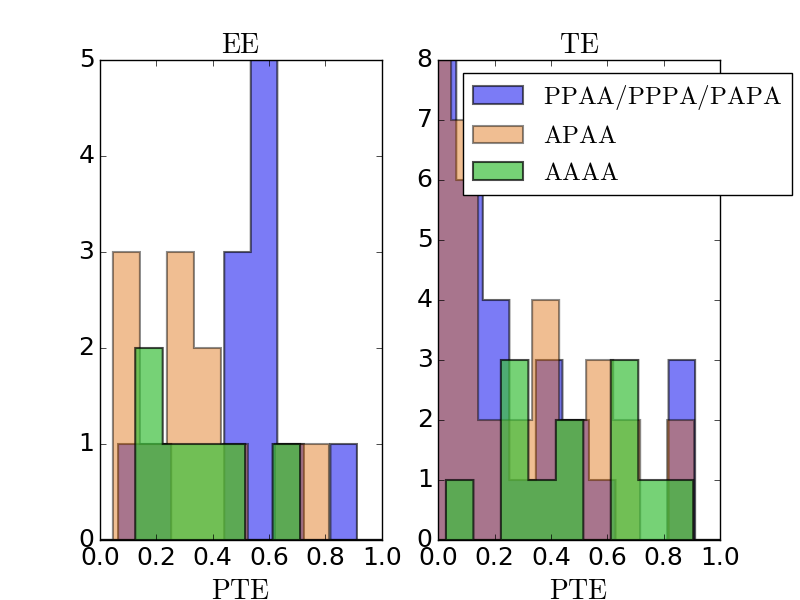}
\caption{The EE and TE distributions of the PTE from Table~\ref{tab:act_planck_ee_pte_matrix} and Table~\ref{tab:act_planck_te_pte_matrix}. The notation ``PPAA" means entries with {\sl Planck}$\times${\sl Planck} from the column heading and  ACT$\times$ACT from the row heading. The overlap of blue and tan is purple.}
\label{fig:TE_pte_hist}
\end{figure}

For TE there are a number of complicating factors. Both ACT and {\sl Planck} have temperature to polarization leakage. ACT's is partially corrected in the mapmaking pipeline and part in the spectrum analysis. {\sl Planck}'s is corrected for the full sky average, and we do not have local corrections for our regions.  
Figure~\ref{fig:AP_TE} shows the AP, PA, AA and PP spectra, where the first entry is T and the second is E. The tightest constraints come from AA and PA. 
The difference spectrum on the bottom is computed as in Figure~\ref{fig:AP_TT_EE} but with AP and PA averaged together. Although the $\chi^2$ is uncomfortably high, the scatter occurs throughout the spectrum. For example, even if the first two points are removed $\chi^2 = 46.0/28$ with a PTE = 0.017.
We investigate the consistency between spectra further in Table~\ref{tab:act_planck_te_pte_matrix} and show the results graphically in Figure~\ref{fig:TE_pte_hist}. We conclude that the ACT TE spectra are consistent with each other, as we would expect from Section~\ref{sec:consistency}, and that the cross spectra with {\sl Planck} are reasonably consistent but there are notable failures, for example 150$\times150-P\times$150 in D56 and P$\times150-98\times$150 in BN, that warrant further investigation with a more detailed analysis involving {\sl Planck} experts. The inconsistency with {\sl Planck} in TE may be related to the difference in the preferred $\Lambda$CDM model but this requires an analysis beyond the scope of this paper.

\smallskip

\smallskip

\section{Discussion and conclusions}
\label{sec:conclude}

We have presented a new CMB data set with significant cosmological constraining power. The data have passed a large number of systematic and null tests of different kinds. The analysis was ``blind," a process that brought with it a considerable time investment. We followed the protocol laid out in June 2017 until we opened the box in February 2020. After unblinding, the primary departure from our protocol was changing the minimum spherical harmonic for TT from $\ell=350.5$ to
$\ell=600.5$. One of our primary results, a new determination of $H_0$, is insensitive to this choice. We also reassessed our temperature to polarization leakage but again this did not have a large effect on the cosmological parameters. 
Three aspects of the analysis, the foreground characterization in the wide region, the TE comparison to {\sl Planck}, and $\ell<600$ deficit in TT spectra require more investigation. However, resolving these open questions is expected to have only a minor impact on the results presented here.

ACT's agreement with the six-parameter $\Lambda$CDM model is comfortably within expectations as shown in Figure~\ref{fig:chi2datasims}. A20 compares the parameter values to those from {\sl WMAP} and {\sl Planck} and investigates how the parameter values depend on the addition of, say, {\sl WMAP} or a prior on the first peak amplitude, which ACT does not yet measure. 
There is some tension between the CMB-only cosmological parameters derived from different measurements. This could be indicative of as yet unidentified low-level systematic errors. We note, though, that ACT's constraining power comes from TE and EE at $\ell>1000$ which are produced by different physics than is TT, the dominant source of information for {\sl WMAP} and {\sl Planck}. Thus the tensions possibly may be suggestive of a missing component of the model.

\begin{figure*}[tp!]
\centering
\includegraphics[width=\textwidth]{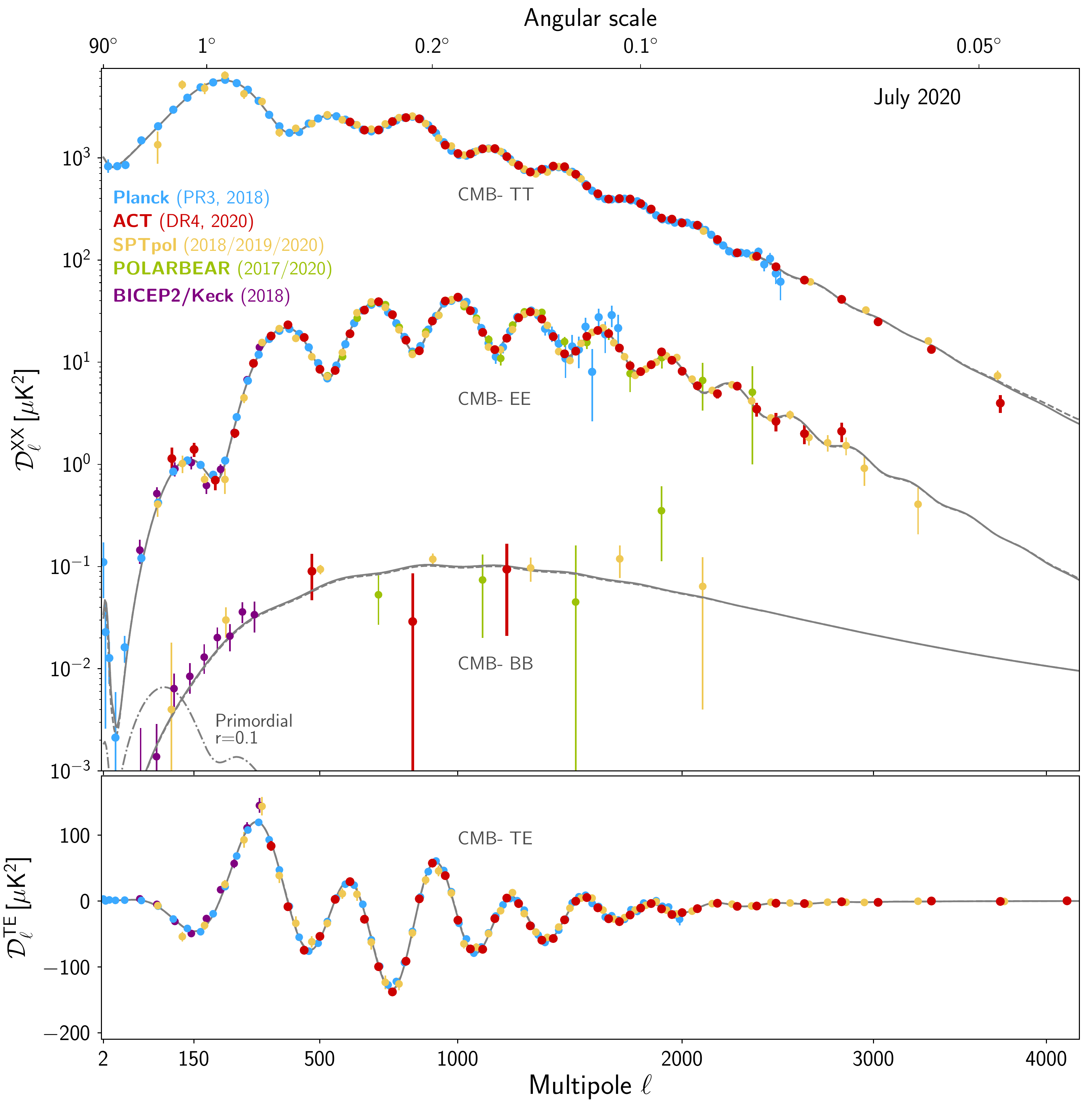}
\caption{Recent measurements of the CMB temperature anisotropy and polarization.
The two models, the thin nearly overlapping grey lines, are from {\sl Planck} (dashed line) and from ACT plus {\sl WMAP} (A20, solid line). The primordial BB signal with $r=0.1$ is also shown with the dot-dashed line.  
For {\sl Planck} we show the 2018 results \citep{planck_spectra:2019}. 
For SPT we show \citet{henning/etal:2018} for 150 GHz TT $\ell<2000$, TE and EE, and \citet{sayre/etal:2019} for BB. For $\ell>2000$ we show the   
SPT spectrum from \citet{george/etal:2015} which has been corrected for point source emission. It is visually indistinguishable from the more precise but uncorrected spectrum in \citet{reichardt/etal:2020}. 
For Polarbear/Simons Array we show EE from \citet{2020arXiv200506168A} and BB from pipeline A in \citet{polarbear:2017}. For BICEP2/Keck we use
\citet{bicep:2018}. All error bars are one sigma and points with no lower bound in TT and EE have been dropped at high $\ell$. 
There is much more to each data set than is plotted here, for example additional frequencies.
For ACT we also show preliminary EE results that were not used in the analysis: for $\ell=[103,150.5,200.5,250.5,300.5]$,  
${\cal D}_\ell^{EE} = [1.14\pm0.32 , 1.40\pm0.22 , 0.70\pm0.14, 2.02\pm0.20, 
9.74\pm0.39]$\,($\mu{\rm K}$)$^2$.}
\label{fig:recent_cmb}
\vspace{0.2in}
\end{figure*}

These results bode well for future measurements from Chile, from which half the sky can be observed. ACT has over four times the data presented here in the process of analysis. ACT is now observing from 30--220 GHz in five frequency bands and annually observes and maps about 40\% of the sky. In the not too distant future the Simons Observatory's Large Aperture Telescope will have more than six times the number of ACT's detectors. This will be complemented by an array of Small Aperture Telescopes that target primordial B-modes. In addition, the CLASS \citep[e.g.,][]{CLASS:2020} and Polarbear/Simons Array experiments \citep[e.g.,][]{polarbear:2019} are taking data at the same site.
Figure~\ref{fig:recent_cmb} gives a snapshot of the field. It is remarkable that to describe all of these data just six cosmological parameters suffice. 

\section{Acknowledgements}
This work was supported by the U.S. National Science Foundation through awards AST-0408698, AST-0965625, and AST-1440226 for the ACT project, as well as awards PHY-0355328, PHY-0855887 and PHY-1214379. Funding was also provided by Princeton University, the University of Pennsylvania, and a Canada Foundation for Innovation (CFI) award to UBC. ACT operates in the Parque Astron\'omico Atacama in northern Chile under the auspices of the Comisi\'on Nacional de Investigaci\'on (CONICYT). 
Computations were performed on the Niagara supercomputer at the SciNet HPC Consortium and on the Simons-Popeye cluster of the Flatiron Institute. SciNet is funded by the CFI under the auspices of Compute Canada, the Government of Ontario, the Ontario Research Fund---Research Excellence, and the University of Toronto.
Cosmological analyses were performed on the Hawk high-performance computing cluster at the Advanced Research Computing at Cardiff (ARCCA). We would like to thank the Scientific Computing Core (SCC) team at the Flatiron Institute, especially Nick Carriero, for their support. Flatiron Institute is supported by the Simons Foundation. Additional computations were performed on Hippo at the University of KwaZulu-Natal, on Tiger as part of Princeton Research Computing resources at Princeton University, on Feynman at Princeton University, and on Cori at NERSC. The development of multichroic detectors and lenses was supported by NASA grants NNX13AE56G and NNX14AB58G. Detector research at NIST was supported by the NIST Innovations in Measurement Science program. We thank Bert Harrop for his extensive efforts on the assembly of the detector arrays. The shops at Penn and Princeton have time and again built beautiful instrumentation on which ACT depends. We also thank Toby Marriage for numerous contributions.

SKC acknowledges support from the Cornell Presidential Postdoctoral Fellowship. RD thanks CONICYT for grant BASAL CATA AFB-170002. ZL, ES and JD are supported through NSF grant AST-1814971. KM and MHi acknowledge support from the National Research Foundation of South Africa. MDN acknowledges support from NSF award AST-1454881.
DH, AM, and NS acknowledge support from NSF grant numbers AST-1513618 and AST-1907657. EC acknowledges support from the STFC Ernest Rutherford Fellowship ST/M004856/2 and STFC Consolidated Grant ST/S00033X/1, and from the Horizon 2020 ERC Starting Grant (Grant agreement No 849169). NB acknowledges support from NSF grant AST-1910021. ML was supported by a Dicke Fellowship. LP gratefully acknowledges support from the Mishrahi and Wilkinson funds. RH acknowledges support as an Azrieli Global Scholar in CIfAR's Gravity \& the Extreme Universe Program and as an Alfred. P. Sloan Research Fellow. RH is also supported by Canada's NSERC Discovery Grants program and the Dunlap Institute, which was established with an endowment by the David Dunlap family and the University of Toronto.
We thank our many colleagues from ALMA, APEX, CLASS, and Polarbear/Simons Array who have helped us at critical junctures. Colleagues at AstroNorte and RadioSky provide logistical support and keep operations in Chile running smoothly.

Lastly, we gratefully acknowledge the many publicly available software packages that were essential for parts of this analysis. They include \texttt{CosmoMC} \citep{Lewis:2013hha,Lewis:2002ah}, \texttt{CAMB}~\citep{CAMB}, 
\texttt{healpy}~\citep{Healpix1}, \texttt{HEALPix}~\citep{Healpix2}, the \texttt{SLATEC}\footnote{http://www.netlib.org/slatec/guide} Fortran subroutine DRC3JJ.F9,  the  \texttt{SOFA} library \citep{SOFA:2019-07-22}, \texttt{libsharp}~\citep{reinecke/2013}, and 
\texttt{pixell}\footnote{https://github.com/simonsobs/pixell}. This research made use of \texttt{Astropy}\footnote{http://www.astropy.org}, a community-developed core Python package for Astronomy \citep{astropy:2013, astropy:2018}. We also acknowledge use of the \texttt{matplotlib}~\citep{Hunter:2007} package and the Python Image Library for producing plots in this paper.

\bibliographystyle{yahapj} 
\bibliography{choi_et_al_bib.bib}

\clearpage
\newpage

\appendix

\section{A. Scan parameters}
\label{appen:scan_param}
\begin{table}[h]
\caption{Summary of scanning parameters}
\vspace{-0.1in}
\begin{center}
\begin{tabular}{c c c c c c c }
\hline
\hline
Season  & Region & Elev, $\alpha$ &Az speed & Turn around & Scan duration & Scan length   \\
   &  & deg  & deg/sec &  s  &   s & deg az   \\
\hline
s13 & D1/D5/D6 &35$^\circ$ & 1.5  & 0.9  &  30.6     &  21.6$^\circ$   \\
  & &60$^\circ$ & 1.5  & 0.9  &  39.7     &    28.4$^\circ$ \\
\hline
s14 & D56 &50$^\circ$    &   1.5   & 0.9   &  33.6 &  23.84$^\circ$     \\
       & D56 &60$^\circ$    & 1.5   & 0.9   & 54.5 &   39.5$^\circ$    \\
\hline
s15 &D56& 50$^\circ$  & 1.5   & 0.9   & 36.0 &  25.62$^\circ$     \\
 &D56& 60$^\circ$  & 1.5   & 0.9   & 57.1 &  41.5$^\circ$     \\
  &BN& 32.5$^\circ$  & 1.5   & 0.9   & 49.5 &  35.76$^\circ$    \\
   &BN& 35$^\circ$  & 1.5   & 0.9   & 51.9 &  37.53$^\circ$    \\
   &BN& 37.5$^\circ$  & 1.5   & 0.9   &  55.8  &  40.5$^\circ$ \\
    &D8& 35$^\circ$  & 1.5   & 0.9   & 29.4 &  20.7$^\circ$     \\
   &D8& 48.5$^\circ$  & 1.5   & 0.9   & 38.8&  27.74$^\circ$   \\
\hline
s16 & wide\_01h\_n & 40$^\circ$   & 1.5   & 0.9   &  87.8 & 64.5$^\circ$     \\
 & wide\_01h\_n & 45$^\circ$   & 1.5   & 0.9   &  106.5  & 78.52$^\circ$     \\
 & wide\_01h\_n & 47.5$^\circ$   & 1.5   & 0.9   & 126.6  & 93.62$^\circ$     \\
 & wide\_01h\_s & 40$^\circ$   & 1.5   & 0.9   & 73.8  & 54.0$^\circ$ \\
 & wide\_01h\_s & 45$^\circ$   & 1.5   & 0.9   & 81.0 & 59.4$^\circ$ \\
 & wide\_01h\_s & 47.5$^\circ$   & 1.5   & 0.9   & 87.0 & 63.92$^\circ$     \\
  &  wide\_12h\_n & 40.0$^\circ$   & 1.5   & 0.9   &61.1  & 44.5$^\circ$     \\
  &  wide\_12h\_n & 45$^\circ$   & 1.5   & 0.9   & 79.1& 57.96$^\circ$     \\
  &  wide\_12h\_n & 47.5$^\circ$   & 1.5   & 0.9   & 93.5& 68.78$^\circ$\\
\end{tabular}
\end{center}
\vspace{-0.1in} 
\small{The elevation angles are targets for the average position of the full array. Depending on the the number of optics tubes present, the telescope bore site, as opposed to the array centroid, is offset from the values above. The scan approximates a triangular wave in azimuth versus time.  The scan duration is the period of the waveform. The scan length is the peak-to-peak amplitude of the waveform. \citet{debernardis/2016} discuss the scan strategy optimization. In s13 several intermediate elevations were also used. In s16 the AA region was scanned at different elevations and scan lengths as denoted by ``wide\_x\_x." }
\label{tab:scans}
\end{table}

\section{B. Noise prescription}
\label{appen:noise_prescription}

\newcommand{\bn}{\boldsymbol{n}}
\newcommand{\bl}{\boldsymbol{\ell}}

Section~\ref{sec:noise_sim} gives an overview of the noise simulations. In the approximation described there, a noise model is built for each combination of season, patch and array. Each combination jointly models an $n$-component map where $n=3$ for the $I$, $Q$, $U$ components of any combination where the array is monochroic (PA1 and PA2) and where $n=6$ for the $I_1$, $Q_1$, $U_1$, $I_2$, $Q_2$, $U_2$ components of combinations with the dichroic PA3 array (with the index 1 corresponding to the \freqb\,GHz channel and the index 2 corresponding to the \freqa\,GHz channel). Correlations between arrays (e.g., between PA1 and PA2 in the same season and region) are negligible and ignored in this model. The noise model we use for our simulations consists of a Gaussian random field with some power spectrum that is subsequently modulated in real space by the inhomogeneous survey coverage in a given region of interest. The inhomogeneity is assumed to be that dictated by the inverse white noise variance $h(\bn)$ in each pixel; proportional to the number of observation hits in each pixel at sky location $\bn$, this quantity represents the inverse variance in the small-scale (high multipole) limit.  We simulate the noise maps as follows:

\begin{enumerate}
    \item For each component indexed by $1 \leq a \leq n$, subtract the coadd map $c^a(\bn)$ from each split $s^a_i(\bn)$ indexed by $1\leq i \leq k$ (this removes all signal including CMB and foregrounds, but also any common systematic like ground pickup) where $k$ is the number of splits ($k=2$ for AA and $k=4$ for all other regions.)
$$
d^a_i(\bn) = s^a_i(\bn) - c^a(\bn)
$$
$$
c^a(\bn) = \sum_{i=1}^k s^a_i(\bn) h^a_i(\bn) / \sum_{i=1}^k h^a_i(\bn),
$$
where $h^a_i(\bn)$ is the inverse variance in each pixel.
\item We use the above signal-nulled maps $d^a_i(\bn)$ to get an estimate of the noise power in the maps. We can build this noise model empirically from the data by calculating the power spectrum after standardizing the high-multipole behavior by multiplying the null map by its effective inverse standard deviation map. The resulting map will be homogeneous and on small scales have white noise with unit variance. To this end, we multiply each of the above maps by an edge-tapered mask $m_t(\bn)$ that selects the well-crosslinked region used in the power spectrum analysis, and by a standardizing weight map $w_i(\bn)$:
$$
m^a_i(\bn) = d^a_i(\bn)  w^a_i(\bn)  m_t(\bn),
$$
where the standardizing weight map is given by
$$
w^a_i(\bn) = \frac{1}{\sqrt{ 1/h^a_i(\bn) - 1/h^a_c(\bn)}},
$$

where $h^a_c(\bn) = \sum_{i=1}^k h^a_i(\bn)$. The above weight map corresponds to the inverse standard deviation map of the signal-nulled map $d^a_i(\bn)$. Since the resulting map is now homogeneous, we may simply calculate its power spectrum, generate realizations consistent with the power spectrum and subsequently divide out the weight map, as described next.
\item We calculate the 2D FFT of each of the above
$$
\tilde{m}^a_i(\bl) = \mathrm{FFT}(d^a_i(\bn)  w^a_i(\bn)  m_t(\bn)),
$$
where $\bl$ denotes 2D Fourier pixels.
\item Next we build the 2D Fourier space power spectra for the noise model of the coadd map in each direction by averaging over the noise powers obtained from each split
$$
P^{ab}(\bl) = \sum_{i=1}^k \tilde{m}_i^a(\bl) \tilde{m}_i^{b*}(\bl) / k / (k - 1) / w_2,
$$
where $w_2 = \langle m_t(\bn)^2 \rangle$ accounts for the loss in power due to the analysis mask and the $k-1$ factor converts the noise power of the difference maps to a noise power estimate of the coadd map under the assumption of uniform splits.
\item The power spectrum $P^{ab}(\bl)$ we obtained above is a noisy estimate of the true underlying covariance matrix $C^{ab}(\bl)$. We can obtain the latter by averaging or smoothing the power spectrum.  We smooth the noise power estimate $P^{ab}(\bl)$ by treating it as a 2D image. This ``image'' is smoothed by applying a low-pass filter that removes high-frequency modes in the Fourier transform of the 2D noise power spectrum. We choose the low-pass filter so that the 2D noise power is effectively block-averaged in blocks of of size $\Delta\ell \times \Delta\ell$, thus downgrading the power spectrum ``image''. We choose $\Delta\ell=200$ for all regions, except for BN and AA, where we use $\Delta\ell=100$. This procedure smooths the 2D Fourier power while preserving anisotropy. For auto-spectra and cross-spectra between intensity components (but not for cross-spectra involving polarization), before smoothing, the 2D power is whitened by fitting (in $\ell>500$) and dividing out a fit to the functional form $[(\frac{\ell_{\rm{knee}}}{\ell})^{-\alpha} + 1]  w^2$ where $w$ is the white noise floor determined at large $\ell$.  In addition, since auto-spectra are non-negative, we perform our smoothing in log-space and apply a correction for the fact that the smoothing is done in log-space.  Since the number of modes available is low for small angular wave-number magnitude $\ell$, for $\ell<300$ we replace the above smoothed power with an isotropic fit of the whitened power to the functional form $A e^{-\ell_0/\ell}+B$, with the whitening factor subsequently multiplied back in.
\item Gaussian random fields with a power spectrum $C^{ab}(\bl)$ can be obtained from the matrix $\sqrt{C^{ab}(\bl)}$. We calculate $\sqrt{C^{ab}(\bl)}$ by diagonalization, i.e. by raising each of its eigenvalues to 0.5. We save this matrix as steps 1--6 do not require repeating for new realizations of the noise.
\item For each requested simulation, random numbers are generated for each Fourier pixel by sampling the normal distribution with unit variance. When multiplied with $\sqrt{C^{ab}(\bl)}$, this returns the FFT of an $n$-component map $r_a(\bl)$ with the desired covariance across the components.
\item The Fourier map $r_a$ is then inverse Fourier transformed resulting in a homogeneous map. This is subsequently divided by the standardizing weight map $w^a_i(\bn)$ for each split (which re-introduces inhomogeneity), and scaled by the square root of the number of splits to obtain a noise simulation of each split of the data:
$$
S^a_i(\bn) = \rm{IFFT}(r_a(\bl) \sqrt{k}) / w^a_i(\bn).
$$
\end{enumerate}

\newpage
\section{C. Additional consistency checks and systematic error summary}
\label{appen:consistency_checks}

Table~\ref{tab:consistency_summary} shows results from the suite of the consistency checks. For the ``Intra-patch" tests, say for D56, there are five measurements at \freqb\,GHz for TT: s14-PA1, s14-PA2, and s15-PA1,2,3 as shown in Table~\ref{tab:obs}. From these five, there are ten differences, each with $n_{\ell,c}$ bins for a total of $\nu=10n_{\ell,c}$. For each measurement there are 500 noise simulations (Section~\ref{sec:noise_sim}) from which the PTE is determined. Since there is only one spectrum for D1 it is not part of these consistency checks. For all spectra except TT, both \freqa and \freqb\,GHz are used leading to the larger number of degrees of freedom for them. Because of the Poisson tail, checks were done only with the \freqb\,GHz TT spectra because there is only a single  \freqa\,GHz TT spectrum per region.

\begin{figure}[th!]
\centering
\includegraphics[width=0.33\columnwidth]{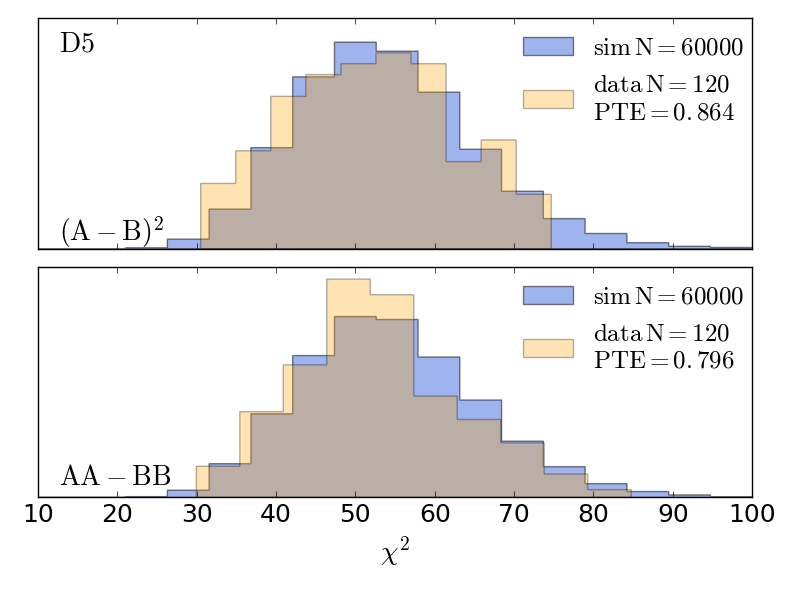}
\includegraphics[width=0.33\columnwidth]{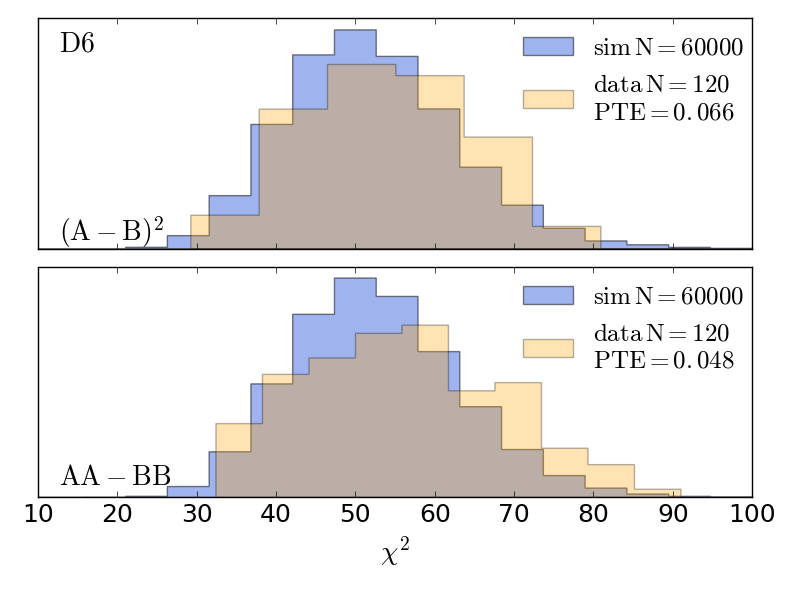}
\includegraphics[width=0.33\columnwidth]{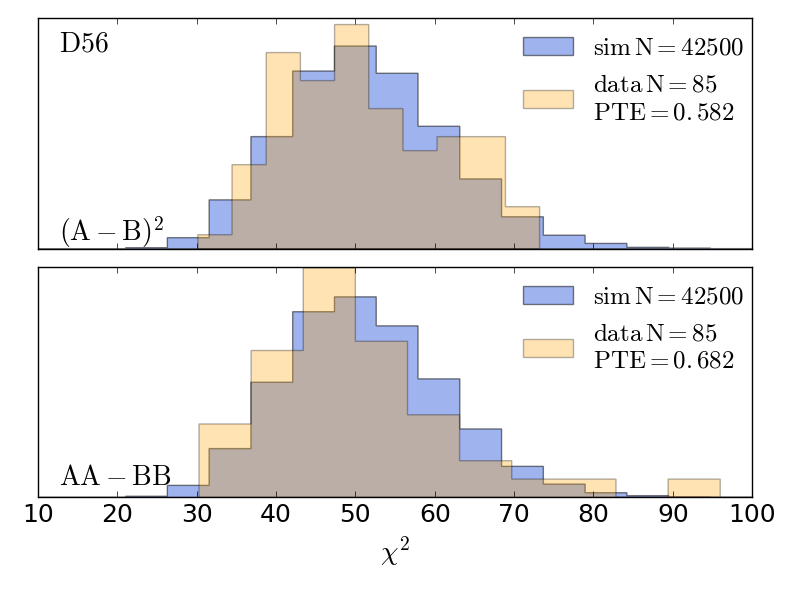}

\includegraphics[width=0.33\columnwidth]{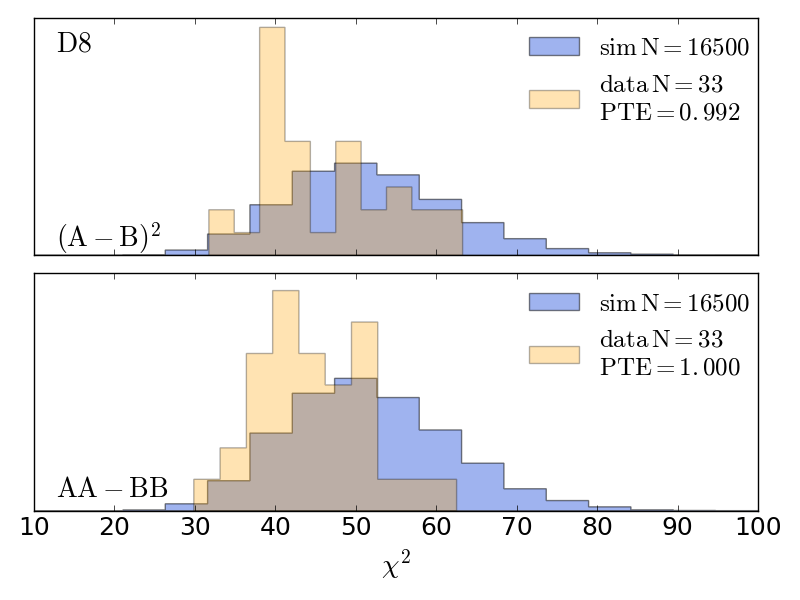}
\includegraphics[width=0.33\columnwidth]{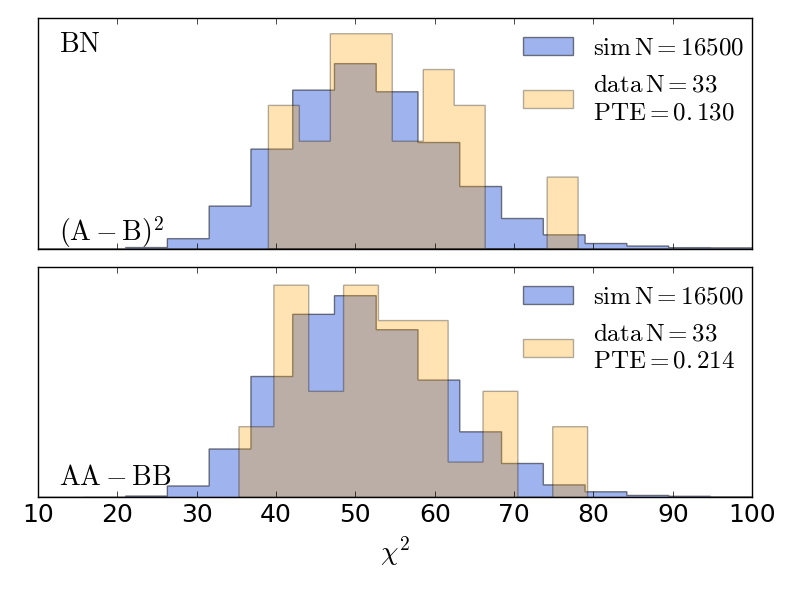}
\includegraphics[width=0.33\columnwidth]{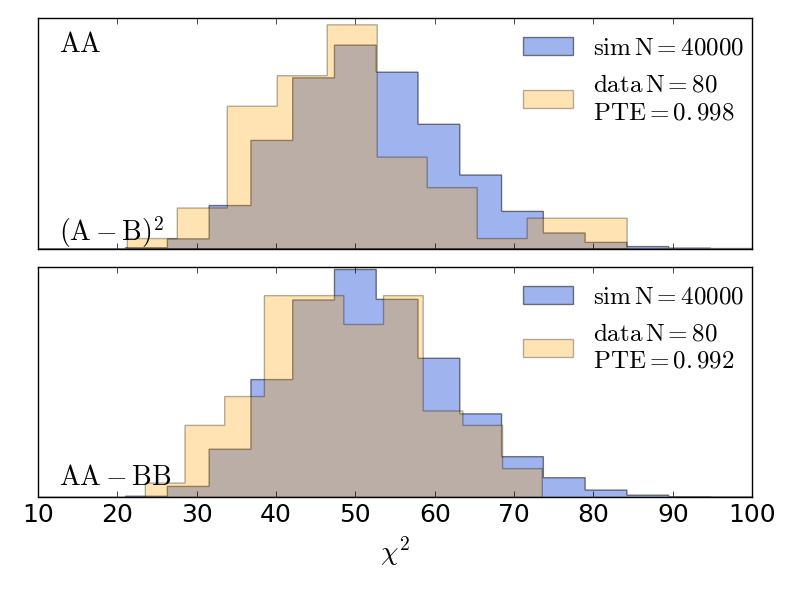}

\caption{Consistency checks for D5, D6, and D56 on the top two rows and D8, BN, and AA on the bottom two rows. The histogram integrals are normalized to unity. The PTE for the data is assessed with the simulations.}
\label{fig:consist-d5-d6-d56}
\end{figure}

A revealing way to assess the consistency is with histograms as shown in Figure~\ref{fig:consist-d5-d6-d56}. Again using D56 as an example (the two top rightmost plots) there are 85 difference spectra in total, 10 for TT as described above and 15 for TE, EE, TB, EB, and BB. Each has 500 simulations for a total of 42500. Based on the number of bins, we expect $\chi^2=n_{\ell,c}$, the central value of the histogram.  

\begin{table*}[htb]
\caption{Intra-patch (left) and Inter-patch (right) consistency tests}
\centering
\begin{adjustbox}{max width=\textwidth}
\begin{tabular}{c c c c c c ||c c c c}
\hline
\hline
Patch & Spectrum & $\chi^2_{(A-B)^2}$/$\nu$ $^a$ & PTE $^b$ & $\chi^2_{(AA-BB)}$/$\nu$ & PTE & Patch & Spectrum & $\chi^2_{(AA-BB)}$/$\nu$ & PTE \\
\hline
D5 & TT & 730/780 & 0.97 & 893/780 & 0.20 & D5$-$D1 & TT & 47/52 & 0.72 \\
& TE & 1170/1092 & 0.35 & 1228/1092 & 0.20 && TE & 64/52 & 0.17 \\ 
& EE & 1082/1092 & 0.79 & 1074/1092 & 0.79 & & EE & 77/52 & 0.03 \\ 
& BB & 1133/1092 & 0.34 & 1068/1092 & 0.79 & & BB & 44/52 & 0.81 \\ 
& EB & 1065/1092 & 0.78 & 1054/1092 & 0.79 & & EB & 38/52 & 0.94 \\ 
& TB & 1087/1092 & 0.71 & 1019/1092 & 0.87 & & TB & 52/52 & 0.56 \\ 
\hline
D6 & TT & 912/780 & 0.05 & 747/780 & 0.74 & D6$-$D1 & TT & 44/52 & 0.80 \\
& TE & 1134/1092 & 0.40 & 1171/1092 & 0.28 && TE & 43/52 & 0.82 \\ 
& EE & 1053/1092 & 0.78 & 1073/1092 & 0.64 && EE & 53/52 & 0.48 \\ 
& BB & 1142/1092 & 0.09 & 1361/1092 & 0.00 && BB & 59/52 & 0.32 \\ 
& EB & 1092/1092 & 0.40 & 1105/1092 & 0.43 && EB & 45/52 & 0.75 \\ 
& TB & 1197/1092 & 0.06 & 1273/1092 & 0.04 && TB & 35/52 & 0.97 \\ 
\hline
D56 & TT & 496/520 & 0.72 & 512/520 & 0.41 & D6$-$D5 & TT & 62/52 & 0.21 \\
& TE & 740/780 & 0.77 & 748/780 & 0.63 & & TE & 43/52 & 0.81 \\ 
& EE & 767/780 & 0.58 & 766/780 & 0.47 && EE & 78/52 & 0.02 \\ 
& BB & 840/780 & 0.05 & 769/780 & 0.54 && BB & 44/52 & 0.82 \\ 
& EB & 736/780 & 0.71 & 651/780 & 0.97 & & EB & 40/52 & 0.93 \\ 
& TB & 770/780 & 0.48 & 836/780 & 0.17 && TB & 60/52 & 0.31 \\ 
\hline
D8 & TT & 121/156 & 0.96 & 125/156 & 0.90 & D8$-$D56 & TT & 190/156 & 0.04 \\
& TE & 320/312 & 0.36 & 288/312 & 0.74 &  & TE & 184/208 & 0.88 \\ 
& EE & 260/312 & 0.95 & 258/312 & 0.91 & & EE & 153/156 & 0.54 \\ 
& BB & 250/312 & 0.99 & 293/312 & 0.69 & & BB & 148/156 & 0.66 \\ 
& EB & 272/312 & 0.91 & 271/312 & 0.86 && EB & 135/156 & 0.87 \\ 
& TB & 298/312 & 0.64 & 238/312 & 0.98 &  & TB & 168/208 & 0.98 \\ 
\hline
BN & TT & 144/156 & 0.69 & 147/156 & 0.54 & BN$-$D56 & TT & 177/156 & 0.12 \\
& TE & 311/312 & 0.47 & 336/312 & 0.21 & & TE & 204/208 & 0.55 \\ 
& EE & 309/312 & 0.50 & 298/312 & 0.55 &  & EE & 155/156 & 0.50 \\ 
& BB & 334/312 & 0.22 & 292/312 & 0.71 &  & BB & 169/156 & 0.22 \\ 
& EB & 359/312 & 0.07 & 350/312 & 0.15 & & EB & 182/156 & 0.07 \\ 
& TB & 331/312 & 0.23 & 339/312 & 0.22 &  & TB & 234/208 & 0.10 \\ 
\hline
AA & TT & 248/260 & 0.71 & 250/260 & 0.63 & BN$-$D8 & TT & 175/156 & 0.14 \\
w01345 & TE & 754/780 & 0.73 & 676/780 & 0.98 & & TE & 189/208 & 0.83 \\ 
& EE & 681/780 & 0.99 & 728/780 & 0.85 && EE & 182/156 & 0.09 \\ 
& BB & 768/780 & 0.60 & 825/780 & 0.18 && BB & 139/156 & 0.82 \\ 
& EB & 696/780 & 0.98 & 697/780 & 0.96 & & EB & 164/156 & 0.33 \\ 
& TB & 757/780 & 0.72 & 718/780 & 0.90 & & TB & 205/208 & 0.57 \\ 
\hline
\end{tabular}
\end{adjustbox}

{\small Notes: $a$) The number of degrees of freedom. $b$) Probability to exceed computed by comparing to simulations.}
\label{tab:consistency_summary}
\end{table*}

\newpage

\begin{figure}[htb!]
\centering
\includegraphics[width=0.5\textwidth]{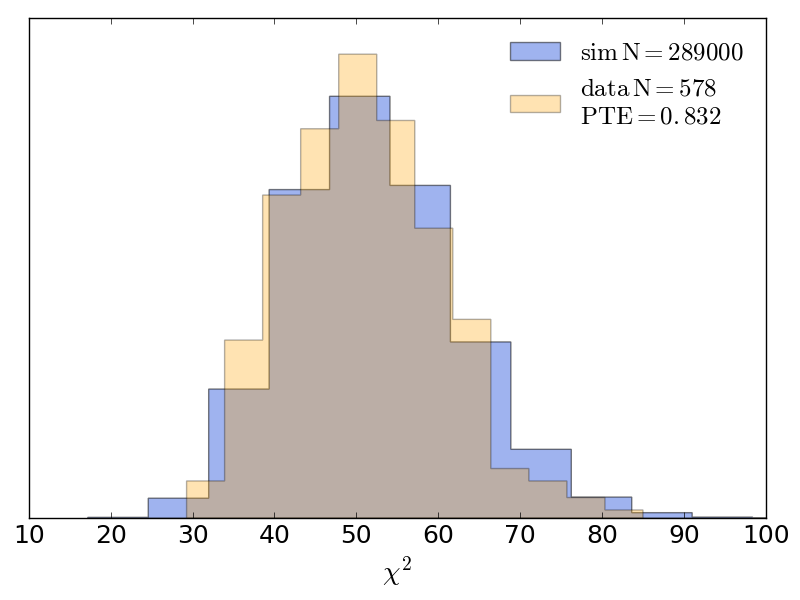}
\caption{Summary of pair-wise consistency checks for inter-patch comparisons for [D56, D8, BN, and AA] and [D1, D5, D6]. We keep the set [D1, D5, D6] separate as these are spectra at \freqb\,GHz only. The expectation is $\chi^2=n_{\ell,c}$, the number of $\ell$ bins. The PTE is computed with the simulations. The histogram integrals are normalized to unity. Some of the elements of this are shown in Table~\ref{tab:consistency_summary}.}
\label{fig:consist-interpatch}
\end{figure}

For the ``Inter-patch'' tests, the number of degrees of freedom were determined with similar considerations to the above. The $\chi^2$s of all inter-patch comparisons were computed and are shown in Figure~\ref{fig:consist-interpatch} of which Table~\ref{tab:consistency_summary} gives a sampling.

\begin{table*}[h]
\caption{Systematic errors not captured by internal consistency tests}
\vspace{-0.15in}
\begin{center}
\begin{tabular}{l l l l}\toprule
Effect & Estimation method & Level & Reference \\
\hline
Calibration & Cross correlation with {\sl Planck} 143 GHz maps & 1\% & Section~\ref{subsec:cal} \\
Beam leakage & Planet maps & $<0.35\%$ & Sections~\ref{sec:inst}~\&~\ref{sec:leak_cor}  \\
Polarization angle & Raytracing, pointing, EB null angle & $<-0.25^\circ$ & Section~\ref{subsec:pol_angle} \\
Mapmaker transfer function & Simulations & $<0.02\sigma$ & Section~\ref{sec:map_tfunc} \\
Detector time constant & Data split null test & To noise limit & Section~\ref{sec:tau_test} \\
Ground contamination & Data split null test & To noise limit & Section~\ref{sec:el_test} \\
PWV & Data split null test & To noise limit & Section~\ref{sec:pwv_test} \\
Galactic foregrounds & Cross correlation with {\sl Planck} and {\sl WMAP} & To noise limit (after subtraction) & Section~\ref{sec:fg_diff} \\
\hline
\end{tabular}
\end{center}
\label{tab:specs}
\vspace{-0.1in}
\end{table*}

\clearpage

\section{D. Foreground modeling}
\label{appen:fgcalcs}

The foreground modeling is done separately for the diffuse Galactic components (Section~\ref{sec:fg_diff}) and secondary anisotropies (Section~\ref{sec:likelihood_foregrounds}). However,
the results from fitting the Galactic components are input to the likelihood which determines the levels of secondary anisotropies. Below we give additional details on how these contributions are determined. 

\subsection{Diffuse Galactic components}

We minimize $\chi^2$ for the fit between maps A and B (e.g., ACT and {\sl Planck}) in $\ell$ space using Equations~\ref{eq:dust_null} and~\ref{eq:null_err2}. To simplify the expression we here set $a_{\rm CIB}=0$ but it is straightforward to add.
\begin{equation}
\chi^2 = \sum_\ell \bigg(\frac{(\mathcal{D}_\ell^{A} + \mathcal{D}_\ell^{B} - 2\mathcal{D}_\ell^{A\times B}) - a_{\rm dust}(\ell/500)^{\alpha+2}({\cal{F}}^d_{\rm{A}}g_1(\nu_{\rm A})\rm{^2}+ {\cal{F}}^d_{\rm{B}}g_1(\nu_{\rm B})\rm{^2} - 2\sqrt{{\cal{F}}^d_{\rm{A}}{\cal{F}}^d_{\rm{B}}}\,g_1(\nu_{\rm A})g_1(\nu_{\rm B}))}{(\Delta \mathcal{D}_\ell^{A-B})}\bigg)^2,
\end{equation}
We restrict the sum over $\ell$ to $600.5\le\ell\le2825.5$ for TT and $350.5\le\ell\le2825.5$ for TE/EE/BB. At high $\ell$, TT is dominated by extragalactic foregrounds that contaminate the fit. 

The diagonal elements of the covariance matrix for are given by 
\begin{equation} 
\label{eq:null_err2}
\begin{split}
(\Delta \mathcal{D}_\ell^{A-B})^2  &= \Big \langle \Big(\mathcal{D}_\ell^{AA} + \mathcal{D}_\ell^{BB} - 2\mathcal{D}_\ell^{AB} \Big)^2\Big \rangle - \Big \langle \mathcal{D}_\ell^{AA} + \mathcal{D}_\ell^{BB} - 2\mathcal{D}_\ell^{AB} \Big \rangle^2 \\
 &= \Theta_\ell^{(AA);(AA)} + \Theta_\ell^{(BB);(BB)} + 2 \Theta_\ell^{(AA);(BB)} \\
 &   \,\,\,\,\,\,\,\,\,\,\,\,\,\,\,\,\,\,\,\,\,\,\,\,\,\,\,\,\,\,\,\,\,\,\,\,\,\,\,\,\,\,\,       - 4 \Theta_\ell^{(AA);(AB)} - 4 \Theta_\ell^{(BB);(AB)} + 4 \Theta_\ell^{(AB);(AB)}, \\
\end{split}
\end{equation}
where the different terms are
\begin{equation}
\begin{split}
\Theta_\ell^{(AA);(AA)} &= \frac{1}{\nu_b}\Big[ 2(\mathcal{D}_\ell^{AA})^2 + 4 \frac{\mathcal{D}_\ell^{AA}}{n_s}N_b^{AA} + 2\frac{(N_b^{AA})^2}{n_d(n_d-1)} \Big] \\
\Theta_\ell^{(BB);(BB)} &= \frac{1}{\nu_b}\Big[ 2(\mathcal{D}_\ell^{BB})^2 + 4 \frac{\mathcal{D}_\ell^{BB}}{n_s}N_b^{BB} + 2\frac{(N_b^{BB})^2}{n_d(n_d-1)} \Big] \\
\Theta_\ell^{(AA);(BB)} &= \frac{1}{\nu_b}\Big[ 2(\mathcal{D}_\ell^{AB})^2 \Big] \\
\Theta_\ell^{(AA);(AB)} &= \frac{1}{\nu_b}\Big[ 2\mathcal{D}_\ell^{AA} \mathcal{D}_\ell^{AB} + 2\frac{N_b^{AA}}{n_d} \mathcal{D}_\ell^{AB} \Big] \\
\Theta_\ell^{(BB);(AB)} &= \frac{1}{\nu_b}\Big[ 2\mathcal{D}_\ell^{BB} \mathcal{D}_\ell^{AB} + 2\frac{N_b^{BB}}{n_d} \mathcal{D}_\ell^{AB} \Big] \\
\Theta_\ell^{(AB);(AB)} &= \frac{1}{\nu_b}\Big[ \mathcal{D}_\ell^{AA} \mathcal{D}_\ell^{BB} + \frac{\mathcal{D}_\ell^{AA} N_b^{BB}}{n_d} + \frac{\mathcal{D}_\ell^{BB} N_b^{AA}}{n_d} + \frac{N_b^{AA} N_b^{BB}}{n_d^2} + (\mathcal{D}_\ell^{AB})^2 \Big]. \\
\end{split}
\end{equation}
The effective number of modes in each bin $\nu_b$ is 
related to that in Equation~\ref{eq:1}, 
although here ``$f^{\mathrm{sky}}$'' 
is given by $f^\mathrm{sky} w_2^2/w_4,$ where $f^\mathrm{sky}$ is the ratio of the number of the nonzero pixels to the total number of pixels for full sky, and $w_i$ is the $i$th moment of the mask. 

To compute the uncertainty, we examine the distribution of fit results to simulations. In particular, we generate 1000 Gaussian spectra from the analytic covariance matrix, which includes pseudo-diagonal elements for correlations across different spectra. These uncertainties on the fit are reported in Table~\ref{tab:dust_per_region}.

For incorporating the diffuse emission into the likelihood, we coadd the fitted power law spectra with the ACT covariance matrix for ``deep" 15 mJy region (D56; D1, D5, D6 are small and do not add information) and the ``wide" 100 mJy region (BN, AA). These results are also reported in Table~\ref{tab:dust_per_region} and shown in Figure~\ref{fig:diff_dust}.

We perform similar fits for Galactic synchrotron emission using {\sl WMAP}'s K-band. One challenge is that K-band's beam window function cuts off at $\ell\sim200$ (for which its beam window function $b_\ell^2<0.2$), where the ACT spectra in this analysis begin. With ACT's baseline 2D Fourier-space filter, most of the low $\ell$ power is filtered out, hence for the synchrotron fit we do not Fourier-space filter the maps. Then, because the maps are unfiltered, we do not include ACT$\times$ACT spectra and use only $\mathcal{D}_\ell^{22} - \mathcal{D}_\ell^{22\times98}$ to estimate $a_{\rm sync}$. 
To compute $\mathcal{D}_\ell^{22}$, we use the {\sl WMAP} weight maps and point source masks, and additionally use the ACT  
weight maps and point source masks to compute $\mathcal{D}_\ell^{22\times98}$. The mean in each region is subtracted before computing the power spectrum and the fit range is restricted to 
$150.5\le\ell\le300.5$.
 
The synchrotron power law index varies over the sky and is steeper in polarization than in temperature
\citep[e.g.,][]{choi/2015}. To obtain a conservatively high estimate of synchrotron contamination, we use $\beta_s=-2.7$.
Table~\ref{tab:sync_per_region} gives the fit results for 
$a_{\rm sync}$ in CMB temperature units at 98\,GHz. There is no measurable synchrotron emission in any of the ACT regions. 

\begin{table*}[tp!]
\caption{Diffuse synchrotron emission by region}
\vspace{-0.15in}
\begin{center}
\begin{tabular}{c c c c c }
\hline
Region & Window & TT  & EE  & BB  \\
\hline\hline
AA & w0 & $17.6 \pm 89$ & $0.01 \pm 0.13$ & $0.01 \pm 0.03$ \\
AA & w1 & $7.6 \pm 98$ & $-0.01 \pm 0.13$ & $0.04 \pm 0.07$ \\
AA & w2 & $17.5 \pm 81$ & $0.01 \pm 0.26$ & $0.00 \pm 0.07$ \\
AA & w3 & $20.0 \pm 99$ & $0.01 \pm 0.49$ & $0.00 \pm 0.18$ \\
AA & w4 & $22.2 \pm 106$ & $0.02 \pm 0.33$ & $0.01 \pm 0.06$ \\
AA & w5 & $-2.8 \pm 443$ & $0.02 \pm 0.81$ & $0.06 \pm 0.13$ \\
BN &  & $1.7 \pm 25$ & $0.04 \pm 0.06$ & $0.00 \pm 0.02$ \\
D56 &  & $-0.6 \pm 51$ & $0.02 \pm 0.13$ & $-0.01 \pm 0.06$ \\
D8 & & $-2.9 \pm 99$ & $0.79 \pm 9.22$ & $-0.05 \pm 1.45$ \\
\hline
deep &  & $-0.5\pm 42$  & $0.02 \pm 0.10$ & $-0.01 \pm 0.05$ \\
wide &  & $8.4\pm 30.4$  & $0.03 \pm 0.05$ & $0.00 \pm 0.02$ \\
\hline
\end{tabular}
\end{center}
\label{tab:sync_per_region}
\vspace{-0.1in}
{\small The synchrotron power level in the ${\cal D}_\ell$ spectrum. All values are relative to the CMB and for a pivot scale of $\ell=500$ and frequency of 98 GHz. }
\end{table*}

Another approach to determining the level of synchrotron contamination is to compute its spectrum in K-band and then 
to simply scale it. Figure 10 in \cite{planck_int_dust} indicates that the contribution will be negligible.

\subsection{Effective frequencies}
As with fitting to any broad-band signal with an intrinsically broad-band instrument, some care must be taken with determining the effective representative frequency, $\nu_{\rm eff}$, for each sky component. Although the current uncertainties on the passbands due to systematic errors in the measurements are $\sigma_{\nu_{\rm eff}}=2.4$\,GHz, the relative uncertainties are smaller. Also, we expect the passband uncertainties to decrease with our increasing knowledge of the instrument. Table~\ref{tab:nu_eff} gives the effective frequencies for each component in this analysis, all of which are treated as beam-filling sources of emission.  The CMB blackbody effective frequencies are determined following the weighted-centroid approach of \citet{jarosik/etal:2003}.  The others are computed by solving the equation
\begin{equation}
\frac{1}{S(\nu_{\rm eff})} \int d\nu \, S(\nu) f(\nu) = g_2(\nu_{\rm eff}) \int d\nu \, g_2^{-1}(\nu) f(\nu) \,,
\label{eq.nu_eff}
\end{equation}
where $S(\nu)$ is the component SED in specific intensity units, $f(\nu)$ is the measured passband transmission, and $g_2(\nu)= (\partial B_\nu (T)/\partial T)^{-1}|_{T_{\rm CMB}}$ $=(2k_B^3T_{\rm CMB}^2x^4e^x/[hc(e^x-1)]^2)^{-1}$ with $x=h\nu/k_BT_{\rm CMB}$ converts flux to thermodynamic temperature. The effective frequencies are computed separately for each passband.  For the passbands centered near 150 GHz, the power spectra from all channels are coadded in the analysis, with separate weights for the deep and wide regions considered in the likelihood.  We apply the appropriate covariance matrices to a generalization of Equation~\ref{eq.nu_eff} to compute weighted effective frequencies after the coaddition.  These weights are slightly different for the deep and wide regions, as denoted by the ``d'' and ``w'' subscripts in Table~\ref{tab:nu_eff}.  We confirm that any scale dependence in the effective frequencies arising from the scale dependence of the coaddition weights is negligible ($<0.1$\% variation over the multipole range used in the likelihood).

\begin{table*}[htb]
\caption{Effective frequencies [GHz]}
\vspace{-0.15in}
\begin{center}
\begin{tabular}{c c c c c c c}
\hline
 & CMB & tSZ  & CIB  & dust & radio & sync  \\
\hline
$\nu_{\rm eff, d}$ & 97.9 & 98.4 & 98.8  & 98.6 & 95.8 & 95.5 \\
$\nu_{\rm eff, w}$ & 97.9 & 98.4 & 98.8  & 98.6 & 95.8 & 95.5 \\
$\nu_{\rm eff, d}$ & 149.7 & 150.1 & 151.2  & 151.1 & 147.2 & 147.1 \\
$\nu_{\rm eff, w}$ & 149.5 & 149.9 & 150.9  & 150.8 & 147.1 & 147.0 \\
\hline
\end{tabular}
\end{center}
\label{tab:nu_eff}
\vspace{-0.1in}
{\small The assumed SED parameters for the different broad-band sources are: CIB follows a modified blackbody SED with $\beta_c=2.1$ and effective dust temperature $T_c = 9.6$ K; Galactic dust follows a modified blackbody SED with $\beta_g=1.5$ and dust temperature $T_d = 19.6$ K as in Equation~\ref{eq:dust_model_1}; radio sources follow a power-law SED in specific intensity units with index $\beta_s=-0.5$; and synchrotron emission follows a power-law SED in specific intensity units with index $\beta=-1$.} 
\end{table*}

\subsection{Likelihood foreground model}

In temperature the CMB plus foreground model in the likelihood is given by 
${\cal D}_\ell^{{\rm th}, ij}={\cal D}_\ell^{\rm CMB}+{\cal D}_\ell^{{\rm sec}, ij}$ 
\citep{dunkley/etal:2013}:
\begin{equation}
{\cal D}_\ell^{{\rm sec},ij} = 
{\cal D}_\ell^{{\rm tSZ},ij}+{\cal D}_\ell^{{\rm kSZ},ij}
+{\cal D}_\ell^{{\rm CIB-P},ij} 
+{\cal D}_\ell^{{\rm CIB-C},ij} 
+{\cal D}_\ell^{{\rm tSZ-CIB},ij}
+{\cal D}_\ell^{{\rm rad},ij}
+{\cal D}_\ell^{{\rm Gal},ij}
\end{equation}
where the terms for the model of the secondary anisotropy are
thermal SZ effect, kinetic SZ effect, cosmic infrared background (CIB) Poission term, cosmic infrared background clustered term, the tSZ-CIB
correlated term, the radio source term, and the diffuse Galactic foreground term respectively, with $i$ and $j$ denoting different frequencies. These terms are quantified as 
\begin{eqnarray}
\label{eq:sec_model}
 {\cal D}_\ell^{{\rm sec}, ij} &=& A_{\rm tSZ}\frac{f(\nu_i)f(\nu_j)}{f^2(\nu_0)}{\cal D}_{0,\ell}^{\rm tSZ}
 +A_{\rm kSZ}{\cal D}_{0,\ell}^{\rm kSZ} 
 +A_{\rm d}\biggl(\frac{\ell}{\ell_0}\biggr)^2 \biggl[ \frac{\mu(\nu_i,\beta_p)\mu(\nu_j,\beta_p)}{\mu^2(\nu_0,\beta_p)}  \biggr]
+A_{\rm c}{\cal D}_{0,\ell}^{\rm CIBc}\ell^{\alpha_{\rm CIB}}\biggl[ \frac{\mu(\nu_i,\beta_c)\mu(\nu_j,\beta_c)}{\mu^2(\nu_0,\beta_c)}  \biggr]
\notag\\
 &-&\xi \sqrt{A_{\rm tSZ}A_{\rm c}}\frac{2f^\prime(\nu_{ij})}{f(\nu_{0})}{\cal D}_{0,\ell}^{\rm tSZ-CIB}
 +A_{\rm s,d\,or\,w}\biggl(\frac{\ell}{\ell_0}\biggr)^{2}\biggl(\frac{\nu_i\nu_j}{\nu_0^2}\biggr)^{\alpha_{ss}}\biggl[ \frac{g_2(\nu_i)g_2(\nu_j)}{g_2^2(\nu_0)} \biggr] \notag \\
&+&A_{\rm dust, d\,or\,w}\bigg(\frac{\ell}{500}\bigg)^{\alpha_d^{\rm TT}+2}\biggl[ \frac{\mu(\nu_i,\beta_g)\mu(\nu_j,\beta_g)}{\mu^2(\nu_0,\beta_g)}  \biggr]
\end{eqnarray}
where $f(\nu) = x \coth (x/2)-4$ with $x = h\nu/k_B T_{\rm CMB}$,  $\mu(\nu,\beta)=\nu^\beta B_\nu(T_{\rm d-eff})g_2(\nu)$ with $T_{\rm d-eff}=9.7\,$K for CIB and $T_{\rm d-eff}=19.6\,$K for Galactic dust, and $f^\prime(\nu_{ij}) = f(\nu_i)\mu(\nu_j,\beta_c) + f(\nu_j)\mu(\nu_i,\beta_c)$. The templates, ${\cal D}_{0,\ell}^{\rm tSZ}$,  ${\cal D}_{0,\ell}^{\rm kSZ}$, ${\cal D}_{0,\ell}^{\rm tSZ-CIB}$, ${\cal D}_{0,\ell}^{\rm CIBc}$ are 
normalized to unity at $\ell=3000$ as shown in \citet{dunkley/etal:2013}.
We take $\nu_0=150$\,GHz as the reference frequency, we fix $\alpha_{ss}=-0.5$ for the radio sources, and $\beta_p=\beta_c $ for the CIB dust index \citep{addison/etal:2012}. The CIB term is a hybrid of {\sl Planck} and \citet{addison/etal:2012} as discussed on page 19 of \cite{planck_int_cos:2016}. It follows the {\sl Planck} model below $\ell=3000$ and scales as $\ell^{0.8}$ for $\ell>3000$.
The subscript ``d or w" denotes the deep or wide regions. When the diffuse Galactic foregrounds are part of the fit we use the parameters given in Section~\ref{sec:fg_diff} as priors on the dust amplitudes. 

In polarization the model includes secondary emission as:
\begin{eqnarray}
{\cal D}_\ell^{{\rm sec}, ij} &=&A^{\rm TE/EE}_{\rm ps,d\,or\,w}\biggl(\frac{\ell}{\ell_0}\biggr)^{2}\biggl(\frac{\nu_i\nu_j}{\nu_0^2}\biggr)^{\alpha_{ss}}\biggl[ \frac{g_2(\nu_i)g_2(\nu_j)}{g_2^2(\nu_0)} \biggr] + A^{\rm TE/EE}_{\rm dust, d\,or\,w}\bigg(\frac{\ell}{500}\bigg)^{\alpha_d^{\rm TE,EE}+2}\biggl[ \frac{\mu(\nu_i,\beta_g)\mu(\nu_j,\beta_g)}{\mu^2(\nu_0,\beta_g)}  \biggr] \,.
\end{eqnarray} 
As with temperature we impose priors on the dust amplitudes in TE/EE as measured in Section~\ref{sec:fg_diff}. The fit of point sources in polarization is described in Section~\ref{sec:likelihood_foregrounds}.

\clearpage
\newpage
\section{E. Tables for the comparison to {\sl Planck}}
\label{appen:planck_compare}

This Appendix contains the tables that support the discussion on the consistency with {\sl Planck} in Section~\ref{sec:planck_compare}.

\begin{table*}[htb]
\caption{Difference spectra and comparisons to $\Lambda$CDM}
\vspace{-0.15in}
\begin{center}
\begin{tabular}{c c c c c c}
\hline
& \multicolumn{2}{c}{BN} & \multicolumn{2}{c}{D56} & BN$-$D56 \\
\hline
& ACT $\Lambda$CDM & {\sl Planck} $\Lambda$CDM &ACT $\Lambda$CDM & {\sl Planck} $\Lambda$CDM  \\
\hline
\hline
TT &  &  & & &  \\
AA & 0.133 (32.9) & 0.104 (34.2) & 0.028 (40.1) & 0.022 (41.1) & 0.057 (39.3) \\
AP & 0.292 (28.4) & 0.375 (26.6) & 0.009 (44.6) & 0.009 (44.6) & 0.043 (40.5) \\
AA$-$AP &  \multicolumn{2}{c}{0.020 (42.9)} &  \multicolumn{2}{c}{0.197 (31.4)}  & - \\
\hline
TE &  &  & & &  \\
AA & 0.875 (21.4) & 0.725 (25.0) & 0.611 (27.2) & 0.443 (30.4) & 0.133 (39.0) \\
AP & 0.678 (26.0) & 0.702 (25.5) & 0.013 (49.9)& 0.011 (50.7) & 0.043 (46.5) \\
PA & 0.054 (43.4) & 0.098 (40.4) & 0.902 (20.5) & 0.881 (21.2) & 0.477 (30.1)  \\
PP & 0.061 (42.8) & 0.137 (38.5) & 0.021 (47.7) & 0.014 (49.4) & 0.010 (47.6)  \\
AA$-$AP &  \multicolumn{2}{c}{0.160 (37.0) } &  \multicolumn{2}{c}{0.100 (41.7) }  & - \\
AA$-$PA &  \multicolumn{2}{c}{0.387 (32.1) } &  \multicolumn{2}{c}{0.000 (64.2) }  & - \\
\hline
EE &  &  & & &  \\
AA & 0.067 ( 42.4)& 0.006 (52.9) & 0.896 (20.7) & (41.8) 0.080 & 0.440 (31.2) \\
AP & 0.482 (29.7) & 0.447 (30.4) & 0.249 (34.8) & (38.5) 0.137 & 0.283 (34.1) \\
PP & 0.465  (30.0)& 0.498 (29.4) & 0.393 (31.5) & (31.2) 0.404 & 0.517 (29.2)  \\
AA$-$AP &  \multicolumn{2}{c}{0.663 (26.2) } &  \multicolumn{2}{c}{0.080 (41.2)}  & - \\
\hline
\end{tabular}
\end{center}
\label{tab:act_planck_chi2_pte}
\vspace{-0.05in}
{\small The comparison TT, TE, EE to the best fit ACT and {\sl Planck} $\Lambda$CDM models in the BN and D56 regions. For TT there are 25 degrees of freedom and for TE and EE there are 30. For each we report PTE ($\chi^2$). Here ``A" corresponds to ACT's \freqb\,GHz map, and ``P" to the {\sl Planck} 143 GHz map. While the comparisons to $\Lambda$CDM have, by necessity, sample variance, the difference between spectra within a region have minimal sample variance by construction. The rightmost column reports the difference between, for example, the ACT TT spectrum in the BN and D56 regions and includes sample variance.}
\end{table*}

\begin{table*}[htb]
\caption{Comparison between ACT and {\sl Planck} EE spectra within regions} 
\vspace{-0.15in}
\begin{center}
\begin{tabular}{c | c c c c c c}
\hline
D56 EE & 98$\times$98 & 98$\times${\bf P} & 98$\times$150 & {\bf P}$\times${\bf P} & {\bf P}$\times$150 & 150$\times$150 \\
\hline
98$\times$98 & - \\
98$\times${\bf P} & 0.380 (31.1) & - \\
98$\times$150 & 0.427 (30.4) & 0.290 (34.0) & - \\
{\bf P}$\times${\bf P} & 0.623 (27.2) & 0.517 (29.7) & 0.523 (29.4) & - \\
{\bf P}$\times$150 & 0.283 (34.2) & 0.083 (42.1) & 0.137 (38.9) & 0.613 (27.9) & - \\
150$\times$150 & 0.343 (32.0) & 0.183 (37.0) & 0.657 (26.1) & 0.520 (29.8) & 0.080 (41.2) & - \\
\hline
BN EE & 98$\times$98 & 98$\times${\bf P} & 98$\times$150 & {\bf P}$\times${\bf P} & {\bf P}$\times$150 & 150$\times$150 \\
\hline
98$\times$98 & - \\
98$\times${\bf P} & 0.753 (24.3) & - \\
98$\times$150 & 0.177 (35.9) & 0.367 (32.2) & - \\
{\bf P}$\times${\bf P} & 0.857 (21.5) & 0.660 (26.8) & 0.607 (27.6) & - \\
{\bf P}$\times$150 & 0.330 (33.3) & 0.240 (35.1) & 0.103 (40.2) & 0.593 (27.2) & - \\
150$\times$150 & 0.150 (37.4) & 0.493 (29.5) & 0.233 (35.6) & 0.597 (27.4) & 0.663 (26.2) & - \\
\hline
\end{tabular}
\end{center}
\label{tab:act_planck_ee_pte_matrix}
\vspace{-0.05in}
{\small Each entry shows the PTE ($\chi^2$) for the difference between the cross spectrum on the top row and the one in the column. Here ``98" corresponds to ACT's 98 GHz map, ``150" to the 150 GHz map, and ``{\bf P}" to the {\sl Planck} 143 GHz map. The number of degrees of freedom for all differences is 30. The PTEs are shown graphically in Figure~\ref{fig:TE_pte_hist}.} 
\end{table*}

\begin{table*}[tp!]
\caption{Comparison between ACT and {\sl Planck} TE spectra within regions} 
\vspace{-0.15in}
\begin{center}
\begin{tabular}{c | c c c c c c c c c}
\hline
D56 TE & 98$\times$98 & 98$\times${\bf P} & 98$\times$150 & {\bf P}$\times$98 & {\bf P}$\times${\bf P} & {\bf P}$\times$150 & 150$\times$98 & 150$\times${\bf P} & 150$\times$150 \\
\hline
98$\times$98 & - \\
98$\times${\bf P} & 0.103 (40.4) & - \\
98$\times$150 & 0.750 (24.4) & 0.080 (42.0) & - \\
{\bf P}$\times$98 & 0.010 (52.6) & 0.130 (39.8) & 0.413 (31.3) & - \\
{\bf P}$\times${\bf P} & 0.033 (46.8) & 0.037 (47.2) & 0.043 (45.7) & 0.033 (45.9) & - \\
{\bf P}$\times$150 & 0.437 (31.3) & 0.080 (41.5) & 0.380 (32.2) & 0.107 (39.5) & 0.067 (44.6) & - \\
150$\times$98 & 0.300 (33.6) & 0.103 (40.1) & 0.307 (34.1) & 0.017 (48.8) & 0.027 (47.9) & 0.017 (46.5) & - \\
150$\times${\bf P} & 0.043 (43.4) & 0.437 (30.2) & 0.030 (47.4) & 0.063 (42.8) & 0.163 (38.2) & 0.030 (46.7) & 0.080 (42.1) & - \\
150$\times$150 & 0.827 (22.3) & 0.160 (38.2) & 0.263 (34.4) & 0.467 (30.2) & 0.057 (45.1) & 0.000 (64.2) & 0.620 (27.2) & 0.100 (41.7) & - \\
\hline
BN TE & 98$\times$98 & 98$\times${\bf P} & 98$\times$150 & {\bf P}$\times$98 & {\bf P}$\times${\bf P} & {\bf P}$\times$150 & 150$\times$98 & 150$\times${\bf P} & 150$\times$150 \\
\hline
98$\times$98 & - \\
98$\times${\bf P} & 0.847 (22.3) & - \\
98$\times$150 & 0.693 (26.0) & 0.633 (26.8) & - \\
{\bf P}$\times$98 & 0.003 (64.1) & 0.853 (21.9) & 0.683 (25.2) & - \\
{\bf P}$\times${\bf P} & 0.537 (28.8) & 0.517 (28.3) & 0.247 (34.6) & 0.823 (22.9) & - \\
{\bf P}$\times$150 & 0.070 (42.7) & 0.437 (30.0) & 0.000 (66.3) & 0.520 (28.2) & 0.420 (31.1) & - \\
150$\times$98 & 0.623 (26.9) & 0.380 (31.4) & 0.360 (31.7) & 0.263 (34.4) & 0.220 (35.0) & 0.120 (39.5) & - \\
150$\times${\bf P} & 0.820 (22.6) & 0.097 (41.2) & 0.577 (28.0) & 0.877 (21.6) & 0.237 (35.2) & 0.313 (33.1) & 0.287 (33.6) & - \\
150$\times$150 & 0.467 (30.1) & 0.557 (27.8) & 0.050 (43.5) & 0.567 (28.1) & 0.117 (38.5) & 0.387 (32.1) & 0.423 (30.9) & 0.160 (37.0) & - \\
\hline
\end{tabular}
\end{center}
\label{tab:act_planck_te_pte_matrix}
\vspace{-0.05in}
{\small Each entry shows the PTE ($\chi^2$) for the difference between the cross spectrum on the top row and the one in the column. Here ``98" corresponds to ACT's \freqa\,GHz map, ``150" to the \freqb\,GHz map, and ``{\bf P}" to the {\sl Planck} 143 GHz map. All $\chi^2$ are computed for $350\leq\ell\leq1800$ (30 degrees of freedom) using the diagonal elements of the covariance matrix. The PTEs are shown graphically in Figure~\ref{fig:TE_pte_hist}. }
\end{table*}

\clearpage
\newpage
\section{F. Composite CMB spectra}
\label{appen:spectra}

\begin{table*}[h]
\caption{CMB only and Coadded TT, TE, and EE spectra}
\vspace{-0.15in}
\begin{center}
\begin{tabular}{cc|cc|cc|cc||cc|cc|cc}\toprule
$\ell$ & $\Delta\ell$ & ${\cal D}^{TT}_b$ & err &
${\cal D}^{TE}_b$ & err &
${\cal D}^{EE}_b$ & err &
${\cal D}^{TT}_b$ & err &
${\cal D}^{TE}_b$ & err &
${\cal D}^{EE}_b$ & err \\
 & & ($\mu$K)$^2$  & ($\mu$K)$^2$  & ($\mu$K)$^2$  & ($\mu$K)$^2$  & ($\mu$K)$^2$   & ($\mu$K)$^2$ &($\mu$K)$^2$& ($\mu$K)$^2$& ($\mu$K)$^2$ &($\mu$K)$^2$ &($\mu$K)$^2$ &($\mu$K)$^2$  \\
 \hline
350.5 & 50 & && $83.5$ & $7.4$ & $18.1$ & $0.8$ & 2057.0 & 126.1 & 86.9 & 6.5 & 17.5 & 0.7 \\
400.5 & 50 & && $-8.7$ & $6.2$ & $23.2$ & $0.8$ & 1641.5 & 90.2 & $-$7.0 & 5.6 & 22.1 & 0.8 \\
450.5 & 50 & && $-74.8$ & $6.0$ & $17.5$ & $0.7$ & 1828.6 & 80.3 & $-$69.7 & 4.9 & 16.4 & 0.6 \\
500.5 & 50 & && $-53.8$ & $5.2$ & $8.5$ & $0.4$ & 2197.5 & 82.9 & $-$52.0 & 4.0 & 8.2 & 0.4 \\
550.5 & 50 & && $2.5$ & $5.2$ & $8.3$ & $0.5$ & 2573.7 & 76.4 & 2.3 & 3.9 & 8.0 & 0.4 \\
600.5 & 50 &  $2239.9$ & $62.4$ & $29.4$ & $4.6$ & $19.0$ & $0.7$ & 2245.1 & 62.0 & 30.4 & 4.4 & 17.9 & 0.7 \\
650.5 & 50 &  $1865.7$ & $50.1$ & $-27.8$ & $4.9$ & $32.4$ & $1.0$ & 1876.6 & 50.0 & $-$26.0 & 4.8 & 30.1 & 0.9 \\
700.5 & 50 &  $1869.9$ & $47.4$ & $-99.9$ & $5.2$ & $39.1$ & $1.1$ & 1847.1 & 47.2 & $-$94.8 & 5.0 & 36.5 & 1.0 \\
750.5 & 50 &  $2260.8$ & $51.1$ & $-138.2$ & $4.8$ & $29.2$ & $0.9$ & 2232.6 & 51.0 & $-$130.3 & 4.7 & 27.2 & 0.8 \\
800.5 & 50 &  $2471.4$ & $52.6$ & $-91.5$ & $4.2$ & $16.5$ & $0.6$ & 2424.5 & 52.4 & $-$86.6 & 4.0 & 15.4 & 0.6 \\
850.5 & 50 &  $2405.1$ & $49.9$ & $3.3$ & $3.9$ & $13.0$ & $0.6$ & 2344.0 & 49.9 & 2.5 & 3.7 & 12.0 & 0.6 \\
900.5 & 50 &  $1891.7$ & $39.7$ & $57.5$ & $4.0$ & $25.4$ & $0.8$ & 1856.1 & 39.6 & 53.6 & 3.8 & 23.3 & 0.8 \\
950.5 & 50 &  $1330.7$ & $29.8$ & $38.5$ & $3.8$ & $39.7$ & $1.1$ & 1302.9 & 29.8 & 36.1 & 3.7 & 36.3 & 1.0 \\
1000.5 & 50 &  $1099.7$ & $23.6$ & $-29.2$ & $3.7$ & $43.3$ & $1.2$ & 1074.3 & 23.6 & $-$27.1 & 3.6 & 39.6 & 1.1 \\
1050.5 & 50 &  $1091.0$ & $23.9$ & $-72.9$ & $3.5$ & $32.0$ & $1.0$ & 1060.8 & 23.9 & $-$68.0 & 3.4 & 29.1 & 1.0 \\
1100.5 & 50 &  $1224.2$ & $23.7$ & $-73.4$ & $3.1$ & $19.6$ & $0.8$ & 1184.8 & 23.6 & $-$69.3 & 3.0 & 18.1 & 0.7 \\
1150.5 & 50 &  $1229.7$ & $23.3$ & $-26.8$ & $2.9$ & $13.3$ & $0.6$ & 1189.2 & 23.3 & $-$25.5 & 2.8 & 12.1 & 0.6 \\
1200.5 & 50 &  $1027.0$ & $20.3$ & $4.4$ & $2.9$ & $17.2$ & $0.8$ & 989.8 & 20.3 & 3.9 & 2.8 & 15.6 & 0.7 \\
1250.5 & 50 &  $841.9$ & $17.1$ & $-3.9$ & $2.9$ & $27.2$ & $0.9$ & 814.5 & 17.1 & $-$3.6 & 2.8 & 24.5 & 0.9 \\
1300.5 & 50 &  $725.5$ & $15.0$ & $-37.8$ & $2.7$ & $31.2$ & $1.0$ & 702.0 & 15.0 & $-$34.3 & 2.6 & 28.0 & 1.0 \\
1350.5 & 50 &  $772.4$ & $14.9$ & $-59.5$ & $2.7$ & $26.5$ & $0.9$ & 741.1 & 14.9 & $-$54.5 & 2.6 & 24.0 & 0.9 \\
1400.5 & 50 &  $828.6$ & $15.1$ & $-57.0$ & $2.6$ & $17.8$ & $0.8$ & 795.1 & 15.1 & $-$51.3 & 2.5 & 16.4 & 0.7 \\
1450.5 & 50 &  $817.2$ & $14.3$ & $-28.7$ & $2.3$ & $12.1$ & $0.7$ & 781.9 & 14.3 & $-$26.3 & 2.3 & 11.1 & 0.7 \\
1500.5 & 50 &  $689.8$ & $13.2$ & $-0.3$ & $2.3$ & $12.8$ & $0.7$ & 665.2 & 13.2 & $-$0.4 & 2.2 & 11.8 & 0.7 \\
1550.5 & 50 &  $530.7$ & $11.1$ & $5.2$ & $2.1$ & $17.9$ & $0.8$ & 512.7 & 11.1 & 4.8 & 2.1 & 16.3 & 0.8 \\
1600.5 & 50 &  $444.6$ & $9.2$ & $-10.4$ & $2.1$ & $20.5$ & $0.9$ & 434.6 & 9.2 & $-$9.6 & 2.0 & 18.7 & 0.9 \\
1650.5 & 50 &  $394.2$ & $8.6$ & $-27.0$ & $2.1$ & $19.1$ & $0.9$ & 388.2 & 8.6 & $-$24.7 & 2.0 & 17.2 & 0.8 \\
1700.5 & 50 &  $398.8$ & $8.3$ & $-31.4$ & $1.9$ & $13.8$ & $0.8$ & 388.4 & 8.3 & $-$28.3 & 1.9 & 12.7 & 0.8 \\
1750.5 & 50 &  $395.0$ & $8.3$ & $-21.3$ & $1.8$ & $9.2$ & $0.7$ & 383.5 & 8.3 & $-$19.4 & 1.7 & 8.7 & 0.7 \\
1800.5 & 50 &  $356.0$ & $7.9$ & $-10.8$ & $1.7$ & $8.1$ & $0.7$ & 351.5 & 7.9 & $-$10.0 & 1.7 & 7.5 & 0.6 \\
1850.5 & 50 &  $314.7$ & $7.0$ & $-3.9$ & $1.7$ & $9.5$ & $0.7$ & 314.5 & 7.0 & $-$3.4 & 1.6 & 8.8 & 0.7 \\
1900.5 & 50 &  $256.8$ & $6.1$ & $-12.0$ & $1.6$ & $12.6$ & $0.8$ & 259.3 & 6.1 & $-$11.1 & 1.5 & 11.6 & 0.7 \\
1950.5 & 50 &  $250.9$ & $6.0$ & $-20.6$ & $1.6$ & $10.5$ & $0.8$ & 256.6 & 5.9 & $-$18.8 & 1.5 & 9.6 & 0.8 \\
2000.5 & 50 &  $230.2$ & $5.7$ & $-17.7$ & $1.5$ & $8.2$ & $0.8$ & 236.5 & 5.7 & $-$15.5 & 1.5 & 7.4 & 0.7 \\
2075.5 & 100 &  $218.8$ & $3.8$ & $-11.7$ & $1.0$ & $5.9$ & $0.5$ & 226.1 & 3.8 & $-$10.4 & 1.0 & 5.6 & 0.5 \\
2175.5 & 100 &  $158.6$ & $3.2$ & $-3.4$ & $0.9$ & $4.9$ & $0.5$ & 170.9 & 3.1 & $-$3.1 & 0.9 & 4.5 & 0.5 \\
2275.5 & 100 &  $117.5$ & $2.7$ & $-8.2$ & $0.9$ & $5.8$ & $0.5$ & 134.6 & 2.6 & $-$7.6 & 0.8 & 5.3 & 0.5 \\
2375.5 & 100 &  $108.6$ & $2.5$ & $-7.7$ & $0.8$ & $3.5$ & $0.5$ & 126.4 & 2.4 & $-$7.2 & 0.8 & 3.3 & 0.5 \\
2475.5 & 100 &  $85.8$ & $2.3$ & $-3.4$ & $0.8$ & $2.6$ & $0.5$ & 105.9 & 2.2 & $-$3.3 & 0.8 & 2.4 & 0.5 \\
2625.5 & 200 &  $63.6$ & $1.4$ & $-3.9$ & $0.5$ & $2.0$ & $0.4$ & 85.6 & 1.3 & $-$3.6 & 0.5 & 1.9 & 0.4 \\
2825.5 & 200 &  $41.2$ & $1.2$ & $-1.2$ & $0.5$ & $2.1$ & $0.5$ & 66.8 & 1.1 & $-$1.1 & 0.5 & 2.0 & 0.4 \\
3025.5 & 200 &  $24.8$ & $1.1$ & $-2.1$ & $0.5$ & $0.4$ & $0.5$ & 51.9 & 1.0 & $-$2.0 & 0.5 & 0.4 & 0.5 \\
3325.5 & 400 &  $13.3$ & $0.8$ &  $0.0$ & $0.4$ & $-0.8$ & $0.4$ & 42.8 & 0.7 & $-$0.4 & 0.4 & $-$0.6 & 0.4 \\
3725.5 & 400 &  $4.0$ & $0.8$ & $-0.7$ & $0.4$ & $-0.2$ & $0.6$ & 37.4 & 0.7 & $-$0.8 & 0.4 & 0.2 & 0.6 \\
4125.5 & 400 &  $0.6$ & $0.8$ & $0.2$ & $0.5$ & $-0.2$ & $0.8$ & 38.3 & 0.7 & 0.4 & 0.5 & $-$0.5 & 0.7 \\
4525.5 & 400 & &&&&& & 42.2 & 0.9 & $-$0.0 & 0.6 & 0.6 & 0.9 \\
4925.5 & 400 & &&&&& & 47.5 & 1.0 & 1.0 & 0.7 & 0.5 & 1.2 \\
5325.5 & 400 & &&&&& & 54.6 & 1.2 & 0.5 & 0.9 & $-$2.3 & 1.5 \\
5725.5 & 400 & &&&&& & 57.8 & 1.5 & $-$1.6 & 1.2 & 0.3 & 1.9 \\
6125.5 & 400 & &&&&& & 65.8 & 1.8 & $-$2.2 & 1.4 & $-$1.6 & 2.5 \\
6725.5 & 800 & &&&&& & 74.4 & 1.7 & $-$3.0 & 1.4 & $-$1.8 & 2.5 \\
7525.5 & 800 & &&&&& & 92.2 & 2.6 & 3.0 & 2.3 & 3.2 & 3.9 \\
\hline
\end{tabular}
\end{center}
\label{tab:specs}
\vspace{-0.1in}
{\small The $\ell$ column refers to the band center and $\Delta\ell$ is the bin width. For example, the first bin is 326$\leq\ell\leq$375, and the second bin is 376$\leq\ell\leq$425. The next three columns are the ``CMB only" spectra from the likelihood. These may be used for cosmological analyses directly. The right three columns are
the fully coadded cosmological data (the inputs to the likelihood) that combine frequencies and combine deep and wide. They are shown for comparison to the CMB only spectra. Although they were combined using the full covariance matrix, no foreground emission was subtracted. For example, the coadded TT spectrum has different levels of extragalactic source contributions for each of the six spectra (including the cross spectra) that comprise the coadd as shown in Figure~\ref{fig:likelihood_input}. }
\end{table*}

\end{document}